\documentclass[11pt,journal,onecolumn]{IEEEtran}
\IEEEoverridecommandlockouts
\usepackage{amsfonts}
\usepackage{amssymb}
\usepackage{amscd}
\usepackage[dvips]{graphicx}
\usepackage{epsfig}
\usepackage{subfigure}
\usepackage[cmex10]{amsmath}
\usepackage{array}
\usepackage{eqparbox}
\usepackage{hyperref}

\newtheorem{theorem}{Theorem}[section]

\newtheorem{corollary}[theorem]{Corollary}

\newenvironment{proof}[1][Proof]{\begin{trivlist}
\item[\hskip \labelsep {\bfseries #1}]}{\end{trivlist}}

\newcommand{\qed}{\nobreak \ifvmode \relax \else
      \ifdim\lastskip<1.5em \hskip-\lastskip
      \hskip1.5em plus0em minus0.5em \fi \nobreak
      \vrule height0.75em width0.5em depth0.25em\fi}

\def\BibTeX{{\rm B\kern-.05em{\sc i\kern-.025em b}\kern-.08em
    T\kern-.1667em\lower.7ex\hbox{E}\kern-.125emX}}

\newcommand{\Hone}{\mathbf{H}_{1}}
\newcommand{\Htwo}{\mathbf{H}_{2}}
\newcommand{\Hthree}{\mathbf{H}_{3}}

\newcommand{\lb}{\left(}
\newcommand{\rb}{\right)}
\newcommand{\lc}{\left\{}
\newcommand{\rc}{\right\}}
\newcommand{\abs}[1]{\left\vert#1\right\vert}

\usepackage{newlfont}
\begin{document}
\title{Multiuser Cognitive Radio Networks:\\ An Information Theoretic Perspective}
\author{\IEEEauthorblockN{K. G. Nagananda, Parthajit Mohapatra, Chandra R. Murthy and Shalinee Kishore \footnote{K. G. Nagananda and Shalinee Kishore are with the Department of Electrical and Computer Engineering, Lehigh University, Bethlehem, PA, USA (e-mail: \{kgn209, skishore\}@lehigh.edu). Parthajit Mohapatra and Chandra R. Murthy are with the Department of Electrical Communication Engineering, Indian Institute of Science, Bangalore, India (e-mail: \{partha, cmurthy\}@ece.iisc.ernet.in)}}
}

\maketitle

\begin{abstract}
Achievable rate regions and outer bounds are derived for three-user interference channels where the transmitters cooperate in a unidirectional manner via a noncausal message-sharing mechanism. The three-user channel facilitates different ways of message-sharing between the primary and secondary (or cognitive) transmitters. Three natural extensions of unidirectional message-sharing from two users to three users are introduced: (i) Cumulative message sharing; (ii) primary-only message sharing; and (iii) cognitive-only message sharing. To emphasize the notion of interference management, channels are classified based on different rate-splitting strategies at the transmitters. Standard techniques, superposition coding and Gel'fand-Pinsker's binning principle, are employed to derive an achievable rate region for each of the cognitive interference channels. Simulation results for the Gaussian channel case are presented; they enable visual comparison of the achievable rate regions for different message-sharing schemes along with the outer bounds. These results also provide useful insights into the effect of rate-splitting at the transmitters, which aids in better interference management at the receivers.
\end{abstract}

\IEEEpeerreviewmaketitle

\section{Introduction}
Cognitive radios (CRs) \cite{refmitola} try to improve spectral efficiency by gathering and using knowledge of their Radio Frequency (RF) environment to adjust their transmission and reception parameters. An overview of the potential benefits offered by the CRs in physical layer research is provided in \cite{refhaykin}. In \cite{refgoldsmith1}, three main CR paradigms have been identified - underlay, overlay and interweave. In the underlay paradigm, CR users are allowed to operate only if their interference to noncognitive (or primary) users is below a certain threshold. While operating in the overlay paradigm, the CRs transmit their data simultaneously with the primary users but employ sophisticated techniques that maintain (or even improve) the performance of  primary users. In the interweave paradigm, the CRs sense unused frequency bands called \emph{spectrum holes} to communicate without disrupting primary transmissions. Of these, the information theoretic research has focused primarily on the overlay paradigm where CR transmitters cooperate using \emph{unidirectional} message sharing in a noncausal manner. Here, the \emph{cognitive} user gains access to messages and the corresponding codewords of the \emph{primary} user before transmission.  Although clairvoyant, such models are popular in establishing performance limits of cooperative multiuser channels.  Then, the primary and cognitive users simultaneously transmit their messages, but the encoding is performed in such a way that the primary user's achievable rates do not suffer. We present first a short survey of recent information theoretic work in this area, followed by a summary of our contributions.

\emph{Related work:} Besides identifying the three CR network paradigms mentioned above, \cite{refgoldsmith1} explored some of the fundamental capacity limits and associated transmission strategies for CR wireless networks. In \cite{refnatasha1}, \cite{refnatasha2}, Devroye et al defined the two-user \emph{genie-aided} CR channel and derived an achievable rate region by employing rate-splitting at both transmitters. The coding scheme comprised a combination of the scheme proposed by Han and Kobayashi for the interference channel \cite{refhan}, and one proposed by Gel'fand and Pinsker (GP) for coding over channels with random parameters \cite{refgelfand}. In \cite{refsriram1}, Wu et al introduced terms like \emph{dumb} and \emph{smart} antennas to refer to primary and cognitive senders, respectively. They employed a combination of GP and superposition coding \cite{refcover1} techniques, without resorting to rate-splitting, to come up with an achievable rate region for the two-user CR channel. In \cite{refjiang}, an achievable rate region for the two-user interference channel with degraded message sets was derived using a combination of superposition and GP coding techniques, where only the CR transmitter employs rate-splitting. In \cite{refpramod1}, Jovi\v{c}i\'{c} et al presented the Gaussian CR channel and derived capacity bounds/results for low and high interference regimes by employing \emph{dirty paper coding} \cite{refcosta}, and joint code design at the two transmitters and multiuser decoding at the primary receiver.

Other prominent information theoretic results in the area of CR networks are as follows.  Capacity bounds for two-user interference channels with cognitive and partially cognitive transmitters were reported in \cite{refmaric5} -  \nocite{refmaric4}\nocite{refmaric2}\nocite{refmaric3}\nocite{refmaric6}\cite{refmaric7}. In \cite{refgarg} - \nocite{refcao1}\nocite{refcao2}\nocite{refmaric8}\cite{refmaric9}, information theoretic results for interference channels with common information were derived. The sum-capacity of the Gaussian MIMO cognitive radio network was presented in \cite{refsriram2}, where the results applied to the single-antenna CR channel as well. Capacity scaling laws for CR networks were presented in \cite{refnatasha3}, while \cite{refnatasha4} considered achievable rates when the encoder non-causally knows different channel states. Multiple access channels with cooperation have been considered in \cite{refmaric10} - \nocite{refwigger2}\cite{refwigger1}. Furthermore, the algebraic structure of random binning schemes of \cite{refgelfand} and \cite{refcosta} have been studied in \cite{refzamir1} - \nocite{refzamir2} \nocite{referez1} \cite{refbennatan1}, paving the way for practical realization of channel codes for CR networks.

\emph{Our contribution:} With increasing interest in CR technology, one is motivated to consider a \emph{network} of CRs sharing the same channel with an incumbent primary user. In particular, how do the primary user and the network of CRs cooperate assuming the overlay network paradigm? In this paper, we consider the case of \emph{three-user} CR interference channels, where two (or one) CRs and one (or two) primary user communicate with three respective receivers. We consider three message sharing mechanisms between the senders, which are extensions of the two-user unidirectional message sharing paradigm to the three-user case. We term these three approaches (i) cumulative message sharing (CuMS); (ii) primary-only message sharing (PrMS); and (iii) cognitive-only message sharing (CoMS). To deal with interference in this three-user channel, we use rate-splitting, which was first reported in \cite{refhan}, to enlarge the achievable rate region for the classical two-user interference channel. The main idea behind rate-splitting is to encode part of the message at a possibly low rate, so that the unintended receiver can decode the interference caused to it by performing simultaneous decoding. To this end, we define five cognitive channel models, two each for CuMS and PrMS, and one for CoMS, with  different rate-splitting strategies. The types of message-sharing mechanisms and rate-splitting strategies will be made precise in the next section. We then employ the standard technique of combining GP's binning principle \cite{refgelfand} and superposition coding \cite{refcover1} to derive an achievable rate region for each of the five channels. As a result, we illustrate the generality of the techniques employed here, and provide useful insights into the rate regions and their characterization. Next, we specialize the achievable rate regions to the Gaussian channel; this enables comparisons of the different rate regions both analytically and through simulations. We also present simple corollaries that help enlarge the rate regions in the Gaussian case. Finally, we compare our achievable regions to some outer bounds, and thereby provide some insight into the optimality of the proposed coding scheme. Initial results of this work have appeared in \cite{refnanda1} -\nocite{refnanda2} \cite{refspcom}.

The outline of the paper is as follows. In Section II, we introduce the discrete memoryless channel models for CuMS, PrMS and CoMS, and review the notation used in the paper. We also present the probability distribution functions characterizing these channels. In Section III, we present the achievability theorem for the channel models and work out the details of the proof for one of the channel models. In Section IV, we consider the Gaussian channel model and construct the framework for numerical evaluation. We also state corollaries that enlarge the rate regions in the Gaussian case and derive some outer bounds. Simulation results and related discussions are presented in Section V. We conclude the paper in Section VI. The achievable rate region equations for the five discrete memoryless channels considered in this paper, the proof of the achievability theorem for one channel model and proofs of corollaries are relegated to the Appendix.

\section{Discrete Memoryless Channel Model and Preliminaries}
The three-user discrete memoryless cognitive interference channel is described by $\left(\mathcal{X}_1, \mathcal{X}_2, \mathcal{X}_3, \mathcal{P}, \mathcal{Y}_1, \mathcal{Y}_2, \mathcal{Y}_3\right)$. For $k=1,2,3$,
\begin{enumerate}
\item the senders and receivers are denoted by $\mathcal{S}_k$ and $\mathcal{R}_k$, respectively;
\item finite sets $\mathcal{X}_k$ and $\mathcal{Y}_k$ denote the channel input and output alphabets, respectively;
\item random variables $X_k \in \mathcal{X}_k$ and $Y_k \in \mathcal{Y}_k$ are the inputs and outputs of the channel respectively; and
\item $\mathcal{P}$ denotes the finite set of conditional probabilities $p\left(y_1, y_2, y_3|x_1, x_2, x_3\right)$, when $\left(x_1, x_2, x_3\right) \in \mathcal{X}_1 \times \mathcal{X}_2 \times \mathcal{X}_3$ are transmitted and $\left(y_1, y_2, y_3\right) \in \mathcal{Y}_1 \times \mathcal{Y}_2 \times \mathcal{Y}_3$ are obtained by the receivers.
\end{enumerate}
The channels are assumed to be memoryless. In the classical three-user interference channel, the messages at the senders are given by $m_k \in \mathcal{M}_k = \{1,\ldots,M_k\}$; $\mathcal{M}_k$ being a finite set with $M_k$ elements. The messages are assumed to be independently generated.

\subsection{Message-Sharing Mechanisms}
We describe now the message-sharing mechanisms considered in this paper.
\begin{enumerate}
\item In the case of cumulative message-sharing (CuMS), sender $\mathcal{S}_2$ has noncausal knowledge of the message $m_1$ and the corresponding codewords of the primary sender, $\mathcal{S}_1$. Sender $\mathcal{S}_3$ has noncausal knowledge of the message $m_1$ of the primary transmitter as well as the message $m_2$ of $\mathcal{S}_2$, and their respective codewords. A schematic of CuMS is shown in Fig. 1.
\item In the case of primary-only message-sharing (PrMS), senders $\mathcal{S}_2$ and $\mathcal{S}_3$ have noncausal knowledge of the message $m_1$ and the corresponding codewords of the primary sender, $\mathcal{S}_1$. There is no message-sharing mechanism between $\mathcal{S}_2$ and $\mathcal{S}_3$ themselves. See Fig. 2 for a channel schematic.
\item In the case of cognitive-only message-sharing (CoMS), sender $\mathcal{S}_3$ has noncausal knowledge of messages $m_1$ and $m_2$, and the corresponding codewords of senders, $\mathcal{S}_1$ and $\mathcal{S}_2$. There is no message-sharing mechanism between the $\mathcal{S}_1$ and $\mathcal{S}_2$. A channel schematic for CoMS is shown in Fig. 3.
\end{enumerate}

An $\lb M_1, M_2, M_3, n, P_e^{(n)}\rb$ code exists for these channels, if there exists the following encoding functions:
\begin{eqnarray}
\nonumber
f_1:~\mathcal{M}_1 \mapsto \mathcal{X}_1^n, & f'_1:~\mathcal{M}_1 \mapsto \mathcal{X}_1^n, & f''_1:~\mathcal{M}_1 \mapsto \mathcal{X}_1^n\\
\nonumber
f_2:~\mathcal{M}_1\times\mathcal{M}_2 \mapsto \mathcal{X}_2^n, & f'_2:~\mathcal{M}_1\times\mathcal{M}_2 \mapsto \mathcal{X}_2^n &  f''_2:~\mathcal{M}_2 \mapsto \mathcal{X}_2^n\\
\nonumber
f_3:~\mathcal{M}_1\times\mathcal{M}_2\times\mathcal{M}_3 \mapsto \mathcal{X}_3^n, & f'_3:~\mathcal{M}_1\times\mathcal{M}_3 \mapsto \mathcal{X}_3^n, & f''_3:~\mathcal{M}_1\times\mathcal{M}_2\times\mathcal{M}_3 \mapsto \mathcal{X}_3^n
\end{eqnarray}
and the following decoding functions, for $k=1,2,3$:
\begin{eqnarray}
\nonumber
g_k:~\mathcal{Y}_k^n \mapsto \mathcal{M}_k, & g'_k:~\mathcal{Y}_k^n \mapsto \mathcal{M}_k, & g''_k:~\mathcal{Y}_k^n \mapsto \mathcal{M}_k,
\end{eqnarray}
such that the decoding error probability $\max\lc P_{e,1}^{(n)}, P_{e,2}^{(n)}, P_{e,3}^{(n)}\rc ~\mbox{is}\leq P_e^{(n)}$. $P_{e,k}^{(n)}$ is the average probability of decoding error computed using:
\begin{eqnarray}
\nonumber
P_{e,k}^{(n)} = \frac{1}{M_1M_2M_3}\sum_{m_1,m_2,m_3}p\left[\hat{m}_k \neq m_k|\lb m_1, m_2, m_3\rb~ \mbox{sent}\right]; k = 1,2,3.
\end{eqnarray}
$f_k$ (or $g_k$) correspond to the encoders (or decoders) used by channels with CuMS, $f'_k$ (or $g'_k$) correspond to the encoders (or decoders) used by channels with PrMS and $f''_k$ (or $g''_k$) correspond to the encoders (or decoders) used by channels with CoMS.

We define two channels denoted $\mathcal{C}^t_{\text{CuMS}}$ , two channels denoted $\mathcal{C}^t_{\text{PrMS}}$ and one channel denoted $\mathcal{C}_{\text{CoMS}}$; $t=1,2$. A non-negative rate triple $(R_1, R_2, R_3)$ is achievable for each of the channels, if for any $0<P_e^{(n)}<1$ there exists a $\lb 2^{\lceil nR_1 \rceil}, 2^{\lceil nR_2\rceil}, 2^{\lceil nR_3\rceil}, n, P_e^{(n)}\rb$ code such that $P_e^{(n)} \rightarrow 0$ as $n \rightarrow \infty$. The capacity region for the channels is the closure of the set of all achievable rate triples $(R_1, R_2, R_3)$. A subset of the capacity region gives an achievable rate region.

\subsection{Rate-Splitting Strategies}
In \cite{refhan}, it has been shown that achievable rate region for the classical two-user interference channel can be enlarged by rate-splitting. Specifically, each transmitter encodes part of the message at a possibly low rate and constructs its codewords using superposition coding. This results in the unintended or non-pairing receiver being able to decode and cancel out the low rate\footnote{In the literature, this is typically called the ``public part'' of the message. Its rate could be large if the cross-channel gains are large.} sub-message from the interfering transmitter using simultaneous decoding, thereby enlarging the achievable rate region. This forms the motivation for employing rate-splitting, as an effective interference management mechanism. In the three-user scenario, however, many more rate-splitting strategies exist compared to the two-user case. For example, sender $\mathcal{S}_1$ can perform rate-splitting in one of the following four ways: (i) it can encode a part of its message such that both unintended receivers, $\mathcal{R}_2$ and $\mathcal{R}_3$, can decode the sub-message; (ii) encode a part of the message such that $\mathcal{R}_2$ can decode it but not $\mathcal{R}_3$; (iii) encode a part of the message such that $\mathcal{R}_3$ can decode it but not $\mathcal{R}_2$; and finally, (iv) encode in a manner such that the sub-message is not decodable at either $\mathcal{R}_2$ or $\mathcal{R}_3$ (i.e., decodable only at the $\mathcal{R}_1$). In this paper, we consider the following rate-splitting strategies:
\begin{enumerate}
\item In $\mathcal{C}^1_{\text{CuMS}}$ and $\mathcal{C}^1_{\text{PrMS}}$, the senders encode part of their respective messages at a rate such that it can be reliably decoded by all the receivers. The other part of the message will be encoded at a rate such that only the intended or pairing receiver can decode it.
\item In  $\mathcal{C}^2_{\text{CuMS}}$ and $\mathcal{C}^2_{\text{PrMS}}$, one part of the message is encoded such that only the intended receiver can decode it, while the other part is encoded at a rate such that it can only be decoded at the intended reciever and the receiver $\mathcal{R}_1$.
\item In $\mathcal{C}_{\text{CoMS}}$, sender $\mathcal{S}_3$ encodes one part of the message at a rate such that all receivers can decode it, while the other part is encoded at a rate such that it can only be decoded at its pairing receiver, $\mathcal{R}_3$. There is no rate-splitting at $\mathcal{S}_1$ and $\mathcal{S}_2$.
\end{enumerate}
Note that, regardless of the manner in which rate-splitting is performed, $\mathcal{R}_t$ should always be able to reliably decode the codewords from $\mathcal{S}_t$, $t = 1, 2, 3$.  

The notation for describing the achievable rates of these sub-messages and their respective description is tabulated in Table \ref{table1}. The decoding capabilities of receivers, resulting from rate-splitting at the transmitters, are summarized in Tables \ref{table2}, \ref{table3} and \ref{table4}.  We also introduce auxiliary random variables defined on finite sets and tabulate them in Table \ref{table5}. Depending on the rate-splitting strategy employed by the senders, only a subset of these sub-messages, their corresponding rates, and the corresponding auxiliary random variables will be used to derive an achievable rate region for each channel model.  Note that we do not consider the practical aspects of the underlying physical realization of such models. Also, the capacity region for a general CR channel still remains an open problem.

\subsection{Channel Modification}
Rate-splitting necessitates modification of  the channels $\mathcal{C}^t_{\text{CuMS}}$, $\mathcal{C}^t_{\text{PrMS}}$ and $\mathcal{C}_{\text{CoMS}}; t=1,2$. Here, we explicitly show the modification for one channel ($\mathcal{C}_{\text{CuMS}}^2$); the modification for the other channel models is similar. Referring to the rate-splitting strategy for the channel $\mathcal{C}_{\text{CuMS}}^2$, the messages at the three senders in the modified channel can be written as:
\begin{center}
\mbox{Sender 1}: $m_{11} \in \mathcal{M}_{11} = \{1,\ldots,M_{11}\}$,\\
\mbox{Sender 2}: $m_{21} \in \mathcal{M}_{21} = \{1,\ldots,M_{21}\}$, $m_{22} \in \mathcal{M}_{22} = \{1,\ldots,M_{22}\}$, \\
\mbox{Sender 3}: $m_{31} \in \mathcal{M}_{31} = \{1,\ldots,M_{31}\}$, $m_{33} \in \mathcal{M}_{33} = \{1,\ldots,M_{33}\}$,
\end{center}
with all messages being defined on sets with finite number of elements. Note that, there is no rate-splitting at sender $\mathcal{S}_1$, but for consistency in notation we write $m_1$ as $m_{11}$.

We define an $\lb M_{11}, M_{21}, M_{22}, M_{31}, M_{33}, n, P_e^{(n)}\rb$ code for the modified channel as a set of $M_{11}$ codewords for $\mathcal{S}_1$, $M_{11}M_{21}M_{22}$ codewords for $\mathcal{S}_2$, and $M_{11}M_{21}M_{22}M_{31}M_{33}$ codewords for $\mathcal{S}_3$, such that the average probability of decoding error is less than $P_e^{(n)}$. We call a tuple $(R_{11}, R_{21}, R_{22}, R_{31}, R_{33})$ achievable if there exists a sequence of $\lb 2^{\lceil nR_{11}\rceil}, 2^{\lceil nR_{21}\rceil}, 2^{\lceil nR_{22}\rceil}, 2^{\lceil nR_{31}\rceil}, 2^{\lceil nR_{33}\rceil}, n, P_e^{(n)}\rb$ codes such that $P_e^{(n)} \rightarrow 0$ as $n \rightarrow \infty$. Here, $R_{11}$ corresponds to $R_1$. The capacity region for the modified channel is the closure of the set of all achievable rate tuples $(R_{11}, R_{21}, R_{22}, R_{31}, R_{33})$. It can be shown that if the rate tuple $(R_{11}, R_{21}, R_{22}, R_{31}, R_{33})$ is achievable for the modified channel, then the rate triple $(R_{11}, R_{21}+R_{22}, R_{31}+R_{33})$ is achievable for the channel $\mathcal{C}_{\text{CuMS}}^2$ (see \cite[Corollary 2.1]{refhan}). In a similar fashion, the remaining channel models can be appropriately modified; the details are omitted to avoid repetition.

\subsection{Probability Distributions}
Here, we present the probability distribution functions which characterize the channels $\mathcal{C}_{\text{CuMS}}^1$, $\mathcal{C}_{\text{CuMS}}^2$, $\mathcal{C}_{\text{PrMS}}^1$, $\mathcal{C}_{\text{PrMS}}^2$ and $\mathcal{C}_{\text{CoMS}}$. Let $\mathcal{P}^t_{\text{CuMS}}$ denote the set of all joint probability distributions $p^t_{\text{CuMS}}(.);~t=1,2$ respectively, that factor as follows:
\begin{eqnarray}
\nonumber p^1_{\text{CuMS}}(q, w_0, w_1, x_1, u_0, u_2, x_2, v_0, v_3, x_3, y_1, y_2, y_3) =\\ \nonumber  p(q)p(w_0,w_1,x_1|q)p(u_0|w_0,w_1,q)p(u_2|w_0,w_1,q) p(x_2|u_0,u_2,w_0,w_1,q)p(v_0|u_0,u_2,w_0,w_1,q)\\ p(v_3|u_0,u_2,w_0,w_1,q)p(x_3|v_0,v_3,u_0,u_2,w_0,w_1,q) p(y_1,y_2,y_3|x_1,x_2,x_3),\label{eq1} \\ \nonumber
\\
\nonumber p_{\text{CuMS}}^2(q, w, x_1, u_1, u_2, x_2, v_1, v_3, x_3, y_1, y_2, y_3)=\\ \nonumber p(q)p(w,x_1|q)p(u_1|w,q)p(u_2|w,q)p(x_2|u_1,u_2,w,q)p(v_1|u_1,u_2,w,q) p(v_3|u_1,u_2,w,q) \\  p(x_3|v_1,v_3,u_1,u_2,w,q)p(y_1,y_2,y_3|x_1,x_2,x_3).\label{eq2}
\end{eqnarray}

Let $\mathcal{P}^t_{\text{PrMS}}$ denote the set of all joint probability distributions $p^t_{\text{PrMS}}(.);~t=1,2$ respectively, that factor as follows:
\begin{eqnarray}
\nonumber
p^1_{\text{PrMS}}(q, w_0, w_1, x_1, u_0, u_2, x_2, v_0, v_3, x_3, y_1, y_2, y_3) =\\ \nonumber  p(q)p(w_0,w_1,x_1|q)p(u_0|w_0,w_1,q)p(u_2|w_0,w_1,q)\\  p(x_2|u_0,u_2,w_0,w_1,q)p(v_0|w_0,w_1,q) p(v_3|w_0,w_1,q) p(x_3|v_0,v_3,w_0,w_1,q)p(y_1,y_2,y_3|x_1,x_2,x_3), \label{eq3} \\\nonumber \\
\nonumber
p_{\text{PrMS}}^2(q, w, x_1, u_1, u_2, x_2, v_1, v_3, x_3, y_1, y_2, y_3)=\\ \nonumber p(q)p(w,x_1|q)p(u_1|w,q)p(u_2|w,q)\\p(x_2|u_1,u_2,w,q)p(v_1|w,q) p(v_3|w,q) p(x_3|v_1,v_3,w,q)p(y_1,y_2,y_3|x_1,x_2,x_3).\label{eq4}
\end{eqnarray}

Let $\mathcal{P}_{\text{CoMS}}$ denote the set of all joint probability distributions $p_{\text{CoMS}}(.)$ respectively, that factor as follows:
\begin{eqnarray}
\nonumber
p_{\text{CoMS}}(q, w_1, x_1, u_2, x_2, v_0, v_3, x_3, y_1, y_2, y_3) = p(q)p(w_1,x_1|q) p(u_2, x_2|q)p(v_0|w_1, u_2, q)p(v_3|w_1, u_2, q)\\ p(x_3|v_0,v_3,u_2,w_1,q) p(y_1,y_2,y_3|x_1,x_2,x_3).
\end{eqnarray}

The lower case letters ($q,w,u_2,v_3,$ etc.) are realizations of their corresponding random variables, and note that for notational simplicity, the same letter ($p$) is used to denote all the different probability distributions above. An achievable rate region for each channel is defined by a set of non-negative real numbers (referred to as rate tuples) that satisfy certain information-theoretic inequalities. An achievable rate region for each of the channels considered in this paper are given in Appendices A, B and C.

\section{Achievability Theorem and Proof} \label{sec:achievability}
\begin{theorem}\label{theorem1}
Let $\mathfrak{C}^t_{\text{CuMS}} (\mbox{or}~ \mathfrak{C}^t_{\text{PrMS}}~\mbox{or}~ \mathfrak{C}_{\text{CoMS}})$ denote the capacity region of the channel $\mathcal{C}_{\text{CuMS}}^t  (\mbox{or} ~\mathcal{C}_{\text{PrMS}}^t~\mbox{or} ~\mathcal{C}_{\text{CoMS}});~t=1,2$.  Let
\begin{eqnarray}
\nonumber \mathfrak{R}_{\text{CuMS}}^t = \!\!\!\!\!\! \bigcup_{p_{\text{CuMS}}^t(.)\in \mathcal{P}_{\text{CuMS}}^t}\!\!\!\!\!\! \mathfrak{R}_{\text{CuMS}}(p^t_{\text{CuMS}}), \mathfrak{R}_{\text{PrMS}}^t = \!\!\!\!\!\! \bigcup_{p_{\text{PrMS}}^t(.)\in \mathcal{P}_{\text{PrMS}}^t} \!\!\!\!\!\! \mathfrak{R}_{\text{PrMS}}(p^t_{\text{PrMS}})~\mbox{and}~\mathfrak{R}_{\text{CoMS}} = \!\!\!\!\!\! \bigcup_{p_{\text{CoMS}}(.)\in \mathcal{P}_{\text{CoMS}}} \!\!\!\!\!\! \mathfrak{R}_{\text{CoMS}}(p_{\text{CoMS}}).
\end{eqnarray}
In the above, $\mathfrak{R}_{\text{CuMS}}(p^t_{\text{CuMS}})$ denotes a set of achievable rates when the channel is characterized by the joint probability distribution function $p^t_{\text{CuMS}}$, and similar definitions apply for the other notations used. The region $\mathfrak{R}_{\text{CuMS}}^t(\mbox{or}~\mathfrak{R}_{\text{PrMS}}^t~\mbox{or}~\mathfrak{R}_{\text{CoMS}})$ is an achievable rate region for the channel $\mathcal{C}^t_{\text{CuMS}}(\mbox{or} ~\mathcal{C}_{\text{PrMS}}^t~\mbox{or} ~\mathcal{C}_{\text{CoMS}})$, i.e., $\mathfrak{R}_{\text{CuMS}}^t(\mbox{or}$ $\mathfrak{R}_{\text{PrMS}}^t$ $\mbox{or}$ $\mathfrak{R}_{\text{CoMS}}) \subseteq$ $\mathfrak{C}^t_{\text{CuMS}}$ $(\mbox{or}$ $\mathfrak{C}^t_{\text{PrMS}}$ $\mbox{or}$ $\mathfrak{C}_{\text{CoMS}}).$
\end{theorem}

\begin{proof}
We employ the standard technique of combining GP's binning principle \cite{refgelfand} and superposition coding \cite{refcover1} to prove the coding theorem and derive a set of achievable rates for each of the channel models. We show the proof for the channels $\mathcal{C}_{\text{CuMS}}^2$ and $\mathcal{C}_{\text{PrMS}}^1$. The proof for the remaining three channels ($\mathcal{C}_{\text{CuMS}}^1$ , $\mathcal{C}_{\text{PrMS}}^2$ and $\mathcal{C}_{\text{CoMS}}$) are along similar lines and are omitted.\\
\\
Proof of achievability for the channel $\mathcal{C}_{\text{CuMS}}^2$:
The proof is presented in four parts, namely, codebook generation, encoding, decoding and analysis of probabilities of decoding errors at the three receivers. We start with the codebook generation scheme.

\subsection{Codebook Generation}Let us fix $p(.) \in \mathcal{P}^2_{\text{CuMS}}$. Generate a random time sharing codeword $\textbf{q}$, of length $n$,  according to the distribution $\prod_{i=1}^np(q_i)$. Generate $2^{nR_{11}}$ independent codewords $\textbf{W}(j)$, according to $\prod_{i=1}^np(w_i|q_i)$. For every $\textbf{w}(j)$, generate one codeword $\textbf{X}_1(j)$ according to $\prod_{i=1}^np(x_{1i}|w_i(j),q_i)$.

For $\tau=1,2$, generate $2^{n(R_{2\tau}+I(W;U_{\tau}|Q)+4\epsilon)}$ independent codewords $\textbf{U}_{\tau}(l_{\tau})$, according to $\prod_{i=1}^np(u_{\tau i}|q_i)$. For every codeword$\!$ triple $\left[\textbf{u}_1(l_1), \textbf{u}_2(l_2), \textbf{w}(j)\right]$, generate one codeword $\textbf{X}_2(l_1, l_2, j)$ according to \\$\prod_{i=1}^np(x_{2i}|u_{1i}(l_1), u_{2i}(l_2), w_i(j), q_i)$. Uniformly distribute the $2^{n(R_{2\tau}+I(W;U_{\tau}|Q)+4\epsilon)}$ codewords $\textbf{U}_{\tau}(l_{\tau})$ into $2^{nR_{2\tau}}$ bins indexed by $k_{\tau} \in \left\{1,\ldots,2^{nR_{2\tau}}\right\}$ such that each bin contains $2^{n(I(W;U_{\tau}|Q)+4\epsilon)}$ codewords.

For $\rho=1,3$, generate $2^{n(R_{3\rho}+ I(W,U_1,U_2;V_{\rho}|Q)+4\epsilon)}$ independent codewords $\textbf{V}_{\rho}(t_{\rho})$, according to $\prod_{i=1}^np(v_{\rho i}|q_i)$. For every codeword quadruple $\left[\textbf{v}_1(t_1), \textbf{v}_3(t_3), \textbf{u}_1(l_1), \textbf{u}_2(l_2), \textbf{w}(j)\right]$, generate one codeword $\textbf{X}_3(t_1, t_3, l_1, l_2, j)$ according to $\prod_{i=1}^np\lb x_{3i}|v_{1i}(t_1), v_{3i}(t_3), u_{1i}(l_1), u_{2i}(l_2), w_i(j), q_i\rb$. Distribute $2^{n(R_{3\rho}+ I(W,U_1,U_2;V_\rho|Q)+4\epsilon)}$ codewords $\textbf{V}_{\rho}(t_{\rho})$ uniformly into $2^{nR_{3\rho}}$ bins indexed by $r_{\rho} \in \!\! \left\{1,\ldots,2^{nR_{3\rho}}\right\}$ such that each bin contains $2^{n(I(W,U_1,U_2;V_{\rho}|Q)+4\epsilon)}$ codewords. The indices are given by $j \in \left\{1,\ldots,2^{nR_{11}}\right\}$, $l_{\tau} \in  \left\{1,\ldots,2^{n(R_{22}+I(W;U_{\tau}|Q)+4\epsilon)}\right\}$ and $t_{\rho} \in\{1,\ldots,$ $2^{n(R_{33}+I(W,U_1,U_2;V_{\rho}|Q)+4\epsilon)}\}$.

\subsection{Encoding $\&$ Transmission}
Let $A_{\epsilon}^{(n)}$ be a typical set. We will be using the notation $A_{\epsilon}^{(n)}$ to describe a typical set over many different random variables, but the definition will be clear from the context.

Let us suppose that the source message vector generated at the three senders is $(m_{11}, m_{21}, m_{22}, m_{31}, m_{33}) = (j, k_1, k_2, r_1, r_3)$. At the encoders, the first component is treated as the message index and the last four components are treated as the bin indices. $\mathcal{S}_2$ looks for a codeword $\textbf{u}_1(l_1)$ in bin $k_1$ and a codeword $\textbf{u}_2(l_2)$ in bin $k_2$ such that $(\textbf{u}_1(l_1), \textbf{w}(j), \textbf{q}) \in A_{\epsilon}^{(n)}$ and $(\textbf{u}_2(l_2), \textbf{w}(j), \textbf{q}) \in A_{\epsilon}^{(n)}$, respectively. $\mathcal{S}_3$ looks for a codeword $\textbf{v}_1(t_1)$ in bin $r_1$ and a codeword $\textbf{v}_3(t_3)$ in bin $r_3$ such that $(\textbf{v}_1(t_1), \textbf{u}_1(l_1), \textbf{u}_2(l_2), \textbf{w}(j), \textbf{q}) \in A_{\epsilon}^{(n)}$ and $(\textbf{v}_3(t_3), \textbf{u}_1(l_1), \textbf{u}_2(l_2), \textbf{w}(j), \textbf{q}) \in A_{\epsilon}^{(n)}$, respectively. $\mathcal{S}_1$, $\mathcal{S}_2$ and $\mathcal{S}_3$ then transmit codewords $\textbf{x}_1(j)$, $\textbf{x}_2(l_1, l_2, j)$ and $\textbf{x}_3(t_1, t_3, l_1, l_2, j)$, respectively, through $n$ channel uses. The transmissions are assumed to be synchronized.

\subsection{Decoding}
Recall that in $\mathcal{C}_{cms}^2$, the primary receiver can decode the public parts of the non-pairing sender's messages, while the secondary receivers can only decode the messages from their pairing transmitters. The three receivers accumulate an $n$-length channel output sequence: $\textbf{y}_1$ at $\mathcal{R}_1$, $\textbf{y}_2$ at $\mathcal{R}_2$ and $\textbf{y}_3$ at $\mathcal{R}_3$. Decoders 1, 2 and 3 look for all indices $(\hat{j}, \hat{\hat{l}}_1, \hat{\hat{t}}_1)$, $(\hat{l}_1, \hat{l}_2)$ and $(\hat{t}_1, \hat{t}_3)$, respectively, such that $(\textbf{w}(\hat{j}), \textbf{u}_1(l_1), \textbf{v}_1(t_1), \textbf{y}_1, \textbf{q}) \in A_{\epsilon}^{(n)}$, $(\textbf{u}_1(\hat{l}_1), \textbf{u}_2(\hat{l}_2), \textbf{y}_2, \textbf{q}) \in A_{\epsilon}^{(n)}$ and $(\textbf{v}_1(\hat{t}_1), \textbf{v}_3(\hat{t}_3), \textbf{y}_3, \textbf{q}) \in A_{\epsilon}^{(n)}$. If $\hat{j}$ in all the index triples found are the same, $\mathcal{R}_1$ declares $m_{11} = \hat{j}$, for some $l_1$ and $t_1$. If $\hat{l}_1$ in all the index pairs found are indices of codewords $\textbf{u}_1(\hat{l}_1)$ from the same bin with index $\hat{k}_1$, and $\hat{l}_2$ in all the index pairs found are indices of codewords $\textbf{u}_2(\hat{l}_2)$ from the same bin with index $\hat{k}_2$, then $\mathcal{R}_2$ determines $(m_{21}, m_{22}) = (\hat{k}_1, \hat{k}_2)$. Similarly, if $\hat{t}_1$ in all the index pairs found are indices of codewords $\textbf{v}_1(\hat{t}_1)$ from the same bin with index $\hat{r}_1$, and $\hat{t}_3$ in all the index pairs found are indices of codewords $\textbf{v}_3(\hat{t}_3)$ from the same bin with index $\hat{r}_3$, then $\mathcal{R}_3$ determines $(m_{31}, m_{33}) = (\hat{r}_1, \hat{r}_3)$. Otherwise, the receivers $\mathcal{R}_1$, $\mathcal{R}_2$ and $\mathcal{R}_3$ declare an error.

\subsection{Analysis of the Probabilities of Error}
In this section we derive upperbounds on the probabilities of error events which could happen during encoding and decoding processes. We assume that a source message vector $\left(m_{11}, m_{21}, m_{22}, m_{31}, m_{33}\right) = (j, k_1, k_2, r_1, r_3)$ is encoded and transmitted. We  consider the analysis of the probability of encoding error at senders $\mathcal{S}_2$ and $\mathcal{S}_3$, and the analysis of the probability of decoding error at each of the three receivers $\mathcal{R}_1$, $\mathcal{R}_2$, and $\mathcal{R}_3$  separately.\\
\\
First, let us define the following events:\\
$(i)$ $E_{jl_1} \triangleq \left\{\left(\textbf{W}(j), \textbf{U}_1(l_1), \textbf{q} \right) \in A_{\epsilon}^{(n)}\right\}$,\\
$(ii)$ $E_{jl_2} \triangleq \left\{\left(\textbf{W}(j), \textbf{U}_2(l_2), \textbf{q} \right) \in A_{\epsilon}^{(n)}\right\}$,\\
$(iii)$ $E_{jl_1l_2t_1} \triangleq \left\{\left(\textbf{W}(j), \textbf{U}_1(l_1), \textbf{U}_2(l_2), \textbf{V}_1(t_1), \textbf{q}\right) \in A_{\epsilon}^{(n)}\right\}$,\\
$(iv)$ $E_{jl_1l_2t_3} \triangleq \left\{\left(\textbf{W}(j), \textbf{U}_1(l_1), \textbf{U}_2(l_2), \textbf{V}_3(t_3), \textbf{q}\right) \in A_{\epsilon}^{(n)}\right\}$,\\
$(v)$ $E_{jl_1t_1} \triangleq \left\{(\textbf{W}(j), \textbf{U}_1(l_1), \textbf{V}_1(t_1), \textbf{Y}_1, \textbf{q}) \in A_{\epsilon}^{(n)}\right\}$,\\
$(vi)$ $E_{l_1l_2} \triangleq \left\{(\textbf{U}_1(l_1), \textbf{U}_2(l_2), \textbf{Y}_2, \textbf{q}) \in A_{\epsilon}^{(n)} \right\}$,\\
$(vii)$ $E_{t_1t_3} \triangleq \left\{(\textbf{V}_1(t_1), \textbf{V}_3(t_3), \textbf{Y}_3, \textbf{q}) \in A_{\epsilon}^{(n)} \right\}$.\\
$E_{(.)}^c \triangleq$ complement of the event $E_{(.)}$. Events $(i) - (iv)$ will be used in the analysis of probability of encoding error while events $(v) - (vii)$ will be used in the analysis of probability of decoding error.

\subsubsection{Probability of Error at the Encoder of $\mathcal{S}_2$}
An error is made if $(a)$ the encoder cannot find a $\textbf{u}_1(l_1)$ in the bin indexed by $k_1$ such that $\left(\textbf{w}(j), \textbf{u}_1(l_1), \textbf{q}\right) \in A_{\epsilon}^{(n)}$ or $(b)$ it cannot find a $\textbf{u}_2(l_2)$ in the bin indexed by $k_2$ such that $\left(\textbf{w}(j), \textbf{u}_2(l_2), \textbf{q}\right) \in A_{\epsilon}^{(n)}$. The probability of encoding error at $\mathcal{S}_2$ can be bounded as
\begin{eqnarray}
\nonumber
\def\argmin{\mathop{\rm \bigcap}}
P_{e,\mathcal{S}_2} \leq P\left(\bigcap_{\textbf{U}_1(l_1) \in \mbox{bin}(k_1)}\left(\textbf{W}(j), \textbf{U}_1(l_1), \textbf{q}\right) \notin A_{\epsilon}^{(n)}\right) + P\left(\bigcap_{\textbf{U}_2(l_2) \in \mbox{bin}(k_2)}\left(\textbf{W}(j), \textbf{U}_2(l_2), \textbf{q}\right) \notin A_{\epsilon}^{(n)}\right),\\
\nonumber
\leq \left(1 - P(E_{jl_1})\right)^{2^{n(I(W;U_1|Q)+4\epsilon)}} + \left(1 - P(E_{jl_2})\right)^{2^{n(I(W;U_2|Q)+4\epsilon)}} ,
\end{eqnarray}
where $P(.)$ is the probability of an event. Since $\textbf{q}$ is predetermined, and $\textbf{w}$ and $\textbf{u}_1$ are independent given $\textbf{q}$,
\begin{eqnarray}
\nonumber P(E_{jl_1}) = \sum_{\left(\textbf{w}, \textbf{u}_1, \textbf{q} \right) \in A_{\epsilon}^{(n)}}P(\textbf{W}(j) = \textbf{w}|\textbf{q})P(\textbf{U}_1(l_1) = \textbf{u}_1|\textbf{q})\\
\nonumber \geq 2^{n(H(W,U_1|Q)-\epsilon)}2^{-n(H(W|Q)+\epsilon)}2^{-n(H(U_1|Q)+\epsilon)}= 2^{-n(I(W;U_1|Q)+3\epsilon)}.
\end{eqnarray}
Similarly, $P(E_{jl_2})\geq 2^{-n(I(W;U_2|Q)+3\epsilon)}$. Therefore,
\begin{eqnarray}
\nonumber P_{e,\mathcal{S}_2} \leq (1-2^{-n(I(W;U_1|Q)+3\epsilon)})^{2^{n(I(W;U_1|Q)+4\epsilon)}} + (1-2^{-n(I(W;U_2|Q)+3\epsilon)})^{2^{n(I(W;U_2|Q)+4\epsilon)}}.
\end{eqnarray}
Now,
\begin{eqnarray}
\nonumber (1-2^{-n(I(W;U_1|Q)+3\epsilon)})^{2^{n(I(W;U_1|Q)+4\epsilon)}} = e^{{2^{n(I(W;U_1|Q)+4\epsilon)}}\ln(1-2^{-n(I(W;U_1|Q)+3\epsilon)})}\\
\nonumber \leq e^{2^{n(I(W;U_1|Q)+4\epsilon)}(-2^{-n(I(W;U_1|Q)+3\epsilon)})}\\
\nonumber = e^{-2^{n\epsilon}}.
\end{eqnarray}
Clearly, $P_{e,\mathcal{S}_2} \rightarrow 0$ as $n \rightarrow \infty$.

\subsubsection{Probability of Error at the Encoder of $\mathcal{S}_3$}
An error is made if $(a)$ the encoder cannot find a $\textbf{v}_1(t_1)$ in the bin indexed by $r_1$ such that $\left(\textbf{w}(j), \textbf{u}_1(l_1), \textbf{u}_2(l_2), \textbf{v}_1(t_1), \textbf{q}\right) \in A_{\epsilon}^{(n)}$ or $(b)$ it cannot find a $\textbf{v}_3(t_3)$ in the bin indexed by $r_3$ such that $\left(\textbf{w}(j), \textbf{u}_1(l_1), \textbf{u}_2(l_2), \textbf{v}_3(t_3), \textbf{q}\right) \in A_{\epsilon}^{(n)}$. The probability of encoding error at $\mathcal{S}_3$ can be bounded as
\begin{eqnarray}
\nonumber
\def\argmin{\mathop{\rm \bigcap}}
P_{e,\mathcal{S}_3} \leq P\left(\bigcap_{\textbf{V}_1(t_1) \in \mbox{bin}(r_1)}\left(\textbf{W}(j), \textbf{U}_1(l_1), \textbf{U}_2(l_2), \textbf{V}_1(t_1), \textbf{q}\right) \notin A_{\epsilon}^{(n)}\right)\\+
\nonumber
P\left(\bigcap_{\textbf{V}_3(t_3) \in \mbox{bin}(r_3)}\left(\textbf{W}(j), \textbf{U}_1(l_1), \textbf{U}_2(l_2), \textbf{V}_3(t_3), \textbf{q}\right) \notin A_{\epsilon}^{(n)}\right)\\
\nonumber
\leq \left(1 - P(E_{jl_1l_2t_1})\right)^{2^{n(I(W,U_1,U_2;V_1|Q)+4\epsilon)}} + \left(1 - P(E_{jl_1l_2t_3})\right)^{2^{n(I(W,U_1,U_2;V_3|Q)+4\epsilon)}}.
\end{eqnarray}
Since $\textbf{q}$ is predetermined, we have,
\begin{eqnarray}
\nonumber P(E_{jl_1l_2t_1}) = \sum_{\left(\textbf{w}, \textbf{u}_1, \textbf{u}_2, \textbf{v}_1, \textbf{q} \right) \in A_{\epsilon}^{(n)}}\!\!\!\!\!\!\!\!\!\!\!\!\!\!\!P(\textbf{W}(j) = \textbf{w}, \textbf{U}_1(l_1) = \textbf{u}_1, \textbf{U}_2(l_2) = \textbf{u}_2|\textbf{q})P(\textbf{V}_1(t_1) = \textbf{v}_1|\textbf{q})\\
\nonumber \geq 2^{n(H(W,U_1,U_2,V_1|Q)-\epsilon))}2^{-n(H(W,U_1,U_2|Q)+\epsilon)}2^{-n(H(V_1|Q)+\epsilon)}\\
\nonumber = 2^{-n(I(W,U_1,U_2;V_1|Q)+3\epsilon)}.
\end{eqnarray}
Similarly, $P(E_{jl_1l_2t_1}) \geq 2^{-n(I(W,U_1,U_2;V_3|Q)+3\epsilon)}$. Therefore,
\begin{eqnarray}
\nonumber P_{e,\mathcal{S}_3} \leq \left(1 - 2^{-n(I(W,U_1,U_2;V_1|Q)+3\epsilon)}\right)^{2^{n(I(W,U_1,U_2;V_1|Q)+4\epsilon)}}\!\!\!\!\!\!\! +  \nonumber \left(1 - 2^{-n(I(W,U_1,U_2;V_3|Q)+3\epsilon)}\right)^{2^{n(I(W,U_1,U_2;V_3|Q)+4\epsilon)}}.
\end{eqnarray}
Proceeding in a way similar to the encoder error analysis at $\mathcal{S}_2$, we can show that $P_{e,\mathcal{S}_3} \rightarrow 0$ as $n \rightarrow \infty$.

\subsubsection{Probability of Error at the Decoder of $\mathcal{R}_1$}
There are two possible events which result in errors: $(a)$The codewords transmitted are not jointly typical i.e., $E_{jl_1t_1}^c$ happens or $(b)$ there exists some $\hat{j} \neq j$ such that $E_{\hat{j}\hat{\hat{l}}_1\hat{\hat{t}}_1}$ happens. Note that $\hat{\hat{l}}_1$ need not equal $l_1$, and $\hat{\hat{t}}_1$ need not equal $t_1$, since $\mathcal{R}_1$ is not required to decode $\hat{\hat{l}}_1$ and $\hat{\hat{t}}_1$ correctly. The probability of decoding error can, therefore, be expressed as
\begin{eqnarray}\label{eq5}
\def\argmin{\mathop{\rm \cup}}
P_{e,\mathcal{R}_1}^{(n)} = P\left(E_{jl_1t_1}^c \bigcup \cup_{\hat{j} \neq j}E_{\hat{j}\hat{\hat{l}}_1\hat{\hat{t}}_1}\right)
\end{eqnarray}
Applying union of events bound, $(\ref{eq5})$ can be written as,
\begin{center}
$P_{e,\mathcal{R}_1}^{(n)} \leq P\left(E_{jl_1t_1}^c\right) + P\left(\cup_{\hat{j} \neq j}E_{\hat{j}\hat{\hat{l}}_1\hat{\hat{t}}_1}\right)$\\
\begin{eqnarray}
\nonumber
\def\argmin{\mathop{\rm \cup}}
= P\left(E_{jl_1t_1}^c\right) + \sum_{\hat{j} \neq j}P\left(E_{\hat{j}l_1t_1}\right) + \sum_{\hat{j} \neq j\hat{\hat{l}}_1 \neq l_1}P\left(E_{\hat{j}\hat{\hat{l}}_1t_1}\right) + \sum_{\hat{j} \neq j \hat{\hat{t}}_1 \neq t_1}\!\!\!\!\!P\left(E_{\hat{j}l_1\hat{\hat{t}}_1}\right) + \sum_{\hat{j} \neq j\hat{\hat{l}}_1 \neq l_1\hat{\hat{t}}_1 \neq t_1}\!\!\!\!\!\!\!P\left(E_{\hat{j}\hat{\hat{l}}_1\hat{\hat{t}}_1}\right)
\end{eqnarray}
\end{center}
\begin{eqnarray}
\nonumber
\def\argmin{\mathop{\rm \cup}}
\leq P\left(E_{jl_1t_1}^c\right) + 2^{nR_{11}}P\left(E_{\hat{j}l_1t_1}\right) + 2^{n(R_{11}+R_{21}+I(W;U_1|Q)+4\epsilon)}P\left(E_{\hat{j}\hat{\hat{l}}_1t_1}\right) + \\ \nonumber 2^{n(R_{11}+R_{31}+I(W,U_1,U_2;V_1|Q)+4\epsilon)}P\left(E_{\hat{j}l_1\hat{\hat{t}}_1}\right) + \\ \nonumber 2^{n(R_{11}+R_{21}+I(W;U_1|Q)+4\epsilon+R_{31}+I(W,U_1,U_2;V_1|Q)+4\epsilon)}P\left(E_{\hat{j}\hat{\hat{l}}_1\hat{\hat{t}}_1}\right).
\end{eqnarray}
$P\left(E_{\hat{j}l_1t_1}\right), P\left(E_{\hat{j}\hat{\hat{l}}_1t_1}\right), P\left(E_{\hat{j}l_1\hat{\hat{t}}_1}\right) ~ \mbox{and}~  P\left(E_{\hat{j}\hat{\hat{l}}_1\hat{\hat{t}}_1}\right)$ can be upper bounded as follows.
\begin{eqnarray}
\nonumber P\left(E_{\hat{j}l_1t_1}\right)\leq 2^{-n(I(W;U_1,V_1,Y_1|Q)-3\epsilon)},\\
\nonumber P\left(E_{\hat{j}\hat{\hat{l}}_1t_1}\right) \leq 2^{-n(I(W,U_1;V_1,Y_1|Q)+I(W;U_1|Q)-4\epsilon)},\\
\nonumber P\left(E_{\hat{j}l_1\hat{\hat{t}}_1}\right) \leq 2^{-n(I(W,V_1;U_1,Y_1|Q)+I(W;V_1|Q)-4\epsilon)},\\
\nonumber P\left(E_{\hat{j}\hat{\hat{l}}_1\hat{\hat{t}}_1}\right) \leq 2^{-n(I(W,U_1,V_1;Y_1|Q)+I(W,U_1;V_1|Q)+I(W;U_1|Q)-5\epsilon)}.
\end{eqnarray}
Substituting these in the probability of decoding error at $\mathcal{R}_1$, we have,
\begin{eqnarray}
\nonumber P_{e,\mathcal{R}_1}^{(n)} = \epsilon + 2^{nR_{11}}2^{-n(I(W;U_1,V_1,Y_1|Q)-3\epsilon)}+ 2^{n(R_{11}+R_{21}+I(W;U_1|Q)+4\epsilon)}2^{-n(I(W,U_1;V_1,Y_1|Q)+I(W;U_1|Q)-4\epsilon)} + \\ \nonumber 2^{n(R_{11}+R_{31}+I(W,U_1,U_2;V_1|Q)+4\epsilon)}2^{-n(I(W,V_1;U_1,Y_1|Q)+I(W;V_1|Q)-4\epsilon)} + \\ \nonumber 2^{n(R_{11}+R_{21}+I(W;U_1|Q)+4\epsilon+R_{31}+I(W,U_1,U_2;V_1|Q)+4\epsilon)} 2^{-n(I(W,U_1,V_1;Y_1|Q)+I(W,U_1;V_1|Q)+I(W;U_1|Q)-5\epsilon)}.
\end{eqnarray}
$P_{e,\mathcal{R}_1}^{(n)} \rightarrow 0$ as $n \rightarrow \infty$ if $R_{11}$, $R_{21}$ and $R_{31}$ satisfy the following constraints:
\begin{eqnarray}
R_{11} \leq I(W;U_1,V_1,Y_1|Q)\label{eq6}\\
R_{11} + R_{21} \leq I(W,U_1;V_1,Y_1|Q)\label{eq7}\\
R_{11}+R_{31} \leq I(W,V_1;U_1,Y_1|Q)+I(W;V_1|Q)-I(W,U_1,U_2;V_1|Q)\label{eq8}\\
R_{11} + R_{21}+R_{31} \leq I(W,U_1,V_1;Y_1|Q)+I(W,U_1;V_1|Q)-I(W,U_1,U_2;V_1|Q).\label{eq9}
\end{eqnarray}

\subsubsection{Probability of Error at the Decoder of $\mathcal{R}_2$}
The two possible error events are: $(a)$ The codewords transmitted are not jointly typical i.e., $E_{l_1l_2}^c$ happens or $(b)$ there exists some $\left(\hat{l}_1 \neq l_1, \hat{l}_2 \neq l_2\right)$ such that $E_{\hat{l}_1\hat{l}_2}$ happens. The probability of decoding error can be written as
\begin{eqnarray}\label{eq10}
\def\argmin{\mathop{\rm \cup}}
P_{e,\mathcal{R}_2}^{(n)} = P\left(E_{l_1l_2}^c \bigcup \cup_{(\hat{l}_1 \neq l_1, \hat{l}_2 \neq l_2)}E_{\hat{l}_1\hat{l}_2}\right)
\end{eqnarray}
Applying union of events bound, $(11)$ can be written as,
\begin{center}
$P_{e,\mathcal{R}_2}^{(n)} \leq P\left(E_{l_1l_2}^c\right) + P\left(\cup_{(\hat{l}_1 \neq l_1, \hat{l}_2 \neq l_2)}E_{\hat{l}_1\hat{l}_2}\right)$
\begin{eqnarray}
\nonumber
\def\argmin{\mathop{\rm \cup}}
= P\left(E_{l_1l_2}^c\right) + \sum_{\hat{l}_1 \neq l_1}P(E_{\hat{l}_1l_2}) + \sum_{\hat{l}_2 \neq l_2}P(E_{l_1\hat{l}_2}) + \sum_{\hat{l}_1 \neq l_1, \hat{l}_2 \neq l_2}P(E_{\hat{l}_1\hat{l}_2})\\
\nonumber \leq P\left(E_{l_1l_2}^c\right) + 2^{n(R_{21}+I(W;U_1|Q)+4\epsilon)}P(E_{\hat{l}_1l_2})+2^{n(R_{22}+I(W;U_2|Q)+4\epsilon)}P(E_{l_1\hat{l}_2})\\ \nonumber + 2^{n(R_{21}+R_{22}+I(W;U_1|Q)+4\epsilon+I(W;U_2|Q)+4\epsilon)}P(E_{\hat{l}_1\hat{l}_2}).
\end{eqnarray}
\end{center}
$P(E_{\hat{l}_1l_2})$, $P(E_{l_1\hat{l}_2})$ and $P(E_{\hat{l}_1\hat{l}_2})$ can be upper bounded as follows.
\begin{eqnarray}
\nonumber P(E_{\hat{l}_1l_2}) \leq 2^{-n(I(U_1;U_2,Y_2|Q)-3\epsilon)},\\
\nonumber P(E_{l_1\hat{l}_2}) \leq 2^{-n(I(U_2;U_1,Y_2|Q)-3\epsilon)},\\
\nonumber P(E_{\hat{l}_1\hat{l}_2}) \leq 2^{-n(I(U_1,U_2;Y_2|Q)+I(U_1;U_2)-4\epsilon)}.
\end{eqnarray}
Substituting these in the probability of decoding error at $\mathcal{R}_2$, we have,
\begin{eqnarray}
\nonumber P_{e,\mathcal{R}_2}^{(n)} = \epsilon + 2^{n(R_{21}+I(W;U_1|Q)+4\epsilon)}2^{-n(I(U_1;U_2,Y_2|Q)-3\epsilon)} + 2^{n(R_{22}+I(W;U_2|Q)+4\epsilon)}2^{-n(I(U_2;U_1,Y_2|Q)-3\epsilon)} + \\ \nonumber 2^{n(R_{21}+R_{22}+I(W;U_1|Q)+4\epsilon+I(W;U_2|Q)+4\epsilon)}2^{-n(I(U_1,U_2;Y_2|Q)+I(U_1;U_2)-4\epsilon)}.
\end{eqnarray}
$P_{e,\mathcal{R}_2}^{(n)} \rightarrow 0$ as $n \rightarrow \infty$ if $R_{21}$ and $R_{22}$ satisfy the following constraints:
\begin{eqnarray}
R_{21} \leq I(U_1;U_2,Y_2|Q)-I(W;U_1|Q)\label{eq10a}\\
R_{22} \leq I(U_2;U_1,Y_2|Q)-I(W;U_2|Q)\label{eq11}\\
R_{21}+R_{22} \leq I(U_1,U_2;Y_2|Q)+I(U_1;U_2|Q)-I(W;U_1|Q)-I(W;U_2|Q).\label{eq12}
\end{eqnarray}

\subsubsection{Probability of Error at the Decoder of $\mathcal{R}_3$}
The two possible error events are: $(a)$ The codewords transmitted are not jointly typical i.e., $E_{t_1t_3}^c$ happens or $(b)$ there exists some $\left(\hat{t}_1 \neq t_1, \hat{t}_3 \neq t_3\right)$ such that $E_{\hat{t}_1\hat{t}_3}$ happens. The probability of decoding error can be written as
\begin{eqnarray}\label{eq25}
\def\argmin{\mathop{\rm \cup}}
P_{e,\mathcal{R}_3}^{(n)} = P\left(E_{t_1t_3}^c \bigcup \cup_{(\hat{t}_1 \neq t_1, \hat{t}_3 \neq t_3)}E_{\hat{t}_1\hat{t}_3}\right)
\end{eqnarray}
Applying union of events bound, $(\ref{eq25})$ can be written as,
\begin{center}
$P_{e,\mathcal{R}_3}^{(n)} \leq P\left(E_{t_1t_3}^c\right) + P\left(\cup_{(\hat{t}_1 \neq t_1, \hat{t}_3 \neq t_3)}E_{\hat{t}_1\hat{t}_3}\right)$
\begin{eqnarray}
\nonumber
\def\argmin{\mathop{\rm \cup}}
\leq P\left(E_{t_1t_3}^c\right) + \sum_{\hat{t}_1 \neq t_1}P(E_{\hat{t}_1t_3}) + \sum_{\hat{t}_3 \neq t_3}P(E_{t_1\hat{t}_3}) + \sum_{\hat{t}_1 \neq t_1, \hat{t}_3 \neq t_3}P(E_{\hat{t}_1\hat{t}_3})\\
\nonumber \leq P\left(E_{t_1t_3}^c\right) + 2^{n(R_{31}+I(W,U_1,U_2;V_1|Q)+4\epsilon)}P(E_{\hat{t}_1t_3})\\ \nonumber + 2^{n(R_{33}+I(W,U_1,U_2;V_3|Q)+4\epsilon)}P(E_{t_1\hat{t}_3}) + 2^{n(R_{31}+I(W,U_1,U_2;V_1|Q)+R_{33}+I(W,U_1,U_2;V_3|Q)+8\epsilon)}P(E_{\hat{t}_1\hat{t}_3})
\end{eqnarray}
\end{center}
$P(E_{\hat{t}_1t_3})$, $P(E_{t_1\hat{t}_3})$ and $P(E_{\hat{t}_1\hat{t}_3})$ can be upper bounded as follows.
\begin{eqnarray}
\nonumber P(E_{\hat{t}_1t_2}) \leq 2^{-n(I(V_1;V_3,Y_3|Q)-3\epsilon)},\\
\nonumber P(E_{t_1\hat{t}_3}) \leq 2^{-n(I(V_3;V_1,Y_3|Q)-3\epsilon)},\\
\nonumber P(E_{\hat{t}_1\hat{t}_3}) \leq  2^{-n(I(V_1,V_3;Y_3|Q)+I(V_1;V_3)-4\epsilon)}.
\end{eqnarray}
Substituting these in the probability of decoding error at $\mathcal{R}_3$, we have,
\begin{eqnarray}
\nonumber P_{e,\mathcal{R}_3}^{(n)} = \epsilon + 2^{n(R_{31}+I(W,U_1,U_2;V_1|Q)+4\epsilon)}2^{-n(I(V_1;V_3,Y_3|Q)-3\epsilon)}\\
\nonumber + 2^{n(R_{33}+I(W,U_1,U_2;V_3|Q)+4\epsilon)}2^{-n(I(V_3;V_1,Y_3|Q)-3\epsilon)}\\
\nonumber + 2^{n(R_{31}+I(W,U_1,U_2;V_1|Q)+R_{33}+I(W,U_1,U_2;V_3|Q)+8\epsilon)}\\ \nonumber \times 2^{-n(I(V_1,V_3;Y_3|Q)+I(V_1;V_3)-4\epsilon)}
\end{eqnarray}
$P_{e,\mathcal{R}_3}^{(n)} \rightarrow 0$ as $n \rightarrow \infty$ if $R_{31}$ and $R_{33}$ satisfy the following constraints:
\begin{eqnarray}
R_{31} \leq I(V_1;V_3,Y_3|Q)-I(W,U_1,U_2;V_1|Q),\label{eq13}\\
R_{33} \leq I(V_3;V_1,Y_3|Q)-I(W,U_1,U_2;V_3|Q),\label{eq14}\\
R_{31}+R_{33} \leq I(V_1,V_3;Y_3|Q)+I(V_1;V_3|Q)-I(W,U_1,U_2;V_3|Q)-I(W,U_1,U_2;V_1|Q).\label{eq15}
\end{eqnarray}
The inequalities (7)-(10), (12)-(14) and (16)-(18) together constitute the achievable rate region for the channel $\mathcal{C}_{cms}^2$.
\end{proof}
The proof of achievability for the channel $\mathcal{C}_{\text{PrMS}}^1$ is relegated to Appendix D.

\section{The Gaussian Case}
In this section, we introduce the Gaussian CR channel to evaluate and plot the rate region for the different channel models considered in this paper. We also describe several extensions, in the form of corollaries, to the achievable rate regions described above. Finally, we derive some outer bounds to help us test the optimality of the coding techniques that we have employed to derive the achievable rate regions. 

\subsection{The Gaussian CR channel} \label{sec:GaussianCRChannel}
The achievable rate regions described for the discrete memoryless channels can be extended to the Gaussian channels by quantizing the channel inputs and outputs \cite{refgallagerbook}.  Let $\mathcal{C}^t_{G,\text{CuMS}}$ denote the cognitive Gaussian channel with cumulative message sharing, $\mathcal{C}^t_{G,\text{PrMS}}$ the cognitive Gaussian channel with primary-only message sharing and $\mathcal{C}^t_{G,\text{CoMS}}$ the cognitive Gaussian channel with cognitive-only message sharing ($G$ for Gaussian, CuMS, PrMS and CoMS are the same as before);$~t=1,2$. We show the extension for only one of the channel models - from $\mathcal{C}_{\text{CuMS}}^2$ to $\mathcal{C}^2_{G,\text{CuMS}}$.

The cognitive Gaussian channel is described by a discrete-time input $\tilde{X}_k$, a corresponding output $\tilde{Y}_k$, and a random variable $\tilde{Z}_k$ denoting noise at the receiver;~$k=1,2,3$.  Following the maximum-entropy theorem  \cite{refcoverbook}, the input random variable $\tilde{X}_k;~k=1,2,3$ is assumed to have a Gaussian distribution. The transmitted codeword $\tilde{\textbf{x}}_k = (\tilde{x}_{k1},\ldots,\tilde{x}_{kn})$ satisfies the average power constraint given by
\begin{eqnarray}
\nonumber \mathbb{E}\{\|\tilde{\textbf{x}}_k\|^{2}\} \leq \tilde{P}_k;~k=1,2,3,
\end{eqnarray}
where $\mathbb{E}\{.\}$ is the expectation operator. The zero-mean random variable $\tilde{Z}_k$ is drawn i.i.d from a Gaussian distribution with variance $\tilde{N}_k;~k=1,2,3$, and is assumed to be independent of the signal $\tilde{X}_k$. The Gaussian CR channel can be converted to a standard from using invertible transformations \cite{refpramod1},\cite{refcarleial}.

For the channel $\mathcal{C}_{G,\text{CuMS}}^2$, we have $W$, $U_1$, $U_2$, $V_1$ and $V_3$ as the random variables (RV) which describe the sources at the transmitters. We also some consider additional RVs - $\tilde{W}$, $\tilde{U}_1$, $\tilde{U}_2$, $\tilde{V}_1$ and $\tilde{V}_3$ - with the following statistics:
\begin{itemize}
\item $\tilde{W} \sim \mathcal{N}(0, P_1)$,
\item $\tilde{U}_1 \sim \mathcal{N}(0,\tau P_2)$, $\tilde{U}_2 \sim \mathcal{N}(0,\bar{\tau} P_2)$, with $\tau+\bar{\tau} = 1$,
\item $\tilde{V}_1 \sim \mathcal{N}(0,\kappa P_3)$, $\tilde{V}_3 \sim \mathcal{N}(0,\bar{\kappa} P_3)$, with $\kappa+\bar{\kappa} = 1$.
\end{itemize}
Further,
\begin{itemize}
\item $W = \tilde{W}$,
\item $U_1 = \tilde{U}_1 + \alpha_1 X_1$, $U_2 = \tilde{U}_2 + \alpha_2 X_1$,
\item $V_1 = \tilde{V}_1 + \alpha_3 X_1 + \beta_1 X_2 $, $V_3 = \tilde{V}_3 + \alpha_4 X_1 + \beta_2 X_2 $,
\end{itemize}
where the input RV's $X_1$, $X_2$ and $X_3$ are given by $X_1  = \tilde{W}$, $X_2 =  \tilde{U}_1 + \tilde{U}_2$ and $X_3 =  \tilde{V}_1 + \tilde{V}_3$. Notice that $\tilde{W}$, $\tilde{U}_1$, $\tilde{U}_2$, $\tilde{V}_1$ and $\tilde{V}_3$ are mutually independent. Therefore, $X_1 \sim \mathcal{N}(0,P_1)$, $X_2 \sim \mathcal{N}(0,P_2)$ and $X_3 \sim \mathcal{N}(0,P_3)$.\\
The values of $\tau$ and $\kappa$ are randomly selected from the interval $[0,1]$. The values of $\alpha_1$, $\alpha_2$, $\alpha_3$, $\alpha_4$, $\beta_1$ and $\beta_2$ are repeatedly generated according to $\mathcal{N}(0,1)$. The channel outputs are
\begin{eqnarray}
\nonumber Y_1  = X_1 + a_{12}X_2 + a_{13}X_3 + Z_1,\\
\nonumber Y_2  = a_{21}X_1 + X_2 + a_{23}X_3 + Z_2,\\
\nonumber Y_3  = a_{31}X_1 + a_{32}X_2 + X_3 + Z_3,
\end{eqnarray}
where $Z_1 \sim \mathcal{N}(0,Q_1)$, $Z_2 \sim \mathcal{N}(0,Q_2)$ and $Z_3 \sim \mathcal{N}(0,Q_3)$ are independent additive noise, and $Q_1$, $Q_2$ and $Q_3$ are noise variances when the input-output relations are represented in the standard form. Substituting for $X_1$, $X_2$ and $X_3$, we get,
\begin{eqnarray}
\nonumber Y_1  = \tilde{W} + a_{12}(\tilde{U}_1 + \tilde{U}_2) + a_{13}(\tilde{V}_1 + \tilde{V}_3) + Z_1,\\
\nonumber Y_2  = a_{21}\tilde{W} + (\tilde{U}_1 + \tilde{U}_2) + a_{23}(\tilde{V}_1 + \tilde{V}_3) + Z_2,\\
\nonumber Y_3  = a_{31}\tilde{W} + a_{32}(\tilde{U}_1 + \tilde{U}_2) + \tilde{V}_1 + \tilde{V}_3 + Z_3,
\end{eqnarray}
where the interference coefficients $a_{12}$, $a_{13}$, $a_{21}$, $a_{23}$, $a_{31}$ and $a_{32}$ are assumed to be real and globally known. The rate region $\mathfrak{R}_{\text{CuMS}}^2$ for the channel $\mathcal{C}_{\text{CuMS}}^2$ can be extended to its respective Gaussian channel model by evaluating the mutual information terms. To this end, we construct a covariance matrix and compute its entries. Let us define a vector $\Theta = (Y_1,~ Y_2,~ Y_3,~ W,~ U_1,~ U_2,~ V_1,~ V_3)$. The covariance matrix is given by
\begin{eqnarray}
\nonumber \mbox{COV}(Y_1,~ Y_2,~ Y_3,~ W,~ U_1,~ U_2,~ V_1,~ V_3) = \mathbb{E}\{\Theta^T\Theta\}.
\end{eqnarray}
The entries of this covariance matrix are used to compute the differential entropy terms, which are further used to evaluate the mutual information.

\begin{theorem}
Let $\Upsilon = (\tau,\kappa,\alpha_1,\alpha_2,\alpha_3,\alpha_4,\beta_1,\beta_2)$. For a fixed $\Upsilon$, let $\mathcal{G}_{\text{CuMS}}^2(\Upsilon)$ be achievable. The rate region $\mathfrak{G}_{\text{CuMS}}^2$ is achievable for the Gaussian channel $\mathcal{C}^2_{G,\text{CuMS}}$ with
\begin{eqnarray}
\nonumber  \mathfrak{G}_{\text{CuMS}}^2 = \bigcup_{\Upsilon}\mathcal{G}_{\text{CuMS}}^2(\Upsilon).
\end{eqnarray}
\end{theorem}
\begin{proof}
Since the computation procedure is cumbersome and lengthy albeit straightforward, we do not provide the proof here.
\end{proof}
The same procedure is followed to compute the mutual information terms for the remaining channel models - $\mathcal{C}^1_{G,\text{CuMS}}$, $\mathcal{C}^t_{G,\text{PrMS}};t=1,2,$ and $\mathcal{C}_{G,\text{CoMS}}$.

\subsection{Extensions}
The corollaries presented in this subsection arise because several important achievable rate tuples for the Gaussian CR channel can be readily identified, which leads to a larger overall achievable region. We have made use of the fact that the cognitive transmitters can be used as relays, depending on their knowledge of the other user's message. It is important to note that the corollaries below only give some examples of rate points that are achievable, and the list is by no means exhaustive. Hence, although it is possible that the rate regions describe below can be further improved upon, a systematic way of doing so seems elusive, and is relegated to future work. Also note that the achievable rate points below are presented as separate corollaries for clarity of presentation; one could state them together as one single result as well. The proofs for some of the corollaries can be found in  Appendix E.

\subsubsection{$\mathcal{C}^t_{G,\text{CuMS}}$}
\begin{corollary}
\label{corol1}
Let $\mathfrak{G}_{\text{CuMS}}^2$ be the set of all points $(R_{1}, R_{21} + R_{22}, R_{31} + R_{33})$  where $(R_{1}, R_{21}, R_{22}, R_{31}, R_{33})$ is an achievable rate tuple of Theorem 4.1. 
Then, the convex hull of the region $\mathfrak{G}_{\text{CuMS}}^2$ with the points $(R_{1}^{*},0,0)$ and $(0,R_{2}^{*},R_{3}^{*})$ is achievable for the $\mathcal{C}^t_{G,\text{CuMS}}$ model, where
\begin{eqnarray}
 R_{1}^{*} &=& \dfrac{1}{2}\log_2\lb1 + \dfrac{\lb\sqrt P_{1} + \abs{a_{12}}\sqrt P_{2} + \abs{a_{13}}\sqrt P_{3}\rb^2}{Q_{1}}\rb, \nonumber \\
 R_{2}^{*} &=& \dfrac{1}{2}\log_2\lb1 + \dfrac{P_{2}}{Q_{2} + \abs{a_{23}}^2P_{3}}\rb, \nonumber \\
 R_{3}^{*} &=& \dfrac{1}{2}\log_2\lb1 + \dfrac{P_{3}}{Q_{3}}\rb. \nonumber
\end{eqnarray}
\end{corollary}

\begin{corollary}
\label{corol2}
Let $\mathfrak{G}_{\text{CuMS}}^2$ be the set of all points $(R_{1}, R_{21} + R_{22}, R_{31} + R_{33})$,  where $(R_{1}, R_{21}, R_{22}, R_{31}, R_{33})$ is an achievable rate tuple of Theorem 4.1. Then the convex hull of the region $\mathfrak{G}_{\text{CuMS}}^2$ with the points $(R_{1}^{*},0,r)$ and $(0,R_{2}^{*},r)$ are achievable for the $\mathcal{C}^t_{G,\text{CuMS}}$ model, where
\begin{eqnarray}
R_{1}^{*} &=& \dfrac{1}{2}\log_2\left(1 + \dfrac{\lb \sqrt{P_1} + |a_{12}|\sqrt{P_{2}} + |a_{13}|\sqrt{P_{3}^{\mathcal{S}_1}} \rb^2 }{Q_{1} + \abs{a_{13}}^{2}P_{3}^{\mathcal{S}_3}}\right), \nonumber \\
R_{2}^{*} &=& \dfrac{1}{2}\log_2\left(1 + \dfrac{\left(\sqrt{P_2} + |a_{23}|\sqrt{P_{3}^{\mathcal{S}_2}}\right)^2}{Q_{2} + \abs{a_{23}}^{2}P_{3}^{\mathcal{S}_3}}\right), \nonumber \\
r &=& \dfrac{1}{2}\log_2\left(1 + \dfrac{P_3^{\mathcal{S}_3}}{Q_{3}}\right), \nonumber
\end{eqnarray}
where $P_{3}^{\mathcal{S}_1} = P_{3}^{\mathcal{S}_2} = P_{3} - P_{3}^{\mathcal{S}_3}$, $\forall \, P_3^{\mathcal{S}_3} \in [0 , P_3]$. 
\end{corollary}

\begin{corollary}
\label{corol3}
The convex hull of the region $\mathfrak{G}_{\text{CuMS}}^2$ with the points $(R_{1}^{*},r,0)$ and $(0,r,R_{3}^{*})$ is achievable for the $\mathcal{C}^t_{G,\text{CuMS}}$ model, where
\begin{eqnarray}
 R_{1}^{*} &=& \dfrac{1}{2}\log_2\left(1 + \dfrac{\left(\sqrt{P1} + |a_{12}|\sqrt{P_{2}^{\mathcal{S}_1}} + |a_{13}|\sqrt{P_{3}}\right)^2}{Q_{1} + |a_{12}|^{2}P_{2}^{\mathcal{S}_2}}\right), \nonumber \\
  r &=& \frac{1}{2}\log_2 \left(1 + \frac{P_2^{\mathcal{S}_2}}{Q_2+|a_{23}|^2P_3} \right), \nonumber \\
 R_{3}^{*} &=& \dfrac{1}{2}\log_2\left(1 + \dfrac{P_{3}}{Q_{3}}\right), \nonumber
\end{eqnarray}
where $P_{2}^{\mathcal{S}_2} = (2^{2r}-1)(Q_{2} + |a_{23}|^{2}P_{3})$, $P_{2}^{\mathcal{S}_1} = P_{2} - P_{2}^{\mathcal{S}_2}$ and $r$ is the minimum rate that $\mathcal{S}_{2}$ is guaranteed to achieve.
\end{corollary}

\begin{corollary}
\label{corol4}
The convex hull of the region $\mathfrak{G}_{\text{CuMS}}^2$ with the points $(0,R_{2}^{*},0)$ and $(0,0,R_{3}^{*})$ is achievable for the $\mathcal{C}^t_{G,\text{CuMS}}$ model, where
\begin{eqnarray}
R_{2}^{*} &=& \dfrac{1}{2}\log_2\lb1 + \dfrac{\lb\sqrt{P}_{2} + \abs{a_{23}}\sqrt{P_{3}}\rb^2}{Q_{2}}\rb, \nonumber \\
R_{3}^{*} &=& \dfrac{1}{2}\log_2\lb1 + \dfrac{P_{3}}{Q_{3}}\rb. \nonumber
\end{eqnarray}
\end{corollary}

\begin{theorem}
The convex hull of the region $\mathfrak{G}_{\text{CuMS}}^2$ with the achievable points in the corollaries \ref{corol1} - \ref{corol4} results in an achievable rate region of the $\mathcal{C}^t_{G,\text{CuMS}}$ channel model.
\end{theorem}
\begin{proof}
The convex hull is achievable by standard time-sharing arguments.
\end{proof}

\subsubsection{$\mathcal{C}^t_{G,\text{PrMS}}$}
\begin{corollary}
\label{corol5}
Let $\mathfrak{G}_{\text{PrMS}}^2$ be the set of all points $(R_{1}, R_{21} + R_{22}, R_{31} + R_{33})$  such that $(R_{1}, R_{21}, R_{22}, R_{31}, R_{33})$ is an achievable rate tuple. Then the convex hull of the region $\mathfrak{G}_{\text{PrMS}}^2$ with the points $(R_{1}^{*},0,0)$ and $(0,R_{2}^{*},R_{3}^{*})$ are achievable for the $\mathcal{C}^t_{G,\text{PrMS}}$ model, where
\begin{eqnarray}
 R_{1}^{*} & = & \dfrac{1}{2}\log \lb 1 + \dfrac{\lb\sqrt P_{1} + \abs{a_{12}}\sqrt P_{2} + \abs{a_{13}}\sqrt P_{3}\rb^2}{Q_{1}}\rb, \nonumber \\
 R_{2}^{*} & = & \dfrac{1}{2}\log\lb 1 + \dfrac{P_{2}}{Q_{2} + \abs{a_{23}}^{2}P_{3}}\rb, \nonumber \\
 R_{3}^{*} & = & \dfrac{1}{2}\log\lb 1 + \dfrac{P_{3}}{Q_{3} + \abs{a_{32}}^{2}P_{2}}\rb.\nonumber
\end{eqnarray}
\end{corollary}

\begin{corollary}
\label{corol6}
The convex hull of the region $\mathfrak{G}_{\text{PrMS}}^2$ with the points $(R_{1}^{*},0,r)$ and $(0,R_{2}^{*},r)$ are achievable for the $\mathcal{C}^t_{G,\text{PrMS}}$ model, where
\begin{eqnarray}
 R_{1}^{*} & = & \dfrac{1}{2}\log_2\lb1 + \dfrac{\lb\sqrt{P_{1}} + \abs{a_{12}}\sqrt{P_{2}} + \abs{a_{13}}\sqrt{P_{3}^{\mathcal{S}_1}}\rb^2}{Q_{1} + \abs{a_{13}}^{2}P_{3}^{\mathcal{S}_3}}\rb, \nonumber \\
 R_{2}^{*} & = & \dfrac{1}{2}\log_2\lb 1 + \dfrac{P_{2}}{Q_{2} + \abs{a_{23}}^{2}P_{3}}\rb, \nonumber \\
 r & = & \dfrac{1}{2}\log_2\lb 1 + \dfrac{P_{3}^{cr2}}{Q_{3} + \abs{a_{32}}^{2}P_{2}} \rb, \nonumber
\end{eqnarray}
where $P_{3}^{\mathcal{S}_1} = P_{3} - P_{3}^{\mathcal{S}_3}$, $\forall P_{3}^{\mathcal{S}_3} \in [0,P_{3}]$.
\end{corollary}

\begin{corollary}
\label{corol7}
The convex hull of the region $\mathfrak{G}_{\text{PrMS}}^2$ with the points $(R_{1}^{*},r,0)$ and $(0,r,R_{3}^{*})$ are achievable for the $\mathcal{C}^t_{G,\text{PrMS}}$ model, where
\begin{eqnarray}
 R_{1}^{*} & = & \dfrac{1}{2}\log_2\lb1 + \dfrac{\lb\sqrt{P_{1}} + \abs{a_{12}}\sqrt{P_{2}^{\mathcal{S}_1}} + \abs{a_{13}}\sqrt{P_{3}}\rb^2}{Q_1 + \abs{a_{12}}^{2}P_{2}^{\mathcal{S}_2}}\rb, \nonumber \\
 r & = & \dfrac{1}{2}\log_2\lb 1 + \dfrac{P_{2}^{\mathcal{S}_2}}{Q_{2} + \abs{a_{23}}^{2}P_{3}}\rb, \nonumber \\
 R_{3}^{*} & = & \dfrac{1}{2}\log_2\lb 1 + \dfrac{P_{3}}{Q_3 + \abs{a_{32}}^{2}P_{2}}\rb, \nonumber
\end{eqnarray}
where $P_{2}^{\mathcal{S}_1} = P_{2} - P_{2}^{\mathcal{S}_2}$, $\forall P_{2}^{\mathcal{S}_2} \in [0,P_{2}]$.
\end{corollary}

\begin{theorem}
The convex hull of the region $\mathfrak{G}_{\text{PrMS}}^2$ with the achievable points in the corollaries \ref{corol5} - \ref{corol7} results in an achievable rate region of the $\mathcal{C}^t_{G,\text{PrMS}}$ channel model.
\end{theorem}
\begin{proof}
The convex hull is achievable by standard time-sharing arguments.
\end{proof}

\subsubsection{$\mathcal{C}_{G,\text{CoMS}}$}
\begin{corollary}
\label{corol8}
Let $\mathfrak{G}_{\text{CoMS}}$ be the set of all points $(R_{1}, R_{2}, R_{31} + R_{33})$  such that $(R_{1}, R_{2}, R_{31}, R_{33})$ is an achievable rate tuple. Then the convex hull of the region $\mathfrak{G}_{\text{CoMS}}$ with the points $(R_{1}^{*},0,0)$, $(0,R_{2}^{*},0)$ and $(0,0,R_{3}^{*})$ are achievable for the $\mathcal{C}_{G,\text{CoMS}}$ model, where
\begin{eqnarray}
 R_{1}^{*} & = & \dfrac{1}{2}\log_2 \lb 1 + \dfrac{\left(\sqrt P_{1} + \abs{a_{13}}\sqrt P_{3}\right)^2}{Q_{1}+\abs{a_{12}}^2 P_{2}}\rb, \nonumber \\
 R_{2}^{*} & = & \dfrac{1}{2}\log_2 \lb 1 + \dfrac{\left(\sqrt P_{2} + \abs{a_{23}}\sqrt P_{3}\right)^2}{Q_{2}+\abs{a_{21}}^2 P_{1}}\rb, \nonumber \\
 R_{3}^{*} & = & \dfrac{1}{2}\log_2\lb 1 + \dfrac{P_{3}}{Q_{3}}\rb.\nonumber
\end{eqnarray}
\end{corollary}

\begin{corollary}
\label{corol9}
The convex hull of the region $\mathfrak{G}_{\text{CoMS}}$ with the points $(R_{1}^{*},0,r)$, $(0,R_{2}^{*},r)$ and $(0,0,r)$ are achievable for the $\mathcal{C}_{G,\text{CoMS}}$ model, where
\begin{eqnarray}
 R_{1}^{*} & = & \dfrac{1}{2}\log_2\lb1 + \dfrac{\lb\sqrt{P_{1}} + \abs{a_{13}}\sqrt{P_{3}^{\mathcal{S}_1}}\rb^2}{Q_{1} + \abs{a_{12}}^{2}P_{2} + \abs{a_{13}}^{2}P_{3}^{\mathcal{S}_3}}\rb, \nonumber \\
 R_{2}^{*} & = & \dfrac{1}{2}\log_2\lb1 + \dfrac{\lb\sqrt{P_{2}} + \abs{a_{13}}\sqrt{P_{3}^{\mathcal{S}_2}}\rb^2}{Q_{2} + \abs{a_{21}}^{2}P_{1} + \abs{a_{13}}^{2}P_{3}^{\mathcal{S}_3}}\rb, \nonumber \\
 r & = & \dfrac{1}{2}\log_2\lb 1 + \dfrac{P_{3}^{\mathcal{S}_3}}{Q_{3}} \rb, \nonumber
\end{eqnarray}
where $P_{3}^{\mathcal{S}_1} = P_{3}^{\mathcal{S}_2} = P_{3} - P_{3}^{\mathcal{S}_3}$, $\forall P_{3}^{\mathcal{S}_3} \in [0,P_{3}]$.
\end{corollary}
\begin{theorem}
The convex hull of the region $\mathfrak{G}_{\text{CoMS}}$ with the achievable points in the corollaries \ref{corol8} and\ref{corol9} results in an achievable rate region of the $\mathcal{C}_{G,\text{CoMS}}$ channel model.
\end{theorem}
\begin{proof}
The convex hull is achievable by standard time-sharing arguments.
\end{proof}

\subsection{Outer Bounds}
The outer bound presented in this paper is inspired by the one found in \cite{refnatasha1}. For the channel models considered in this paper, let us consider the scenario where the transmitters cooperate in a bidirectional manner, i.e., every sender knows the message of every other sender in a noncausal manner. In such a scenario, our channel models reduce to a multiple antenna broadcast channel (MIMO-BC) with one sender having three antennas and three receivers with one antenna each. Since bidirectional message-sharing is tantamount to additional information at the transmitters, the achievable rate regions can be enlarged. Further, this enlarged region also turns out to be the capacity region of the MIMO-BC (see \cite{refshlomo1}) and is an outer bound to our achievable rate regions. Unfortunately, the capacity region for the MIMO-BC is neither concave nor convex, making its computation difficult. We therefore resort to duality results of the broadcast (BC) and the multiple access channels (MAC), reported first in \cite{refsriram3}.

Let $P$ be the total power constraint for the MIMO-BC and $P_{1}$, $P_{2}$ and $P_{3}$ be the individual power constraint for the MAC.  On the MAC channel, the rate achieved by user $j$ is given by
\begin{equation}
 R_{\text{MAC},j} = \log_2\dfrac{\left|\mathbf{I} + \displaystyle\sum_{i=j}^{K}\mathbf{H}_{i}^{H}P_{i}\mathbf{H}_{i}\right|}{\left|\mathbf{I} + \displaystyle\sum_{i=j+1}^{K}\mathbf{H}_{i}^{H}P_{i}\mathbf{H}_{i}\right|},
\end{equation}
where $\abs{A}$ denotes the determinant of $A$; and the channel matrices are $\Hone = [1 \: a_{12} \: a_{13}]$, $\Htwo = [a_{21} \: 1 \: a_{23}]$ and $\Hthree = [a_{31} \: a_{32} \: 1]$; and $\mathbf{I} + \displaystyle\sum_{i=j+1}^{K}\mathbf{H}_{i}^{H}P_{i}\mathbf{H}_{i}$ is the interference experienced by the $j^{th}$ user. The MIMO-BC capacity region with power constraint $P$ is equal to the union of capacity regions of the dual MAC, where the union is taken over all individual power constraint, $P_{1}$, $P_{2}$ and $P_{3}$, such that $P = P_1+P_2+P_3$. Therefore,
\begin{equation} \label{eq:eq20}
C_{\text{BC}}(P,H) = \displaystyle\bigcup_{P_{1},P_{2},P_{3}:\sum_{j=1}^{3}P_{j} = P}C_{\text{MAC}}(P_{1}, P_{2}, P_{3};\mathbf{H}^{T}),
\end{equation}
where
\begin{equation}\label{eq:eq21}
 C_{\text{MAC}}(P_{1}, P_{2}, P_{3};\mathbf{H}^{T}) = \displaystyle\bigcup_{j \in \{1,2,3\}} R_{\text{MAC},j},
\end{equation}
where $R_{\text{MAC},j}$ is given by (19). We thus obtain the capacity region of the MIMO-BC, which forms an outer bound for the channel models considered in this paper. Generally, this outer bound tends to be loose, since the MIMO-BC capacity region was obtained by allowing bidirectional (or complete) transmitter cooperation. As in \cite{refnatasha1}, the rates of individual users can be further bounded depending on the model (\text{CuMS}, \text{PrMS} or \text{CoMS}).
\begin{enumerate}
\item In the case of \text{CuMS}, senders $\mathcal{S}_2$ and $\mathcal{S}_3$ have complete knowledge of the $\mathcal{S}_1$'s message and $\mathcal{S}_3$ has knowledge of $\mathcal{S}_2$'s message but not vice-versa. Note that, the rate of $\mathcal{S}_1$ cannot be bounded by the interference-free case where $a_{12} = 0$ and $a_{13} = 0$. This is because unidirectional message sharing enables $\mathcal{S}_2$ and $\mathcal{S}_3$ to transmit the message of $\mathcal{S}_1$, thereby increasing the rate of $\mathcal{S}_1$ beyond what is achievable with the $\mathcal{S}_1$ alone transmitting its message. Hence, rate $R_1$ can upper bounded as follows.
    \begin{equation}\label{eq:eq22}
    R_{1} \leq \frac{1}{2}\log_2\left(1+\dfrac{(\sqrt{P_{1}} + |a_{12}|\sqrt{P_{2}} + |a_{13}|\sqrt{P_{3}})^{2}}{Q_{1}}\right).
    \end{equation}
    Similarly, the rate of $\mathcal{S}_2$ cannot be bounded by the interference free rate as $\mathcal{S}_3$ can use its knowledge of $\mathcal{S}_2$'s message to enable $\mathcal{S}_2$ increase its rate. Hence, the rate of $\mathcal{S}_2$ can upper bounded as
    \begin{equation}\label{eq:eq23}
    R_{2} \leq \frac{1}{2}\log_2\left(1 + \dfrac{(\sqrt{P_{2}} + |a_{23}|\sqrt{P_{3}})^{2}}{Q_{2}}\right).
    \end{equation}
    Finally, the rate of $\mathcal{S}_3$ can be upper bounded by the interference free case.
    \begin{eqnarray}\label{eq:eq24}
    R_{3} \leq \frac{1}{2}\log_2\left(1 + \dfrac{P_{3}}{Q_{3}}\right).
    \end{eqnarray}

\item In the case of \text{PrMS}, although $\mathcal{S}_2$ and $\mathcal{S}_3$ have complete knowledge of $\mathcal{S}_1$'s message, they do not have each other's message. Therefore, the bound on the $\mathcal{S}_1$'s rate given by (22) remains valid, as the $\mathcal{S}_2$ and $\mathcal{S}_3$ can use their knowledge of $\mathcal{S}_1$'s message to increase its rate. The bound on $\mathcal{S}_3$'s rate is same as in the case of \text{CuMS} and is given by (24). Lastly, the $\mathcal{S}_2$'s rate can be upper bounded by the interference-free case as follows.
    \begin{equation}\label{eq:eq25}
    R_{2} \leq \frac{1}{2}\log_2\left(1 +\dfrac{P_{2}}{Q_{2}}\right).
    \end{equation}

\item We upper bound now the sum rate of $\mathcal{S}_2$ and $\mathcal{S}_3$ by allowing full cooperation between their transmitters and pairing receivers. This results in a point-to-point MIMO channel, whose capacity is expressed as follows.
    \begin{equation}\label{eq:eq26}
    C_{\text{MIMO}} = \displaystyle\max_{i, \sum_{i}P_{i} \leq P}\frac{1}{2}\displaystyle\sum_{i=1}^{N}\log_2\left(1 + \dfrac{P_{i}\sigma_{i}^{2}}{Q}\right),
    \end{equation}
    where $\dfrac{P_i\sigma_{i}^{2}}{Q}$ is the signal-to-noise ratio associated with the $i^{th}$ channel, $\sigma_{i}$s are the singular values and $N$ represents the number of singular values of the MIMO channel. The optimum power allocation $P_i$ can be obtained by the water-filling algorithm \cite{refcoverbook}.

\item In the case of \text{CoMS}, sender $\mathcal{S}_3$  has noncausal knowledge of $\mathcal{S}_1$ and $\mathcal{S}_2$. Therefore, the rates of $\mathcal{S}_1$ and $\mathcal{S}_2$ cannot be bounded by the interference free scenario. The rate of $\mathcal{S}_1$ can be upper bounded as follows:
    \begin{equation}\label{eq:eq27}
    R_{1} \leq \frac{1}{2}\log_2\left(1+\dfrac{(\sqrt{P_{1}} + |a_{13}|\sqrt{P_{3}})^{2}}{Q_{1}}\right).
    \end{equation}
    The rates of $\mathcal{S}_2$ and $\mathcal{S}_3$ can be upper bounded as in (23) and (24), respectively. To bound the sum rate of $\mathcal{S}_1$ and $\mathcal{S}_2$ we allow full cooperation between the transmitters and pairing receivers, resulting in a point-to-point MIMO channel. The capacity of this channel is given by \eqref{eq:eq26}.
\end{enumerate}

\section{Simulation Results and Discussion}
\subsection{Simulation Setup}
We consider a $3$-user Gaussian cognitive channel with \text{CuMS}, \text{PrMS} and \text{CoMS} for the simulations. We generate the source and channel symbols as described in Section \ref{sec:GaussianCRChannel}.
\begin{enumerate}
\item The direct channel gains are $a_{11} = a_{22} = a_{33} = 1$.
\item The interference coefficients $a_{12} = a_{13} = a_{21} = a_{23} = a_{31} = a_{32} = 0.55$.
\item The values of $\tau$ and $\kappa$ are assumed to be randomly selected from the interval $[0,1]$.
\item The values of $\alpha_1$, $\alpha_2$, $\alpha_3$, $\alpha_4$, $\beta_1$ and $\beta_2$ are repeatedly generated according to $\mathcal{N}(0,1)$.
\item The noise variances $Q_1 = Q_2 = Q_3 = 1$.
\item The transmit powers $P_1 = P_2 = P_3 = 7.8$dB or $10$dB, as specified.
\end{enumerate}

\subsection{Simulation Results and Discussion}
We now present the simulation results for the two and three-user scenario and draw several interesting observations.
\begin{enumerate}
\item Two-user channels:
\begin{enumerate}
\item Figure \ref{fig:fig4} shows the plot of rate regions for the $2$-user interference channels with various rate-splitting strategies.  For convenience, we introduce the following notation. The senders are denoted $\mathrm{Tx}_1$ and $\mathrm{Tx}_2$, and the pairing receivers are denoted $\mathrm{Rx}_1$ and $\mathrm{Rx}_2$, respectively. In Fig. \ref{fig:fig4}(a), we consider the case where both $\mathrm{Tx}_1$ and $\mathrm{Tx}_2$ do not perform rate-splitting. In Fig. \ref{fig:fig4}(b), $\mathrm{Tx}_2$ performs rate-splitting allowing $\mathrm{Rx}_1$ to decode the public part of $\mathrm{Tx}_2$'s message and cancels the interfering signals. In Fig. \ref{fig:fig4}(c), both $\mathrm{Tx}_1$ and $\mathrm{Tx}_2$ perform rate-splitting which allows $\mathrm{Rx}_1$ and $\mathrm{Rx}_2$ to decode the public part of the non-pairing transmitter's message and cancel out the interference by employing successive decoding. Therefore, it achieves the biggest rate region. Note that, Fig. \ref{fig:fig4}(c) is the Han-Kobayashi achievable rate region for the two-user interference channel \cite{refhan}.

\item Figure \ref{fig:fig5} shows the achievable rate regions for the 2-user CR and interference channels. Here, $\mathrm{Tx}_2$ is assumed to be cognitive in the sense that it has noncausal knowledge of the messages and codewords of $\mathrm{Tx}_1$. We consider two rate-splitting scenarios. In the first scenario, only $\mathrm{Tx}_2$ performs rate-splitting  (Fig. \ref{fig:fig5}(b)), while in the second scenario both $\mathrm{Tx}_1$ and $\mathrm{Tx}_2$ perform rate-splitting  (Fig. \ref{fig:fig5}(c)). We show the Han-Kobayashi rate region for the classical interference channel in (Fig. \ref{fig:fig5}(a)). Note that, in both the CR channels, $\mathrm{Tx}_2$ employs \emph{dirty paper coding} and cancels out known interference from $\mathrm{Tx}_1$. Therefore, it can transmit assuming that there is no interference due to $\mathrm{Tx}_1$. Further, if $\mathrm{Tx}_1$ performs rate-splitting and encode part of its message at a possibly low rate, then $\mathrm{Rx}_2$ can decode and cancel out that part of interference, thus enlarging the rate region (Fig. \ref{fig:fig5}(c)). The region shown in Fig. \ref{fig:fig5}(b) is the one presented in \cite{refjiang}. The outer bound that we have shown is the capacity region for the MIMO-BC.
\end{enumerate}

\item Three-user channels with \text{CuMS} (channel $\mathcal{C}_{\text{CuMS}}^2$): Here, we use the notation defined earlier for the three-user channels.
\begin{enumerate}
\item In Fig. \ref{fig:fig6}, we plot the rate of sender $\mathcal{S}_1$ ($R_1$) versus the sum of the  rates of $\mathcal{S}_2$ and $\mathcal{S}_3$ (i.e., $R_2+R_3$) for the channel $\mathcal{C}_{\text{CuMS}}^2$. In the figure, the outer bound, labeled $\mathrm{BC-Outerbound}$, is the intersection of (\ref{eq:eq20}), (\ref{eq:eq22})-(\ref{eq:eq24}) and (\ref{eq:eq26}). The innermost region corresponds to the achievable region given in Theorem 4.1. The second largest region corresponds to Corollary \ref{corol1}.
    Note that our inner bound is for a specific rate-splitting strategy at the transmitters, which the outer bounds do not account for, due to which the outer bound may be loose in the examples considered in this paper. More insight on the $R_2$ and $R_3$ achievable via our scheme, and how it compares with the outer bound, can be obtained from the plots presented later in the discussion. 

\item Figure \ref{fig:fig7} shows plots of the rate of $\mathcal{S}_2$ ($R_{2}$) versus the rate of $\mathcal{S}_3$ ($R_{3}$) when $\mathcal{S}_1$ achieves a minimum rate of $0$, $1$ and $1.5$ bps/Hz. Although there is a gap between the inner bound and the outer bound, Corollary \ref{corol2} coincides with the outer bound at the corner points. Note that due to the noncausal knowledge of $\mathcal{S}_{1}$'s message, by employing dirty paper coding, the interference from $\mathcal{S}_{1}$ can be eliminated at $\mathcal{R}_3$. Owing to this, with increase in rate $R_{1}$, the rate $R_{2}$ does not decrease much for relatively small values of $R_1$. On the other hand, as the rate $R_1$ increases, sender $\mathcal{S}_1$ cannot achieve the required rate without senders $\mathcal{S}_2$ and $\mathcal{S}_3$ using their noncausal message knowledge to help $\mathcal{S}_1$. Due to this, for higher values of $R_1$, the achievable rates of $R_2$ and $R_3$ decrease, as expected. Similar observations can also be made in the remaining cases presented below.

\item In Fig. \ref{fig:fig8}, we plot the rate of $\mathcal{S}_1$ ($R_1$), versus that of $\mathcal{S}_2$ ($R_2$), when $\mathcal{S}_3$ achieves a minimum rate of $R_{3}=0, 1 \text{ and } 1.5$ bps/Hz. The gap between the inner bound and the outer bound is relatively small. The rate of $\mathcal{S}_{2}$ does not decrease much as it employs dirty paper coding to eliminate interference when $\mathcal{S}_{1}$ and $\mathcal{S}_{3}$ achieve relatively smaller rates. It can be observed that as $\mathcal{S}_3$ achieves higher rates, the achievable rate region of the $\mathcal{S}_1$ and $\mathcal{S}_2$ shrinks. Also, when $R_3 > 0$, the rates achievable using the extensions provided by the corollaries lies completely above the rates achievable by the coding scheme in Sec. \ref{sec:achievability}, which is due to the suboptimality of that scheme with respect to the achievable rates of $\mathcal{S}_1$ and $\mathcal{S}_2$ for a fixed $R_3$. The rate of $\mathcal{S}_1$ has a larger relative reduction compared to that of $\mathcal{S}_2$, yet $\mathcal{S}_1$ achieves a higher rate than $\mathcal{S}_2$, as expected. Figure \ref{fig:fig9} shows a similar plot, but the rate of the $\mathcal{S}_1$ is compared with that of $\mathcal{S}_3$ instead of with $\mathcal{S}_2$. As $\mathcal{S}_2$ achieves a higher and higher rate, the rates of $\mathcal{S}_1$ and $\mathcal{S}_3$ decrease, but the reduction is smaller than that in Fig. \ref{fig:fig8}. Note that, in this case, the rate achieved by $\mathcal{S}_3$ matches the outer bound at the corner points when $R_2 = 0$.
\end{enumerate}

\item Three-user channels with \text{PrMS} (channel $\mathcal{C}_{\text{PrMS}}^2$):
\begin{enumerate}
\item In Fig. \ref{fig:fig10}, we plot the rate achieved by $\mathcal{S}_1$ versus the sum rate of $\mathcal{S}_2$ and $\mathcal{S}_3$ along with the outer bound. Here, the outer bound is different from the $\mathcal{C}_{\text{CuMS}}^{2}$ as the cutoff value used to bound $R_{2}$ is different for $\mathcal{S}_2$. The plot labeled \texttt{Outer bound} is the intersection of the capacity region given by \eqref{eq:eq20}, \eqref{eq:eq22}, \eqref{eq:eq24} - \eqref{eq:eq26}. Also shown is the plot of Corollary \ref{corol5}.

\item In Fig. \ref{fig:fig11}, we plot the rates of $\mathcal{S}_2$ and $\mathcal{S}_3$, when $\mathcal{S}_1$ achieves a minimum rate of $R_1=0,1,1.5$ bps/Hz along with the plots of Corollary \ref{corol6}. As $\mathcal{S}_1$ achieves a higher and higher minimum rate, the rates achieved by $\mathcal{S}_2$ and $\mathcal{S}_3$ decrease, as expected.

\item Fig. \ref{fig:fig12} shows the plot of the rate of $\mathcal{S}_1$ versus that of $\mathcal{S}_2$, when $\mathcal{S}_3$ achieves a minimum rates of 0, 0.4 5and 0.8 bps/Hz. Here again, we see that the rates of $\mathcal{S}_1$ and $\mathcal{S}_2$ decrease with increasing rate of $\mathcal{S}_3$. However, the decrease in $\mathcal{S}_2$'s rate is relatively smaller than that of $\mathcal{S}_1$, but $\mathcal{S}_1$ achieves a higher maximum rate compared to $\mathcal{S}_2$. Figure \ref{fig:fig13} is similar to Fig. \ref{fig:fig12}, except that the rate of $\mathcal{S}_3$ is plotted versus that of $\mathcal{S}_1$, with a constraint on the minimum rate achieved by $\mathcal{S}_2$. Similar trends as in Fig. \ref{fig:fig12} can be observed.
\end{enumerate}

\item Three-user channel with CoMS:
\begin{enumerate}
\item In Fig. \ref{fig:fig14}, we plot the sum rate of senders $\mathcal{S}_1$ and $\mathcal{S}_1$, $R_1+R_2$, versus the rate of $\mathcal{S}_3$, along with the outer bound and the plot of Corollary \ref{corol8}. The outer bound is the intersection of \eqref{eq:eq20}, \eqref{eq:eq23}, \eqref{eq:eq24}, \eqref{eq:eq26} and \eqref{eq:eq27}.

\item Figure \ref{fig:fig15} shows the plots of the rates of $\mathcal{S}_1$ and $\mathcal{S}_2$, when $\mathcal{S}_3$ achieves a minimum rate of $0.5, 1$ and $1.5$ bps/Hz, along with the plot of Corollary \ref{corol9}. Here again, we see that the rates of $\mathcal{S}_1$ and $\mathcal{S}_2$ decrease with increasing rate of $\mathcal{S}_3$. However, compared to Figs. \ref{fig:fig12} and \ref{fig:fig13}, the reduction in the size of the region is more symmetric i.e., both $R_1$ and $R_2$ simultaneously decrease, and roughly speaking, by the same relative amount.
\end{enumerate}
\end{enumerate}

The inner bounds for the $C_{\text{CuMS}}^{1}$ and $C_{\text{PrMS}}^{1}$ have not been plotted here. This is mainly because applying the Fourier-Motzkin elimination procedure on the rate region is a formidable task, given the number of inequalities involved. Nevertheless, one can expect (i) the achievable rate regions for $C_{\text{CuMS}}^{1}$ and $C_{\text{PrMS}}^{1}$ to be larger than that for $C_{\text{CuMS}}^{2}$ and $C_{\text{PrMS}}^{2}$ and (ii) the gap between the achievable rate region and the outer bound for $C_{\text{CuMS}}^{1}$ and $C_{\text{PrMS}}^{1}$ to be smaller than that to $C_{\text{CuMS}}^{2}$ and $C_{\text{PrMS}}^{2}$ respectively, because $\mathcal{S}_1$ also employs rate-splitting strategy in the former case.

As a concluding remark, note that, as mentioned above, the is a gap between the inner and outer bounds in all the cases plotted, although it is within two bits. There are a couple reasons for this. First, in the case of $C_{\text{CuMS}}^{2}$ and $C_{\text{PrMS}}^{2}$, $\mathcal{S}_1$ does not perform rate-splitting, thereby rendering the receivers of $\mathcal{S}_2$ and $\mathcal{S}_3$ vulnerable to interference caused due to $\mathcal{S}_1$'s transmissions. In the case of $C_{\text{CoMS}}$, neither $\mathcal{S}_1$ nor $\mathcal{S}_2$ performs rate-splitting, leading to poor interference management at all the receivers.
However, several corollaries were derived based on the idea of allowing senders to dedicate (part of) their power for transmitting the primary sender's message, which expanded the achievable rate regions, and it was shown that the achievable rates matched with the outer bounds at several corner points. A systematic way of expanding the rate region by including the different coding schemes is an open problem, which can be explored by future researchers.
Second, the outer bounds were derived by taking the intersection of  the capacity region with bidirectional sharing and the individual user rates with unidirectional sharing, and hence have a natural advantage over the purely-unidirectional model assumed in deriving the rate regions. Thus, the outer bound is general in the sense that it makes no explicit assumption about the decoding ability of the receivers resulting from rate-splitting at the transmitters; and in fact, the duality result implicitly assumes that the receivers can successfully decode the interfering signals to a large extent.

\section{Conclusions}
In this paper, we introduced multiuser channels with noncausal transmitter cooperation and presented three different ways of message sharing which we termed cumulative message sharing (CuMS), primary-only message sharing (PrMS) and cognitive-only message sharing (CoMS). We modified the channel model to introduce rate-splitting to enable better interference management at the receivers. We then derived an achievable rate region for each of the channels by employing a coding scheme which comprised a combination of superposition and Gel'fand-Pinsker coding techniques. Numerical evaluation of the Gaussian case enabled a visual comparison between the rate regions and some simple outer bounds. We also presented some corollaries using which several achievable rate tuples for the Gaussian channel were readily identified, thereby enlarging the rate regions.  Thus, we have demonstrated the effect of noncausal cooperation and rate-splitting in multiuser networks; the former aims at improving the throughput capacity by conforming itself to the overlay cognitive radio network paradigm, while the latter addresses the issue of interference management at the receivers. Rate-constrained cooperation, wherein the cognitive radio estimates the message index transmitted by the primary user in a causal manner, is a more discernible formulation of the practical scenario and is an interesting open problem.

\section*{Acknowledgments}
K. G. Nagananda and Chandra R. Murthy would like to thank Rajesh Sundaresan at Indian Institute of Science for valuable discussions. The work of K. G. Nagananda and Shalinee Kishore at Lehigh University was partly supported by the National Science Foundation under
Grant CNS $0721433/0721445$. The work of Parthajit Mohapatra, Chandra R. Murthy and the initial work of K. G. Nagananda at the Indian Institute of Science was partly supported by a research grant from the Aerospace Network Research Consortium.

\appendices
\section{}
The channel model $\mathcal{C}_{\text{CuMS}}^1$ is symmetric, in the sense that all transmitters perform rate-splitting so that each receiver can decode and cancel out the interfering signals from the non-pairing senders. However, the receivers are not required to decode the public part of the non-pairing transmitter's message \emph{correctly}. Considering these, for the channel $\mathcal{C}_{\text{CuMS}}^1$, we can derive a total of 36 inequalities, as given below.
\begin{eqnarray}
\nonumber R_{10} \leq I(W_0;W_1,U_0,V_0,Y_1|Q),\\
\nonumber R_{11} \leq I(W_1;W_0,U_0,V_0,Y_1|Q),\\
\nonumber R_{10}+R_{11} \leq I(W_0,W_1;U_0,V_0,Y_1|Q)+I(W_0;W_1),\\
\nonumber R_{10}+R_{20} \leq I(W_0,U_0;W_1,V_0,Y_1|Q)+I(W_0;U_0|Q) - I(W_0,W_1;U_0|Q),\\
\nonumber R_{10}+R_{30} \leq I(W_0,V_0;W_1,U_0,Y_1|Q)+I(W_0;V_0|Q) - I(W_0,W_1,U_0,U_2;V_0|Q),\\
\nonumber R_{11}+R_{20} \leq I(W_1,U_0;W_0,V_0,Y_1|Q)+I(W_1;U_0|Q) - I(W_0,W_1;U_0|Q),\\
\nonumber R_{11}+R_{30} \leq I(W_1,V_0;W_0,U_0,Y_1|Q)+I(W_1;V_0|Q) - I(W_0,W_1,U_0,U_2;V_0|Q),\\
\nonumber R_{10}+R_{11}+R_{20} \leq I(W_0,W_1,U_0;V_0,Y_1|Q)+I(W_0,W_1;U_0|Q)+I(W_0;W_1|Q) - I(W_0,W_1;U_0|Q),\\
\nonumber R_{10}+R_{11}+R_{30} \leq I(W_0,W_1,V_0;U_0,Y_1|Q)+I(W_0,W_1;V_0|Q)+I(W_0;W_1|Q) - I(W_0,W_1,U_0,U_2;V_0|Q),\\
\nonumber R_{10}+R_{20}+R_{30} \leq I(W_0,U_0,V_0;W_1,Y_1|Q)+I(W_0,U_0;V_0|Q)+I(W_0;U_0|Q)\\ \nonumber  - I(W_0,W_1;U_0|Q) - I(W_0,W_1,U_0,U_2;V_0|Q)\\
\nonumber R_{11}+R_{20}+R_{30} \leq I(W_1,U_0,V_0;W_0,Y_1|Q)+I(W_1,U_0;V_0|Q)+I(W_1;U_0|Q)\\ \nonumber  - I(W_0,W_1;U_0|Q) - I(W_0,W_1,U_0,U_2;V_0|Q)\\
\nonumber R_{10}+R_{11}+R_{20}+R_{30} \leq I(W_0,W_1,U_0,V_0;Y_1|Q)+I(W_0,W_1,U_0;V_0|Q)+I(W_0,W_1;U_0|Q)+I(W_0,W_1|Q)\\ \nonumber - I(W_0,W_1;U_0|Q) - I(W_0,W_1,U_0,U_2;V_0|Q),\\
\nonumber R_{20} \leq I(U_0;W_0,U_2,V_0,Y_2|Q)-I(W_0,W_1;U_0|Q),\\
\nonumber R_{22} \leq I(U_2;W_0,U_0,V_0,Y_2|Q)-I(W_0,W_1;U_2|Q),\\
\nonumber R_{20}+R_{22} \leq I(U_0,U_2;W_0,V_0,Y_2|Q)+I(U_0;U_2|Q)-I(W_0,W_1;U_0|Q)-I(W_0,W_1;U_2|Q),\\
\nonumber R_{10}+R_{20} \leq I(W_0,U_0;U_2,V_0,Y_2|Q)+I(W_0;U_0|Q) - I(W_0,W_1;U_0|Q),\\
\nonumber R_{10}+R_{22} \leq I(W_0,U_2;U_0,V_0,Y_2|Q)+I(W_0;U_2|Q) - I(W_0,W_1;U_2|Q),\\
\nonumber R_{20}+R_{30} \leq I(U_0,V_0;W_0,U_2,Y_2|Q)+I(U_0;V_0|Q) - I(W_0,W_1;U_0|Q) - I(W_0,W_1,U_0,U_2;V_0|Q),\\
\nonumber R_{22}+R_{30} \leq I(U_2,V_0;W_0,U_0,Y_2|Q)+I(U_2;V_0|Q) - I(W_0,W_1;U_2|Q) - I(W_0,W_1,U_0,U_2;V_0|Q),\\
\nonumber R_{10}+R_{20}+R_{22} \leq I(W_0,U_0,U_2;V_0,Y_2|Q)+I(W_0,U_0;U_2|Q)+I(W_0;U_0|Q)\\ \nonumber  - I(W_0,W_1;U_0|Q)-I(W_0,W_1;U_2|Q)\\
\nonumber R_{10}+R_{20}+R_{30} \leq I(W_0,U_0,V_0;U_2,Y_2|Q)+I(W_0,U_0;V_0|Q)+I(W_0;U_0|Q) - \\ \nonumber  I(W_0,W_1;U_0|Q) - I(W_0,W_1,U_0,U_2;V_0|Q)\\
\nonumber R_{10}+R_{22}+R_{30} \leq I(W_0,U_2,V_0;U_0,Y_2|Q)+I(W_0,U_2;V_0|Q)+I(W_0;U_2|Q) - \\ \nonumber I(W_0,W_1;U_2|Q) - I(W_0,W_1,U_0,U_2;V_0|Q)\\
\nonumber R_{20}+R_{22}+R_{30} \leq I(U_0,U_2,V_0;W_0,Y_2|Q)+I(U_0,U_2;V_0|Q)+I(U_0;U_2|Q) \\ \nonumber -I(W_0,W_1;U_0|Q)-I(W_0,W_1;U_2|Q) -I(W_0,W_1,U_0,U_2;V_0|Q)\\
\nonumber R_{10}+R_{20}+R_{22}+R_{30} \leq I(W_0,U_0,U_2,V_0;Y_2|Q)+I(W_0,U_0,U_2;V_0|Q)+I(W_0,U_0;U_2|Q)\\ \nonumber +I(W_0,U_0|Q) - -I(W_0,W_1;U_0|Q)-I(W_0,W_1;U_2|Q) -I(W_0,W_1,U_0,U_2;V_0|Q),\\
\nonumber R_{30} \leq I(V_0;W_0,U_0,V_3,Y_3|Q)-I(W_0,W_1,U_0,U_2;V_0|Q),\\
\nonumber R_{33} \leq I(V_3;W_0,U_0,V_0,Y_3|Q)-I(W_0,W_1,U_0,U_2;V_3|Q),\\
\nonumber R_{30}+R_{33} \leq I(V_0,V_3;W_0,U_0,Y_3|Q)+I(V_0;V_3|Q)\\ \nonumber -I(W_0,W_1,U_0,U_2;V_0|Q)-I(W_0,W_1,U_0,U_2;V_3|Q),\\
\nonumber R_{10}+R_{30} \leq I(W_0,V_0;U_0,V_3,Y_3|Q)+I(W_0;V_0|Q) - I(W_0,W_1,U_0,U_2;V_0|Q),\\
\nonumber R_{10}+R_{33} \leq I(W_0,V_3;U_0,V_0,Y_3|Q)+I(W_0;V_3|Q) - I(W_0,W_1,U_0,U_2;V_3|Q),\\
\nonumber R_{20}+R_{30} \leq I(U_0,V_0;W_0,V_3,Y_3|Q)+I(U_0;V_0|Q) - I(W_0,W_1;U_0|Q) - I(W_0,W_1,U_0,U_2;V_0|Q),\\
\nonumber _{20}+R_{33} \leq I(U_0,V_3;W_0,V_0,Y_3|Q)+I(U_0;V_3|Q) - I(W_0,W_1;U_0|Q) - I(W_0,W_1,U_0,U_2;V_3|Q),\\
\nonumber R_{10}+R_{20}+R_{30} \leq I(W_0,U_0,V_0;V_3,Y_3|Q)+I(W_0,U_0;V_0|Q)+I(W_0;U_0|Q) \\ \nonumber  - I(W_0,W_1;U_0|Q) - I(W_0,W_1,U_0,U_2;V_0|Q),\\
\nonumber R_{10}+R_{20}+R_{33} \leq I(W_0,U_0,V_3;V_0,Y_3|Q)+I(W_0,U_0;V_3|Q)+I(W_0;U_0|Q) \\ \nonumber - I(W_0,W_1;U_0|Q) - I(W_0,W_1,U_0,U_2;V_3|Q),\\
\nonumber R_{10}+R_{30}+R_{33} \leq I(W_0,V_0,V_3;U_0,Y_3|Q)+I(W_0,V_0;V_3|Q)+I(W_0;V_0|Q) \\ \nonumber  -I(W_0,W_1,U_0,U_2;V_0|Q)-I(W_0,W_1,U_0,U_2;V_3|Q),\\
\nonumber R_{20}+R_{30}+R_{33} \leq I(U_0,V_0,V_3;W_0,Y_3|Q)+I(U_0,V_0;V_3|Q)+I(U_0;V_0|Q)\\ \nonumber  - I(W_0,W_1;U_0|Q)- I(W_0,W_1,U_0,U_2;V_0|Q)-I(W_0,W_1,U_0,U_2;V_3|Q),\\
\nonumber R_{10}+R_{20}+R_{30}+R_{33} \leq I(W_0,U_0,V_0,V_3;Y_3|Q)+I(W_0,U_0,V_0;V_3|Q)+I(W_0,U_0;V_0|Q)+I(W_0;U_0|Q)\\ \nonumber  - I(W_0,W_1;U_0|Q)- I(W_0,W_1,U_0,U_2;V_0|Q)-I(W_0,W_1,U_0,U_2;V_3|Q).
\end{eqnarray}
\\
The channel model $\mathcal{C}_{\text{CuMS}}^2$ is not symmetric, in the sense that only senders $\mathcal{S}_2$ and $\mathcal{S}_3$ perform rate-splitting. This results in receiver $\mathcal{R}_1$ being able to decode the public part of the non-pairing sender's message, while receivers $\mathcal{R}_2$ and $\mathcal{R}_3$ can only decode the message from the pairing transmitter. We have a total of 10 inequalities which are given below.
\begin{eqnarray}
\nonumber R_{11} \leq I(W;U_1,V_1,Y_1|Q),\\
\nonumber R_{11} + R_{21} \leq I(W,U_1;V_1,Y_1|Q),\\
\nonumber R_{11} +R_{31} \leq I(W,V_1;U_1,Y_1|Q)+I(W;V_1|Q) -I(W,U_1,U_2;V_1|Q),\\
\nonumber R_{11} + R_{21}+R_{31} \leq I(W,U_1,V_1;Y_1|Q)I(W,U_1;V_1|Q)-I(W,U_1,U_2;V_1|Q),\\
\nonumber R_{21} \leq I(U_1;U_2,Y_2|Q)-I(W;U_1|Q),\\
\nonumber R_{22} \leq I(U_2;U_1,Y_2|Q)-I(W;U_2|Q),\\
\nonumber R_{21}+R_{22} \leq I(U_1,U_2;Y_2|Q)+I(U_1;U_2|Q)-I(W;U_1|Q)-I(W;U_2|Q),\\
\nonumber R_{31} \leq I(V_1;V_3,Y_3|Q)-I(W,U_1,U_2;V_1|Q),\\
\nonumber R_{33} \leq I(V_3;V_1,Y_3|Q)-I(W,U_1,U_2;V_3|Q),\\
\nonumber R_{31}+R_{33} \leq I(V_1,V_3;Y_3|Q)+I(V_1;V_3|Q)-I(W,U_1,U_2;V_3|Q)-I(W,U_1,U_2;V_1|Q).
\end{eqnarray}

\section{}
An achievable rate region for the channel $\mathcal{C}_{\text{PrMS}}^1$ is given by the following inequalities. The number of inequalities is the same as with the case of $\mathcal{C}_{\text{CuMS}}^1$. Note that, the only difference between the two channel models is that they have different message-sharing schemes.
\begin{eqnarray}
\nonumber R_{10} \leq I(W_0;W_1,U_0,V_0,Y_1|Q),\\
\nonumber R_{11} \leq I(W_1;W_0,U_0,V_0,Y_1|Q),\\
\nonumber R_{10}+R_{11} \leq I(W_0,W_1;U_0,V_0,Y_1|Q)+I(W_0;W_1|Q),\\
\nonumber R_{10}+R_{20} \leq I(W_0,U_0;W_1,V_0,Y_1|Q)+I(W_0;U_0|Q) - I(W_0,W_1;U_0|Q),\\
\nonumber R_{10}+R_{30} \leq I(W_0,V_0;W_1,U_0,Y_1|Q)+I(W_0;V_0|Q) - I(W_0,W_1;V_0|Q),\\
\nonumber R_{11}+R_{20} \leq I(W_1,U_0;W_0,V_0,Y_1|Q)+I(W_1;U_0|Q) - I(W_0,W_1;U_0|Q),\\
\nonumber R_{11}+R_{30} \leq I(W_1,V_0;W_0,U_0,Y_1|Q)+I(W_1;V_0|Q) - I(W_0,W_1;V_0|Q),\\
\nonumber R_{10}+R_{11}+R_{20} \leq I(W_0,W_1,U_0;V_0,Y_1|Q)+I(W_0,W_1;U_0|Q)+I(W_0;W_1|Q) - I(W_0,W_1;U_0|Q),\\
\nonumber R_{10}+R_{11}+R_{30} \leq I(W_0,W_1,V_0;U_0,Y_1|Q)+I(W_0,W_1;V_0|Q)+I(W_0;W_1|Q) - I(W_0,W_1;V_0|Q),\\
\nonumber R_{10}+R_{20}+R_{30} \leq I(W_0,U_0,V_0;W_1,Y_1|Q)+I(W_0,U_0;V_0|Q)+I(W_0;U_0|Q)\\ \nonumber  - I(W_0,W_1;U_0|Q) - I(W_0,W_1;V_0|Q)\\
\nonumber R_{11}+R_{20}+R_{30} \leq I(W_1,U_0,V_0;W_0,Y_1|Q)+I(W_1,U_0;V_0|Q)+I(W_1;U_0|Q)\\ \nonumber  - I(W_0,W_1;U_0|Q) - I(W_0,W_1;V_0|Q)\\
\nonumber R_{10}+R_{11}+R_{20}+R_{30} \leq I(W_0,W_1,U_0,V_0;Y_1|Q)+I(W_0,W_1,U_0;V_0|Q)+I(W_0,W_1;U_0|Q)\\ \nonumber  +I(W_0,W_1|Q) - I(W_0,W_1;U_0|Q) - I(W_0,W_1;V_0|Q)\\
\nonumber R_{20} \leq I(U_0;W_0,U_2,V_0,Y_2|Q)-I(W_0,W_1;U_0|Q),\\
\nonumber R_{22} \leq I(U_2;W_0,U_0,V_0,Y_2|Q)-I(W_0,W_1;U_2|Q),\\
\nonumber R_{20}+R_{22} \leq I(U_0,U_2;W_0,V_0,Y_2|Q)+I(U_0;U_2|Q)-I(W_0,W_1;U_0|Q)-I(W_0,W_1;U_2|Q),\\
\nonumber R_{10}+R_{20} \leq I(W_0,U_0;U_2,V_0,Y_2|Q)+I(W_0;U_0|Q) - I(W_0,W_1;U_0|Q),\\
\nonumber R_{10}+R_{22} \leq I(W_0,U_2;U_0,V_0,Y_2|Q)+I(W_0;U_2|Q) - I(W_0,W_1;U_2|Q),\\
\nonumber R_{20}+R_{30} \leq I(U_0,V_0;W_0,U_2,Y_2|Q)+I(U_0;V_0|Q) - I(W_0,W_1;U_0|Q) - I(W_0,W_1;V_0|Q),\\
\nonumber R_{22}+R_{30} \leq I(U_2,V_0;W_0,U_0,Y_2|Q)+I(U_2;V_0|Q) - I(W_0,W_1;U_2|Q) - I(W_0,W_1;V_0|Q),\\
\nonumber R_{10}+R_{20}+R_{22} \leq I(W_0,U_0,U_2;V_0,Y_2|Q)+I(W_0,U_0;U_2|Q)+I(W_0;U_0|Q)\\ \nonumber - I(W_0,W_1;U_0|Q)-I(W_0,W_1;U_2|Q)\\
\nonumber R_{10}+R_{20}+R_{30} \leq I(W_0,U_0,V_0;U_2,Y_2|Q)+I(W_0,U_0;V_0|Q)+I(W_0;U_0|Q) - \\ \nonumber  I(W_0,W_1;U_0|Q) - I(W_0,W_1;V_0|Q)\\
\nonumber R_{10}+R_{22}+R_{30} \leq I(W_0,U_2,V_0;U_0,Y_2|Q)+I(W_0,U_2;V_0|Q)+I(W_0;U_2|Q) - \\ \nonumber I(W_0,W_1;U_2|Q) - I(W_0,W_1;V_0|Q)\\
\nonumber R_{20}+R_{22}+R_{30} \leq I(U_0,U_2,V_0;W_0,Y_2|Q)+I(U_0,U_2;V_0|Q)+I(U_0;U_2|Q) \\ \nonumber -I(W_0,W_1;U_0|Q)-I(W_0,W_1;U_2|Q) -I(W_0,W_1;V_0|Q)\\
\nonumber R_{10}+R_{20}+R_{22}+R_{30} \leq I(W_0,U_0,U_2,V_0;Y_2|Q)+I(W_0,U_0,U_2;V_0|Q)+I(W_0,U_0;U_2|Q)\\ \nonumber +I(W_0,U_0|Q) - -I(W_0,W_1;U_0|Q)-I(W_0,W_1;U_2|Q) -I(W_0,W_1;V_0|Q)\\
\nonumber R_{30} \leq I(V_0;W_0,U_0,V_3,Y_3|Q)-I(W_0,W_1;V_0|Q),\\
\nonumber R_{33} \leq I(V_3;W_0,U_0,V_0,Y_3|Q)-I(W_0,W_1;V_3|Q),\\
\nonumber R_{30}+R_{33} \leq I(V_0,V_3;W_0,U_0,Y_3|Q)+I(V_0;V_3|Q)\\ \nonumber -I(W_0,W_1;V_0|Q)-I(W_0,W_1;V_3|Q),\\
\nonumber R_{10}+R_{30} \leq I(W_0,V_0;U_0,V_3,Y_3|Q)+I(W_0;V_0|Q) - I(W_0,W_1;V_0|Q),\\
\nonumber R_{10}+R_{33} \leq I(W_0,V_3;U_0,V_0,Y_3|Q)+I(W_0;V_3|Q) - I(W_0,W_1;V_3|Q),\\
\nonumber R_{20}+R_{30} \leq I(U_0,V_0;W_0,V_3,Y_3|Q)+I(U_0;V_0|Q) - I(W_0,W_1;U_0|Q) - I(W_0,W_1;V_0|Q),\\
\nonumber R_{20}+R_{33} \leq I(U_0,V_3;W_0,V_0,Y_3|Q)+I(U_0;V_3|Q) - I(W_0,W_1;U_0|Q) - I(W_0,W_1;V_3|Q),\\
\nonumber R_{10}+R_{20}+R_{30} \leq I(W_0,U_0,V_0;V_3,Y_3|Q)+I(W_0,U_0;V_0|Q)+I(W_0;U_0|Q) \\ \nonumber - I(W_0,W_1;U_0|Q) - I(W_0,W_1;V_0|Q),\\
\nonumber R_{10}+R_{20}+R_{33} \leq I(W_0,U_0,V_3;V_0,Y_3|Q)+I(W_0,U_0;V_3|Q)+I(W_0;U_0|Q) \\ \nonumber - I(W_0,W_1;U_0|Q) - I(W_0,W_1;V_3|Q),\\
\nonumber R_{10}+R_{30}+R_{33} \leq I(W_0,V_0,V_3;U_0,Y_3|Q)+I(W_0,V_0;V_3|Q)+I(W_0;V_0|Q) \\ \nonumber -I(W_0,W_1;V_0|Q)-I(W_0,W_1;V_3|Q),\\
\nonumber R_{20}+R_{30}+R_{33} \leq I(U_0,V_0,V_3;W_0,Y_3|Q)+I(U_0,V_0;V_3|Q)+I(U_0;V_0|Q)\\ \nonumber  - I(W_0,W_1;U_0|Q)- I(W_0,W_1;V_0|Q)-I(W_0,W_1;V_3|Q),\\
\nonumber R_{10}+R_{20}+R_{30}+R_{33} \leq I(W_0,U_0,V_0,V_3;Y_3|Q)+I(W_0,U_0,V_0;V_3|Q)+I(W_0,U_0;V_0|Q)+I(W_0;U_0|Q)\\ \nonumber  - I(W_0,W_1;U_0|Q)- I(W_0,W_1;V_0|Q)-I(W_0,W_1;V_3|Q).
\end{eqnarray}
\\
An achievable rate region for the channel $\mathcal{C}_{\text{PrMS}}^2$ is given by the following inequalities. As before, the number of inequalities is same as that for the channel $\mathcal{C}_{\text{CuMS}}^2$. One should also guard against direct comparison of the rate region equations of various channel models, since these channels are governed by joint distributions with different underlying factorizations.
\begin{eqnarray}
\nonumber R_{11} \leq I(W;U_1,V_1,Y_1|Q)\\
\nonumber R_{11} + R_{21} \leq I(W,U_1;V_1,Y_1|Q)\\
\nonumber R_{11}+R_{31} \leq I(W,V_1;U_1,Y_1|Q)\\
\nonumber R_{11} + R_{21}+R_{31} \leq I(W,U_1,V_1;Y_1|Q)+I(W,U_1;V_1|Q)-I(W;V_1|Q),\\
\nonumber R_{21} \leq I(U_1;U_2,Y_2|Q)-I(W;U_1|Q)\\
\nonumber R_{22} \leq I(U_2;U_1,Y_2|Q)-I(W;U_2|Q)\\
\nonumber R_{21}+R_{22} \leq I(U_1,U_2;Y_2|Q)+I(U_1;U_2|Q)-I(W;U_1|Q)-I(W;U_2|Q),\\
\nonumber R_{31} \leq I(V_1;V_3,Y_3|Q)-I(W;V_1|Q),\\
\nonumber R_{33} \leq I(V_3;V_1,Y_3|Q)-I(W;V_3|Q),\\
\nonumber R_{31}+R_{33} \leq I(V_1,V_3;Y_3|Q)+I(V_1;V_3|Q)-I(W;V_3|Q)-I(W;V_1|Q).
\end{eqnarray}

\section{}
An achievable rate region for the channel $\mathcal{C}_{\text{CoMS}}$ is given by the following inequalities. Here, sender $\mathcal{S}_3$ has noncausal knowledge of the messages and codewords of $\mathcal{S}_1$ and $\mathcal{S}_2$, and performs rate-splitting. There is no rate-splitting at senders $\mathcal{S}_1$ and $\mathcal{S}_2$.
\begin{eqnarray}
\nonumber R_1 \leq I(W;V_0,Y_1|Q), \\
\nonumber R_1 + R_{31} \leq I(W,V_0;Y_1|Q) + I(W;V_0|Q) - I(W,U;V_0|Q),\\
\nonumber R_2 \leq I(U;V_0,Y_2|Q), \\
\nonumber R_2 + R_{31} \leq I(U,V_0;Y_2|Q) + I(U;V_0|Q) - I(W,U;V_0|Q),\\
\nonumber R_{31} \leq I(V_0;V_3,Y_3|Q) - I(W,U;V_0|Q),\\
\nonumber R_{33} \leq I(V_3;V_0,Y_3|Q) - I(W,U;V_3|Q),\\
\nonumber R_{31}+R_{33} \leq I(V_0,V_3;Y_3|Q) + I(V_0;V_3|Q) - I(W,U;V_0|Q) - I(W,U;V_3|Q).
\end{eqnarray}

\section{}
Proof of achievability for the channel $\mathcal{C}_{\text{PrMS}}^1$:
\subsection{Codebook Generation}
Let us fix $p(.) \in \mathcal{P}$. Generate a random time sharing codeword $\textbf{q}$ of length $n$, according to the distribution $\prod_{i=1}^np(q_i)$. For $\gamma =0,1$, $\tau=0,2$ and $\rho=0,3$:\\
generate $2^{nR_{1\gamma}}$ independent codewords $\textbf{W}_{\gamma}(j_{\gamma})$, $j_{\gamma} \in \{1,\ldots,2^{nR_{1\gamma}}\}$ according to $\prod_{i=1}^np(w_{\gamma i}|q_i)$. For every codeword pair
$(\textbf{w}_0(j_0),\textbf{w}_1(j_1))$, generate one codeword $\textbf{X}_1(j_0,j_1)$ according to $\prod_{i=1}^np(x_{1i}|w_i(j),q_i)$.\\ \\
Generate $2^{n(R_{2\tau}+I(W_0,W_1;U_{\tau}|Q)+4\epsilon)}$ independent code words $\textbf{U}_{\tau}(l_{\tau})$, according to $\prod_{i=1}^np(u_{\tau i}|q_i)$. For every codeword tuple $(\textbf{u}_0(l_0), \textbf{u}_2(l_2), \textbf{w}_0(j_0),\textbf{w}_1(j_1))$, generate one code word $\textbf{X}_2(l_0, l_2, j_0,j_1)$ according to \\$\prod_{i=1}^np(x_{2i}|u_{0i}(l_0), u_{2i}(l_2), w_{0i}(j_0), w_{1i}(j_1) q_i)$. Uniformly distribute the $2^{n(R_{2\tau}+I(W_0,W_1;U_{\tau}|Q)+4\epsilon)}$ code words $\textbf{U}_{\tau}(l_{\tau})$ into $2^{nR_{2\tau}}$ bins indexed by $k_{\tau} \in \{1,\ldots,2^{nR_{2\tau}}\}$ such that each bin contains $2^{n(I(W_0,W_1;U_{\tau}|Q)+4\epsilon)}$ codewords. \\
\\
Generate $2^{n(R_{3\rho}+I(W_0,W_1;V_{\rho}|Q)+4\epsilon)}$ independent code words $\textbf{V}_{\rho}(t_{\rho})$, according to $\prod_{i=1}^np(v_{\rho i}|q_i)$. For every code word tuple $(\textbf{v}_0(t_0), \textbf{v}_3(t_3),  \textbf{w}_0(j_0),\textbf{w}_1(j_1))$, generate one codeword $\textbf{X}_3(t_0, t_3,j_0, j_1)$ according to \\$\prod_{i=1}^np(x_{3i}|v_{0i}(t_0), v_{3i}(t_3), w_{0i}(j_0), w_{1i}(j_1) q_i)$. Distribute $2^{n(R_{3\rho}+I(W_0,W_1;V_{\rho}|Q)+4\epsilon)}$ code words $\textbf{V}_{\rho}(t_{\rho})$ uniformly into $2^{nR_{3\rho}}$ bins indexed by $r_{\rho} \in \!\! \{1,\ldots,2^{nR_{3\rho}}\}$ such that each bin contains $2^{n(I(W_0,W_1;V_{\rho}|Q)+4\epsilon)}$ code words. The indices are given by $j_{\gamma} \in \{1,\ldots,2^{nR_{1\gamma}}\}$, $l_{\tau} \in  \{1,\ldots,2^{n(R_{2\tau}+I(W_0,W_1;U_{\tau}|Q)+4\epsilon)}\}$, $t_{\rho} \in \{1,\ldots,$ $2^{n(R_{3\rho}+I(W_0,W_1;V_{\rho}|Q)+4\epsilon)}\}$.

\subsection{Encoding $\&$ Transmission} Let us suppose that the source message vector generated at the three senders is $\left(m_{10},m_{11},m_{20}, m_{22}, m_{30}, m_{33}\right) = \left(j_0, j_1, k_0, k_2, r_0, r_3\right)$. $\mathcal{S}_1$ transmits codeword $\textbf{x}_1(j_0,j_1)$ with $n$ channel uses. $\mathcal{S}_2$ first looks for a codeword $\textbf{u}_0(l_0)$ in bin $k_0$ such that $(\textbf{u}_0(l_0), \textbf{w}_0(j_0), \textbf{w}_1(j_1), \textbf{q}) \in A_{\epsilon}^{(n)}$, and a codeword $\textbf{u}_2(l_2)$ in bin $k_2$ such that \\$(\textbf{u}_2(l_2), \textbf{w}_0(j_0), \textbf{w}_1(j_1),\textbf{q}) \in A_{\epsilon}^{(n)}$. It then transmits $\textbf{x}_2(l_0, l_2, j_0, j_1)$ through $n$ channel uses. Otherwise, $\mathcal{S}_2$ declares an error. $\mathcal{S}_3$ first looks for a codeword $\textbf{v}_0(t_0)$ in bin $r_0$ such that $(\textbf{v}_0(t_0), \textbf{w}_0(j_0), \textbf{w}_1(j_1), \textbf{q}) \in A_{\epsilon}^{(n)}$, and a codeword $\textbf{v}_3(t_3)$ in bin $r_3$ such that $(\textbf{v}_3(t_3), \textbf{w}_0(j_0), \textbf{w}_1(j_1), \textbf{q}) \in A_{\epsilon}^{(n)}$. It then transmits $\textbf{x}_3(t_0, t_3, j_0,j_1)$ through $n$ channel uses. Otherwise, $\mathcal{S}_3$ declares an error. The transmissions are assumed to be perfectly synchronized.

\subsection{Decoding}
Recall the decoding capability assumed here (see Table II). The three receivers accumulate an $n$-length channel output sequence: $\textbf{y}_1$ at $\mathcal{R}_1$, $\textbf{y}_2$ at $\mathcal{R}_2$ and $\textbf{y}_3$ at $\mathcal{R}_3$. Decoder 1 looks for all index tuples $(\hat{j}_0, \hat{j}_1, \hat{\hat{l}}_0, \hat{\hat{t}}_0)$ such that \\ $(\textbf{w}_0(\hat{j}_0), \textbf{w}_1(\hat{j}_1), \textbf{u}_0(l_0), \textbf{v}_0(t_0), \textbf{y}_1, \textbf{q}) \in A_{\epsilon}^{(n)}$. If $\hat{j}_0$ and $\hat{j}_1$ in all the index tuples found are the same, $\mathcal{R}_1$ determines $(m_{10},m_{11}) = (\hat{j}_0,\hat{j}_1)$ for some $l_0$ and $t_0$. Otherwise, it declares an error.
Decoder 2 looks for all index tuples $(\hat{l}_0, \hat{l}_2,\hat{\hat{j}}_0,\hat{\hat{t}}_0 )$ such that $(\textbf{w}_0(\hat{\hat{j}}_0),\textbf{u}_0(\hat{l}_0), \textbf{u}_2(\hat{l}_2), \textbf{v}_0(\hat{\hat{t}}_0), \textbf{y}_2, \textbf{q}) \in A_{\epsilon}^{(n)}$. If $\hat{l}_0$ in all the index pairs found are indices of codewords $\textbf{u}_0(\hat{l}_0)$ from the same bin with index $\hat{k}_0$, and $\hat{l}_2$ in all the index pairs found are indices of codewords $\textbf{u}_2(\hat{l}_2)$ from the same bin with index $\hat{k}_2$, then $\mathcal{R}_2$ determines $(m_{20}, m_{22}) = (\hat{k}_0, \hat{k}_2)$. Otherwise, it declares an error.
Decoder 3 looks for all index pairs $(\hat{t}_0, \hat{t}_3,\hat{\hat{l}}_0,\hat{\hat{j}}_0)$ such that $(\textbf{w}_0(\hat{\hat{j}}_0), \textbf{u}_0(\hat{\hat{l}}_0),\textbf{v}_0(\hat{t}_0), \textbf{v}_3(\hat{t}_3), \textbf{y}_3, \textbf{q}) \in A_{\epsilon}^{(n)}$. If $\hat{t}_0$ in all the index pairs found are indices of codewords $\textbf{v}_0(\hat{t}_0)$ from the same bin with index $\hat{r}_0$, and $\hat{t}_3$ in all the index pairs found are indices of codewords $\textbf{v}_3(\hat{t}_3)$ from the same bin with index $\hat{r}_3$, then $\mathcal{R}_3$ determines $(m_{30}, m_{33}) = (\hat{r}_0, \hat{r}_3)$. Otherwise, it declares an error.

\subsection{Analysis of the Probabilities of Error}
In this subsection we derive upper bounds on the probabilities of error events which happen during encoding and decoding processes. We assume that a source message vector $\left(m_{10},m_{11}, m_{20}, m_{22}, m_{30}, m_{33}\right)$ is encoded and transmitted. As before, we consider the analysis of probability of encoding error at senders $\mathcal{S}_2$ and $\mathcal{S}_3$, and the analysis of probability of decoding error at each of the three receivers $\mathcal{R}_1$, $\mathcal{R}_2$, and $\mathcal{R}_3$  separately.\\
\\
First, let us define the following events:\\
$(i)$ $E_{j_0j_1l_0} \triangleq \left\{\left(\textbf{W}_0(j_0), \textbf{W}_1(j_1), \textbf{U}_0(l_0), \textbf{q} \right) \in A_{\epsilon}^{(n)}\right\}$,\\
$(ii)$ $E_{j_0j_1l_2} \triangleq \left\{\left(\textbf{W}_0(j_0), \textbf{W}_1(j_1), \textbf{U}_2(l_2), \textbf{q} \right) \in A_{\epsilon}^{(n)}\right\}$,\\
$(iii)$ $E_{j_0j_1t_0} \triangleq \left\{\left(\textbf{W}_0(j_0), \textbf{W}_1(j_1), \textbf{V}_0(t_0), \textbf{q}\right) \in A_{\epsilon}^{(n)}\right\}$,\\
$(iv)$ $E_{j_0j_1t_3} \triangleq \left\{\left(\textbf{W}_0(j_0), \textbf{W}_1(j_1), \textbf{V}_3(t_3), \textbf{q}\right) \in A_{\epsilon}^{(n)}\right\}$,\\
$(v)$ $E_{j_0j_1l_0t_0} \triangleq \left\{(\textbf{W}_0(j_0), \textbf{W}_1(j_1), \textbf{U}_0(l_0), \textbf{V}_0(t_0), \textbf{Y}_1, \textbf{q}) \in A_{\epsilon}^{(n)}\right\}$,\\
$(vi)$ $E_{j_0l_0l_2t_0} \triangleq \left\{(\textbf{W}_0(j_0), \textbf{U}_0(l_0), \textbf{U}_2(l_2), \textbf{V}_0(t_0), \textbf{Y}_2, \textbf{q}) \in A_{\epsilon}^{(n)} \right\}$,\\
$(vii)$ $E_{j_0l_0t_0t_3} \triangleq \left\{(\textbf{W}_0(j_0), \textbf{U}_0(l_0), \textbf{V}_0(t_0), \textbf{V}_3(t_3), \textbf{Y}_3, \textbf{q}) \in A_{\epsilon}^{(n)} \right\}$.\\
$E_{(.)}^c \triangleq$ complement of the event $E_{(.)}$. Events $(i) - (iv)$ will be used in the analysis of probability of encoding error while events $(v) - (vii)$ will be used in the analysis of probability of decoding error.

\subsubsection{Probability of Error at the Encoder of $\mathcal{S}_2$}
An error is made if $(a)$ the encoder cannot find $\textbf{u}_0(l_0)$ in bin indexed by $k_0$ such that $\left(\textbf{w}_0(j_0), \textbf{w}_1(j_1), \textbf{u}_0(l_0), \textbf{q}\right) \in A_{\epsilon}^{(n)}$ or $(b)$ it cannot find $\textbf{u}_2(l_2)$ in bin indexed by $k_2$ such that $\left(\textbf{w}_0(j_0), \textbf{w}_1(j_1), \textbf{u}_2(l_2), \textbf{q}\right) \in A_{\epsilon}^{(n)}$. The probability of encoding error at $\mathcal{S}_2$ can be bounded as
\begin{eqnarray}
\nonumber
\def\argmin{\mathop{\rm \bigcap}}
P_{e,\mathcal{S}_2} \leq P\left(\bigcap_{\textbf{U}_0(l_0) \in \mbox{bin}(k_0)}\left(\textbf{W}_0(j_0), \textbf{W}_1(j_1), \textbf{U}_0(l_0), \textbf{q}\right) \notin A_{\epsilon}^{(n)}\right)\\ \nonumber + P\left(\bigcap_{\textbf{U}_2(l_2) \in \mbox{bin}(k_2)}\left(\textbf{W}_0(j_0), \textbf{W}_1(j_1), \textbf{U}_2(l_2), \textbf{q} \right) \notin A_{\epsilon}^{(n)}\right),\\
\nonumber
\leq \left(1 - P(E_{j_0j_1l_0})\right)^{2^{n(I(W_0,W_1;U_0|Q)+4\epsilon)}} + \left(1 - P(E_{j_0j_1l_2})\right)^{2^{n(I(W_0,W_1;U_2|Q)+4\epsilon)}} ,
\end{eqnarray}
Since $\textbf{q}$ is predetermined,
\begin{eqnarray}
\nonumber P(E_{j_0j_1l_0}) = \sum_{\left(\textbf{w}_0,\textbf{w}_1, \textbf{u}_0, \textbf{q} \right) \in A_{\epsilon}^{(n)}}\!\!\!\!\!\!\!\!\!P(\textbf{W}_0(j_0) = \textbf{w}_0,\textbf{W}_1(j_1) = \textbf{w}_1|\textbf{q})P(\textbf{U}_0(l_0) = \textbf{u}_0|\textbf{q})\\
\nonumber \geq 2^{n(H(W_0,W_1,U_0|Q)-\epsilon))}2^{-n(H(W_0,W_1|Q)+\epsilon))}2^{-n(H(U_0|Q)+\epsilon)}= 2^{-n(I(W_0,W_1;U_0|Q)+3\epsilon)}.
\end{eqnarray}
Similarly, $P(E_{j_0j_1l_2})\geq 2^{-n(I(W_0,W_1;U_2|Q)+3\epsilon)}$. Therefore,
\begin{eqnarray}
\nonumber P_{e,\mathcal{S}_2} \leq (1-2^{-n(I(W_0,W_1;U_0|Q)+3\epsilon)})^{2^{n(I(W_0,W_1;U_0|Q)+4\epsilon)}} + (1-2^{-n(I(W_0,W_1;U_2|Q)+3\epsilon)})^{2^{n(I(W_0,W_1;U_2|Q)+4\epsilon)}}.
\end{eqnarray}
Now,
\begin{eqnarray}
\nonumber (1-2^{-n(I(W_0,W_1;U_0|Q)+3\epsilon)})^{2^{n(I(W_0,W_1;U_0|Q)+4\epsilon)}} = e^{{2^{n(I(W_0,W_1;U_0|Q)+4\epsilon)}}\ln(1-2^{-n(I(W_0,W_1;U_0|Q)+3\epsilon)})}\\
\nonumber \leq e^{2^{n(I(W_0,W_1;U_0|Q)+4\epsilon)}(-2^{-n(I(W_0,W_1;U_0|Q)+3\epsilon)})}\\
\nonumber = e^{-2^{n\epsilon}}.
\end{eqnarray}Clearly, $P_{e,\mathcal{S}_2} \rightarrow 0$ as $n \rightarrow \infty$.

\subsubsection{Probability of Error at the Encoder of $\mathcal{S}_3$}
An error is made if $(a)$ the encoder cannot find $\textbf{v}_0(t_0)$ in bin indexed by $r_0$ such that $\left(\textbf{w}_0(j_0),\textbf{w}_1(j_1),\textbf{v}_0(t_0), \textbf{q}\right) \in A_{\epsilon}^{(n)}$ or $(b)$ it cannot find $\textbf{v}_3(t_3)$ in bin indexed by $r_3$ such that $\left(\textbf{w}_0(j_0),\textbf{w}_1(j_1), \textbf{v}_3(t_3), \textbf{q}\right) \in A_{\epsilon}^{(n)}$. The probability of encoding error at $\mathcal{S}_3$ can be bounded as
\begin{eqnarray}
\nonumber
P_{e,\mathcal{S}_3} \leq P\left(\bigcap_{\textbf{V}_0(t_0) \in \mbox{bin}(r_0)}\left(\textbf{W}_0(j_0), \textbf{W}_1(j_1),  \textbf{V}_0(t_0), \textbf{q}\right) \notin A_{\epsilon}^{(n)}\right)\\+
\nonumber
P\left(\bigcap_{\textbf{V}_3(t_3) \in \mbox{bin}(r_3)}\left(\textbf{W}_0(j_0), \textbf{W}_1(j_1), \textbf{V}_3(t_3), \textbf{q}\right) \notin A_{\epsilon}^{(n)}\right)\\
\nonumber
\leq \left(1 - P(E_{j_0j_1t_0})\right)^{2^{n(I(W_0,W_1;V_0|Q)+4\epsilon)}} + \left(1 - P(E_{j_0j_1t_3})\right)^{2^{n(I(W_0,W_1;V_3|Q)+4\epsilon)}}.
\end{eqnarray}
Since $\textbf{q}$ is predetermined, we have,
\begin{eqnarray}
\nonumber P(E_{j_0j_1t_0}) = \!\!\!\!\!\!\!\!\!\!\!\!\!\!\! \sum_{\left(\textbf{w}_0,\textbf{w}_1, \textbf{v}_0, \textbf{q} \right) \in A_{\epsilon}^{(n)}}\!\!\!\!\!\!\!\!\!\!\!\!\!\!\!P(\textbf{W}_0(j_0) = \textbf{w}_0,\textbf{W}_1(j_1) = \textbf{w}_1|\textbf{q})P(\textbf{V}_0(t_0) = \textbf{v}_0|\textbf{q})\\
\nonumber \geq 2^{n(H(W_0,W_1,V_0|Q)-\epsilon))}2^{-n(H(W_0,W_1|Q)+\epsilon)}2^{-n(H(V_0|Q)+\epsilon)}
= 2^{-n(I(W_0,W_1;V_0|Q)+3\epsilon)}.
\end{eqnarray}
Similarly, $P(E_{j_0j_1t_3}) \geq 2^{-n(I(W_0,W_1;V_3|Q)+3\epsilon)}$. Therefore,
\begin{eqnarray}
\nonumber P_{e,\mathcal{S}_3} \leq \left(1 - 2^{-n(I(W_0,W_1;V_0|Q)+3\epsilon)}\right)^{2^{n(I(W_0,W_1;V_0|Q)+4\epsilon)}}\!\!\!\!\!\!\! + \left(1 - 2^{-n(I(W_0,W_1;V_3|Q)+3\epsilon)}\right)^{2^{n(I(W_0,W_1;V_3|Q)+4\epsilon)}}.
\end{eqnarray}
Proceeding in a way similar to the encoder error analysis at $\mathcal{S}_2$, we get $P_{e,\mathcal{S}_3} \rightarrow 0$ as $n \rightarrow \infty$.

\subsubsection{Probability of Error at the Decoder of $\mathcal{R}_1$}
There are two possible events which can be classified as errors: $(a)$The codewords transmitted are not jointly typical i.e., $E_{j_0j_1l_0t_0}^c$ happens or $(b)$ there exists some $\hat{j}_0 \neq j_0$ and $\hat{j}_1 \neq j_1$ such that $E_{\hat{j}_0\hat{j}_1\hat{\hat{l}}_0\hat{\hat{t}}_0}$ happens. The probability of decoding error can, therefore, be expressed as
\begin{eqnarray}\label{eq29}
P_{e,\mathcal{R}_1}^{(n)} = P\left(E_{j_0j_1l_0t_0}^c \bigcup \cup_{\hat{j}_0 \neq j_0,\hat{j}_1 \neq j_1}E_{\hat{j}_0\hat{j}_1\hat{\hat{l}}_0\hat{\hat{t}}_0}\right)
\end{eqnarray}
Applying union of events bound, (\ref{eq29}) can be written as,
\begin{center}
$P_{e,\mathcal{R}_1}^{(n)} \leq P\left(E_{j_0j_1l_0t_0}^c\right) + P\left(\cup_{\hat{j}_0 \neq j_0,\hat{j}_1 \neq j_1}E_{\hat{j}_0\hat{j}_1\hat{\hat{l}}_0\hat{\hat{t}}_0}\right)$\\
\begin{eqnarray}
\nonumber
\def\argmin{\mathop{\rm \cup}}
= P\left(E_{j_0j_1l_0t_0}^c\right) + \sum_{\hat{j}_0 \neq j_0}P\left(E_{\hat{j}_0j_1l_0t_0}\right) + \sum_{\hat{j}_1 \neq j_1}P\left(E_{j_0\hat{j}_1l_0t_0}\right) + \sum_{\hat{j}_0 \neq j_0,\hat{j}_1 \neq j_1}\!\!\!\!\!\!\!\!P\left(E_{\hat{j}_0\hat{j}_1l_0t_0}\right)+
\sum_{\hat{j}_0 \neq j_0,\hat{l}_0 \neq l_0}\!\!\!\!\!\!\!\!P\left(E_{\hat{j}_0j_1\hat{l}_0t_0}\right)\\ \nonumber
+\sum_{\hat{j}_0 \neq j_0,\hat{t}_0 \neq t_0}\!\!\!\!\!\!\!\!P\left(E_{\hat{j}_0j_1l_0\hat{t}_0}\right)+\sum_{\hat{j}_1 \neq j_1,\hat{l}_0 \neq l_0}\!\!\!\!\!\!\!\!P\left(E_{j_0\hat{j}_1\hat{l}_0t_0}\right)+\sum_{\hat{j}_1 \neq j_1,\hat{t}_0 \neq t_0}\!\!\!\!\!\!\!\!P\left(E_{j_0\hat{j}_1l_0\hat{t}_0}\right)+\sum_{\hat{j}_0 \neq j_0,\hat{j}_1 \neq j_1,\hat{l}_0 \neq l_0}\!\!\!\!\!\!\!\!P\left(E_{\hat{j}_0\hat{j}_1\hat{l}_0t_0}\right)+\\\nonumber
\sum_{\hat{j}_0 \neq j_0,\hat{j}_1 \neq j_1,\hat{t}_0 \neq t_0}\!\!\!\!\!\!\!\!P\left(E_{\hat{j}_0\hat{j}_1l_0\hat{t}_0}\right)+\sum_{\hat{j}_0 \neq j_0,\hat{l}_0 \neq l_0,\hat{t}_0 \neq t_0}\!\!\!\!\!\!\!\!P\left(E_{\hat{j}_0j_1\hat{l}_0\hat{t}_0}\right)+\sum_{\hat{j}_1 \neq j_1,\hat{l}_0 \neq l_0,\hat{t}_0 \neq t_0}\!\!\!\!\!\!\!\!P\left(E_{j_0\hat{j}_1\hat{l}_0\hat{t}_0}\right)
+\!\!\!\!\!\!\!\!\sum_{\hat{j}_0 \neq j_0,\hat{j}_1 \neq j_1,\hat{l}_0 \neq l_0,\hat{t}_0 \neq t_0}\!\!\!\!\!\!\!\!\!\!\!\!\!\!\!\!P\left(E_{\hat{j}_0\hat{j}_1\hat{l}_0\hat{t}_0}\right)
\end{eqnarray}
\end{center}

\begin{eqnarray}
\nonumber
\def\argmin{\mathop{\rm \cup}}
\leq P\left(E_{j_0j_1l_0t_0}^c\right) + 2^{nR_{10}}P\left(E_{\hat{j}_0j_1l_0t_0}\right) + 2^{nR_{11}}P\left(E_{j_0\hat{j}_1l_0t_0}\right)\\ \nonumber + 2^{n(R_{10}+R_{11})}P\left(E_{\hat{j}_0\hat{j}_1l_0t_0}\right) + \\ \nonumber 2^{n(R_{10}+R_{20}+I(W_0,W_1;U_0|Q)+4\epsilon)}P\left(E_{\hat{j}_0j_1\hat{l}_0t_0}\right)\\ \nonumber + 2^{n(R_{10}+R_{30}+I(W_0,W_1;V_0|Q)+4\epsilon)}P\left(E_{\hat{j}_0j_1l_0\hat{t}_0}\right) + \\ \nonumber 2^{n(R_{11}+R_{20}+I(W_0,W_1;U_0|Q)+4\epsilon)}P\left(E_{j_0\hat{j}_1\hat{l}_0t_0}\right)\\ \nonumber
+ 2^{n(R_{11}+R_{30}+I(W_0,W_1;V_0|Q)+4\epsilon)}P\left(E_{j_0\hat{j}_1l_0\hat{t}_0}\right)+ \\ \nonumber 2^{n(R_{10}+R_{11}+R_{20}+I(W_0,W_1;U_0|Q)+4\epsilon)}P\left(E_{\hat{j}_0\hat{j}_1\hat{l}_0t_0}\right)\\ \nonumber + 2^{n(R_{10}+R_{11}+R_{30}+I(W_0,W_1;V_0|Q)+4\epsilon)}P\left(E_{\hat{j}_0\hat{j}_1l_0\hat{t}_0}\right)+ \\ \nonumber 2^{n(R_{10}+R_{20}+I(W_0,W_1;U_0|Q)+4\epsilon)+R_{30}+I(W_0,W_1;V_0|Q)+4\epsilon)}P\left(E_{\hat{j}_0j_1\hat{l}_0\hat{t}_0}\right)+ \\ \nonumber
2^{n(R_{11}+R_{20}+I(W_0,W_1;U_0|Q)+4\epsilon)+R_{30}+I(W_0,W_1;V_0|Q)+4\epsilon)}P\left(E_{j_0\hat{j}_1\hat{l}_0\hat{t}_0}\right)+ \\ \nonumber
2^{n(R_{10}+R_{11}+R_{20}+I(W_0,W_1;U_0|Q)+4\epsilon)+R_{30}+I(W_0,W_1;V_0|Q)+4\epsilon)}P\left(E_{\hat{j}_0\hat{j}_1\hat{l}_0\hat{t}_0}\right).
\end{eqnarray}

The probability of error events can be upper bounded as follows.
\begin{eqnarray}
\nonumber P\left(E_{\hat{j}_0j_1l_0t_0}\right)\leq 2^{-n(I(W_0;W_1,U_0,V_0,Y_1|Q)-3\epsilon)},\\
\nonumber P\left(E_{j_0\hat{j}_1l_0t_0}\right) \leq 2^{-n(I(W_1;W_0,U_0,V_0,Y_1|Q)-3\epsilon)},\\
\nonumber P\left(E_{\hat{j}_0\hat{j}_1l_0t_0}\right)\leq 2^{-n(I(W_0,W_1;U_0,V_0,Y_1|Q)+I(W_0;W_1|Q)-4\epsilon)},\\
\nonumber P\left(E_{\hat{j}_0j_1\hat{l}_0t_0}\right)\leq 2^{-n(I(W_0,U_0;W_1,V_0,Y_1|Q)+I(W_0;U_0|Q)-4\epsilon)},\\
\nonumber P\left(E_{\hat{j}_0j_1l_0\hat{t}_0}\right)\leq 2^{-n(I(W_0,V_0;W_1,U_0,Y_1|Q)+I(W_0;V_0|Q)-4\epsilon)},\\
\nonumber P\left(E_{j_0\hat{j}_1\hat{l}_0t_0}\right)\leq 2^{-n(I(W_1,U_0;W_0,V_0,Y_1|Q)+I(W_1;U_0|Q)-4\epsilon)},\\
\nonumber P\left(E_{j_0\hat{j}_1l_0\hat{t}_0}\right)\leq 2^{-n(I(W_1,V_0;W_0,U_0,Y_1|Q)+I(W_1;V_0|Q)-4\epsilon)},\\
\nonumber P\left(E_{\hat{j}_0\hat{j}_1\hat{l}_0t_0}\right)\leq 2^{-n(I(W_0,W_1,U_0;V_0,Y_1|Q)+I(W_0,W_1;U_0|Q)+I(W_0;W_1|Q)-5\epsilon)},\\
\nonumber P\left(E_{\hat{j}_0\hat{j}_1l_0\hat{t}_0}\right)\leq 2^{-n(I(W_0,W_1,V_0;U_0,Y_1|Q)+I(W_0,W_1;V_0|Q)+I(W_0;W_1|Q)-5\epsilon)},\\
\nonumber P\left(E_{\hat{j}_0j_1\hat{l}_0\hat{t}_0}\right) \leq 2^{-n(I(W_0,U_0,V_0;W_1,Y_1|Q)+I(W_0,U_0;V_0|Q)+I(W_0;U_0|Q)-5\epsilon)},\\
\nonumber P\left(E_{j_0\hat{j}_1\hat{l}_0\hat{t}_0}\right)\leq 2^{-n(I(W_1,U_0,V_0;W_0,Y_1|Q)+I(W_1,U_0;V_0|Q)+I(W_1;U_0|Q)-5\epsilon)},\\
\nonumber P\left(E_{\hat{j}_0\hat{j}_1\hat{l}_0\hat{t}_0}\right)\leq  2^{-n(I(W_0,W_1,U_0,V_0;Y_1|Q)+I(W_0,W_1,U_0;V_0|Q)+I(W_0,W_1;U_0|Q)+I(W_0,W_1|Q)-6\epsilon)}.
\end{eqnarray}
Substituting these in the probability of decoding error at $\mathcal{R}_1$, we note that $P_{e,\mathcal{R}_1}^{(n)} \rightarrow 0$ as $n \rightarrow \infty$ if the following constraints are satisfied:
\begin{align*}
R_{10} \leq I(W_0;W_1,U_0,V_0,Y_1|Q),\\
R_{11} \leq I(W_1;W_0,U_0,V_0,Y_1|Q),\\
R_{10}+R_{11} \leq I(W_0,W_1;U_0,V_0,Y_1|Q)+I(W_0;W_1|Q),\\
R_{10}+R_{20} \leq I(W_0,U_0;W_1,V_0,Y_1|Q)+I(W_0;U_0|Q) - I(W_0,W_1;U_0|Q),\\
R_{10}+R_{30} \leq I(W_0,V_0;W_1,U_0,Y_1|Q)+I(W_0;V_0|Q) - I(W_0,W_1;V_0|Q),\\
R_{11}+R_{20} \leq I(W_1,U_0;W_0,V_0,Y_1|Q)+I(W_1;U_0|Q) - I(W_0,W_1;U_0|Q),\\
R_{11}+R_{30} \leq I(W_1,V_0;W_0,U_0,Y_1|Q)+I(W_1;V_0|Q) - I(W_0,W_1;V_0|Q),\\
R_{10}+R_{11}+R_{20} \leq I(W_0,W_1,U_0;V_0,Y_1|Q)+I(W_0,W_1;U_0|Q)+I(W_0;W_1|Q) - I(W_0,W_1;U_0|Q),\\
R_{10}+R_{11}+R_{30} \leq I(W_0,W_1,V_0;U_0,Y_1|Q)+I(W_0,W_1;V_0|Q)+I(W_0;W_1|Q) - I(W_0,W_1;V_0|Q),\\
\nonumber R_{10}+R_{20}+R_{30} \leq I(W_0,U_0,V_0;W_1,Y_1|Q)+I(W_0,U_0;V_0|Q)+I(W_0;U_0|Q)\\ - I(W_0,W_1;U_0|Q) - I(W_0,W_1;V_0|Q)\\
\nonumber R_{11}+R_{20}+R_{30} \leq I(W_1,U_0,V_0;W_0,Y_1|Q)+I(W_1,U_0;V_0|Q)+I(W_1;U_0|Q)\\ - I(W_0,W_1;U_0|Q) - I(W_0,W_1;V_0|Q)\\
R_{10}+R_{11}+R_{20}+R_{30} \leq I(W_0,W_1,U_0,V_0;Y_1|Q)+I(W_0,W_1,U_0;V_0|Q)+I(W_0,W_1;U_0|Q)\\ +I(W_0,W_1|Q) - I(W_0,W_1;U_0|Q) - I(W_0,W_1;V_0|Q)
\end{align*}

\subsubsection{Probability of Error at the Decoder of $\mathcal{R}_2$}
There are two possible events which can be classified as errors: $(a)$ The codewords transmitted are not jointly typical i.e., $E_{j_0l_0l_2t_0}^c$ happens or $(b)$ there exists some $\hat{l}_0 \neq l_0$ and $\hat{l}_2 \neq l_2$ such that $E_{\hat{\hat{j}}_0\hat{l}_0\hat{l}_2\hat{\hat{t}}_0}$ happens. The probability of decoding error can, therefore, be expressed as
\begin{eqnarray}\label{eq30}
\def\argmin{\mathop{\rm \cup}}
P_{e,\mathcal{R}_2}^{(n)} = P\left(E_{j_0l_0l_2t_0}^c \bigcup \cup_{(\hat{l}_0 \neq l_0, \hat{l}_2 \neq l_2)}E_{\hat{\hat{j}}_0\hat{l}_0\hat{l}_2\hat{\hat{t}}_0}\right)
\end{eqnarray}
Applying union of events bound, (\ref{eq30}) can be written as,
\begin{center}
$P_{e,\mathcal{R}_2}^{(n)} \leq P\left(E_{j_0l_0l_2t_0}^c\right) + P\left(\cup_{(\hat{l}_0 \neq l_0, \hat{l}_2 \neq l_2)}E_{\hat{\hat{j}}_0\hat{l}_0\hat{l}_2\hat{\hat{t}}_0}\right)$
\begin{eqnarray}
\nonumber
\def\argmin{\mathop{\rm \cup}}
= P\left(E_{j_0l_0l_2t_0}^c\right) + \sum_{\hat{l}_0 \neq l_0}P\left(E_{j_0\hat{l}_0l_2t_0}\right) + \sum_{\hat{l}_2 \neq l_2}P\left(E_{j_0l_0\hat{l}_2t_0}\right) + \sum_{\hat{l}_0 \neq l_0,\hat{l}_2 \neq l_2}\!\!\!\!\!\!\!\!P\left(E_{j_0\hat{l}_0\hat{l}_2t_0}\right)+
\sum_{\hat{j}_0 \neq j_0,\hat{l}_0 \neq l_0}\!\!\!\!\!\!\!\!P\left(E_{\hat{j}_0\hat{l}_0l_2t_0}\right)\\ \nonumber
+\sum_{\hat{j}_0 \neq j_0,\hat{l}_2 \neq l_2}\!\!\!\!\!\!\!\!P\left(E_{\hat{j}_0l_0\hat{l}_2t_0}\right)+\sum_{\hat{l}_0 \neq l_0,\hat{t}_0 \neq t_0}\!\!\!\!\!\!\!\!P\left(E_{j_0\hat{l}_0l_2\hat{t}_0}\right)+\sum_{\hat{l}_2 \neq l_2,\hat{t}_0 \neq t_0}\!\!\!\!\!\!\!\!P\left(E_{j_0l_0\hat{l}_2\hat{t}_0}\right)+\sum_{\hat{j}_0 \neq j_0,\hat{l}_0 \neq l_0,\hat{l}_2 \neq l_2}\!\!\!\!\!\!\!\!P\left(E_{\hat{j}_0\hat{l}_0\hat{l}_2t_0}\right)+\\ \nonumber
\sum_{\hat{j}_0 \neq j_0,\hat{l}_0 \neq l_0,\hat{t}_0 \neq t_0}\!\!\!\!\!\!\!\!P\left(E_{\hat{j}_0\hat{l}_0l_2\hat{t}_0}\right)+\sum_{\hat{j}_0 \neq j_0,\hat{l}_2 \neq l_2,\hat{t}_0 \neq t_0}\!\!\!\!\!\!\!\!P\left(E_{\hat{j}_0l_0\hat{l}_2\hat{t}_0}\right)+\sum_{\hat{l}_0 \neq l_0,\hat{l}_2 \neq l_2,\hat{t}_0 \neq t_0}\!\!\!\!\!\!\!\!P\left(E_{j_0\hat{l}_0\hat{l}_2\hat{t}_0}\right)
+\sum_{\hat{j}_0 \neq j_0,\hat{l}_0 \neq l_0,\hat{l}_2 \neq l_2,\hat{t}_0 \neq t_0}\!\!\!\!\!\!\!\!\!\!\!\!\!\!\!\!P\left(E_{\hat{j}_0\hat{l}_0\hat{l}_2\hat{t}_0}\right)\\ \nonumber \\ 
\nonumber \leq P\left(E_{j_0l_0l_2t_0}^c\right) + 2^{n(R_{20}+I(W_0,W_1;U_0|Q)+4\epsilon)}P(E_{j_0\hat{l}_0l_2t_0})+\\ \nonumber 2^{n(R_{22}+I(W_0,W_1;U_2|Q)+4\epsilon)}P(E_{j_0l_0\hat{l}_2t_0})\\ \nonumber + 2^{n(R_{20}+R_{22}+I(W_0,W_1;U_0|Q)+4\epsilon+I(W_0,W_1;U_2|Q)+4\epsilon)}P(E_{j_0\hat{l}_0\hat{l}_2t_0}) + \\ \nonumber 2^{n(R_{10}+R_{20}+I(W_0,W_1;U_0|Q)+4\epsilon)}P(E_{\hat{j}_0\hat{l}_0l_2t_0})\\ \nonumber + 2^{n(R_{10}+R_{22}+I(W_0,W_1;U_2|Q)+4\epsilon)}P(E_{\hat{j}_0l_0\hat{l}_2t_0}) \\ \nonumber +
2^{n(R_{20}+R_{30}+I(W_0,W_1;U_0|Q)+4\epsilon+I(W_0,W_1;V_0|Q)+4\epsilon)}P(E_{j_0\hat{l}_0l_2\hat{t}_0}) \\ \nonumber +
2^{n(R_{22}+R_{30}+I(W_0,W_1;U_2|Q)+4\epsilon+I(W_0,W_1;V_0|Q)+4\epsilon)}P(E_{j_0l_0\hat{l}_2\hat{t}_0}) \\ \nonumber +
2^{n(R_{10}+R_{20}+R_{22}+I(W_0,W_1;U_0|Q)+4\epsilon+I(W_0,W_1;U_2|Q)+4\epsilon)}P(E_{\hat{j}_0\hat{l}_0\hat{l}_2t_0}) + \\ \nonumber
2^{n(R_{10}+R_{20}+R_{30}+I(W_0,W_1;U_0|Q)+4\epsilon+I(W_0,W_1;V_0|Q)+4\epsilon)}P(E_{\hat{j}_0\hat{l}_0l_2\hat{t}_0}) \\ \nonumber +2^{n(R_{10}+R_{22}+R_{30}+I(W_0,W_1;U_2|Q)+4\epsilon+I(W_0,W_1;V_0|Q)+4\epsilon)}P(E_{\hat{j}_0l_0\hat{l}_2\hat{t}_0}) \\ \nonumber +2^{n(R_{20}+R_{22}+R_{30}+I(W_0,W_1;U_0|Q)+4\epsilon+I(W_0,W_1;U_2|Q)+4\epsilon+I(W_0,W_1;V_0|Q)+4\epsilon)}P(E_{j_0\hat{l}_0\hat{l}_2\hat{t}_0}) \\ \nonumber+2^{n(R_{10}+R_{20}+R_{22}+R_{30}+I(W_0,W_1;U_0|Q)+4\epsilon+I(W_0,W_1;U_2|Q)+4\epsilon+I(W_0,W_1;V_0|Q)+4\epsilon)}P(E_{\hat{j}_0\hat{l}_0\hat{l}_2\hat{t}_0}) \end{eqnarray}
\end{center}

The upper bounds on the probability of error events follow.
\begin{eqnarray}
\nonumber P(E_{j_0\hat{l}_0l_2t_0})\\ \nonumber \leq 2^{-n(I(U_0;W_0,U_2,V_0,Y_2|Q)-3\epsilon)},\\
\nonumber P(E_{j_0l_0\hat{l}_2t_0})\leq 2^{-n(I(U_2;W_0,U_0,V_0,Y_2|Q)-3\epsilon)},\\
\nonumber P(E_{j_0\hat{l}_0\hat{l}_2t_0})\leq 2^{-n(I(U_0,U_2;W_0,V_0,Y_2|Q)+I(U_0;U_2|Q)-4\epsilon)},\\
\nonumber P(E_{\hat{j}_0\hat{l}_0l_2t_0})\leq 2^{-n(I(W_0,U_0;U_2,V_0,Y_2|Q)+I(W_0;U_0|Q)-4\epsilon)},\\
\nonumber P(E_{\hat{j}_0l_0\hat{l}_2t_0})\leq 2^{-n(I(W_0,U_2;U_0,V_0,Y_2|Q)+I(W_0;U_2|Q)-4\epsilon)},\\
\nonumber P(E_{j_0\hat{l}_0l_2\hat{t}_0})\leq 2^{-n(I(U_0,V_0;W_0,U_2,Y_2|Q)+I(U_0;V_0|Q)-4\epsilon)},\\
\nonumber P(E_{j_0l_0\hat{l}_2\hat{t}_0})\leq 2^{-n(I(U_2,V_0;W_0,U_0,Y_2|Q)+I(U_2;V_0|Q)-4\epsilon)},\\
\nonumber P(E_{\hat{j}_0\hat{l}_0\hat{l}_2t_0})\leq 2^{-n(I(W_0,U_0,U_2;V_0,Y_2|Q)+I(W_0,U_0;U_2|Q)+I(W_0;U_0|Q)-5\epsilon)},\\
\nonumber P(E_{\hat{j}_0\hat{l}_0l_2\hat{t}_0})\leq 2^{-n(I(W_0,U_0,V_0;U_2,Y_2|Q)+I(W_0,U_0;V_0|Q)+I(W_0;U_0|Q)-5\epsilon)},\\
\nonumber P(E_{\hat{j}_0l_0\hat{l}_2\hat{t}_0})\leq 2^{-n(I(W_0,U_2,V_0;U_0,Y_2|Q)+I(W_0,U_2;V_0|Q)+I(W_0;U_2|Q)-5\epsilon)},\\
\nonumber P(E_{j_0\hat{l}_0\hat{l}_2\hat{t}_0})\leq 2^{-n(I(U_0,U_2,V_0;W_0,Y_2|Q)+I(U_0,U_2;V_0|Q)+I(U_0;U_2|Q)-5\epsilon)},\\
\nonumber P(E_{j_0\hat{l}_0\hat{l}_2\hat{t}_0})\leq 2^{-n(I(W_0,U_0,U_2,V_0;Y_2|Q)+I(W_0,U_0,U_2;V_0|Q)+I(W_0,U_0;U_2|Q)+I(W_0,U_0|Q)-6\epsilon)}.
\end{eqnarray}
Substituting these in the probability of decoding error at $\mathcal{R}_2$, we note that $P_{e,\mathcal{R}_2}^{(n)} \rightarrow 0$ as $n \rightarrow \infty$ if the following constraints are satisfied:
\begin{align*}
R_{20} \leq I(U_0;W_0,U_2,V_0,Y_2|Q)-I(W_0,W_1;U_0|Q),\\
R_{22} \leq I(U_2;W_0,U_0,V_0,Y_2|Q)-I(W_0,W_1;U_2|Q),\\
R_{20}+R_{22} \leq I(U_0,U_2;W_0,V_0,Y_2|Q)+I(U_0;U_2|Q)-I(W_0,W_1;U_0|Q)-I(W_0,W_1;U_2|Q),\\
R_{10}+R_{20} \leq I(W_0,U_0;U_2,V_0,Y_2|Q)+I(W_0;U_0|Q) - I(W_0,W_1;U_0|Q),\\
R_{10}+R_{22} \leq I(W_0,U_2;U_0,V_0,Y_2|Q)+I(W_0;U_2|Q) - I(W_0,W_1;U_2|Q),\\
R_{20}+R_{30} \leq I(U_0,V_0;W_0,U_2,Y_2|Q)+I(U_0;V_0|Q) - I(W_0,W_1;U_0|Q) - I(W_0,W_1;V_0|Q),\\
R_{22}+R_{30} \leq I(U_2,V_0;W_0,U_0,Y_2|Q)+I(U_2;V_0|Q) - I(W_0,W_1;U_2|Q) - I(W_0,W_1;V_0|Q),\\
\nonumber R_{10}+R_{20}+R_{22} \leq I(W_0,U_0,U_2;V_0,Y_2|Q)+I(W_0,U_0;U_2|Q)+I(W_0;U_0|Q)\\ - I(W_0,W_1;U_0|Q)-I(W_0,W_1;U_2|Q)\\
\nonumber R_{10}+R_{20}+R_{30} \leq I(W_0,U_0,V_0;U_2,Y_2|Q)+I(W_0,U_0;V_0|Q)+I(W_0;U_0|Q) - \\ I(W_0,W_1;U_0|Q) - I(W_0,W_1;V_0|Q)\\
\nonumber R_{10}+R_{22}+R_{30} \leq I(W_0,U_2,V_0;U_0,Y_2|Q)+I(W_0,U_2;V_0|Q)+I(W_0;U_2|Q) - \\ I(W_0,W_1;U_2|Q) - I(W_0,W_1;V_0|Q)\\
\nonumber R_{20}+R_{22}+R_{30} \leq I(U_0,U_2,V_0;W_0,Y_2|Q)+I(U_0,U_2;V_0|Q)+I(U_0;U_2|Q) \\-I(W_0,W_1;U_0|Q)-I(W_0,W_1;U_2|Q) -I(W_0,W_1;V_0|Q)\\
\nonumber R_{10}+R_{20}+R_{22}+R_{30} \leq I(W_0,U_0,U_2,V_0;Y_2|Q)+I(W_0,U_0,U_2;V_0|Q)+I(W_0,U_0;U_2|Q)\\+I(W_0,U_0|Q) - I(W_0,W_1;U_0|Q)-I(W_0,W_1;U_2|Q) -I(W_0,W_1;V_0|Q)
\end{align*}

\subsubsection{Probability of Error at the Decoder of $\mathcal{R}_3$}
There are two possible events which can be classified as errors: $(a)$ The codewords transmitted are not jointly typical i.e., $E_{j_0l_0t_0t_3}^c$ happens or $(b)$ there exists some $\hat{t}_0 \neq t_0$ and $\hat{t}_3 \neq t_3$ such that $E_{\hat{\hat{j}}_0\hat{\hat{l}}_0\hat{t}_0\hat{t}_3}$ happens. The probability of decoding error can, therefore, be expressed as
\begin{eqnarray}\label{eq31}
\def\argmin{\mathop{\rm \cup}}
P_{e,\mathcal{R}_3}^{(n)} = P\left(E_{j_0l_0t_0t_3}^c \bigcup \cup_{(\hat{t}_0 \neq t_0, \hat{t}_3 \neq t_3)}E_{\hat{\hat{j}}_0\hat{\hat{l}}_0\hat{t}_0\hat{t}_3}\right)
\end{eqnarray}
Applying union of events bound, (\ref{eq31}) can be written as,
\begin{center}
$P_{e,\mathcal{R}_3}^{(n)} \leq P\left(E_{j_0l_0t_0t_3}^c\right) + P\left(\cup_{(\hat{t}_0 \neq t_0, \hat{t}_3 \neq t_3)}E_{\hat{\hat{j}}_0\hat{\hat{l}}_0\hat{t}_0\hat{t}_3}\right)$
\begin{eqnarray}
\nonumber
\def\argmin{\mathop{\rm \cup}}
= P\left(E_{j_0l_0t_0t_3}^c\right) + \sum_{\hat{t}_0 \neq t_0}P\left(E_{j_0l_0\hat{t}_0t_3}\right) + \sum_{\hat{t}_3 \neq t_3}P\left(E_{j_0l_0t_0\hat{t}_3}\right) + \sum_{\hat{t}_0 \neq t_0,\hat{t}_3 \neq t_3}\!\!\!\!\!\!\!\!P\left(E_{j_0l_0\hat{t}_0\hat{t}_3}\right)+
\sum_{\hat{j}_0 \neq j_0,\hat{t}_0 \neq t_0}\!\!\!\!\!\!\!\!P\left(E_{\hat{j}_0l_0\hat{t}_0t_3}\right)\\ \nonumber
+\sum_{\hat{j}_0 \neq j_0,\hat{t}_3 \neq t_3}\!\!\!\!\!\!\!\!P\left(E_{\hat{j}_0l_0t_0\hat{t}_3}\right)+\sum_{\hat{l}_0 \neq l_0,\hat{t}_0 \neq t_0}\!\!\!\!\!\!\!\!P\left(E_{j_0\hat{l}_0\hat{t}_0t_3}\right)+\sum_{\hat{l}_0 \neq l_0,\hat{t}_3 \neq t_3}\!\!\!\!\!\!\!\!P\left(E_{j_0\hat{l}_0t_0\hat{t}_3}\right)+\sum_{\hat{j}_0 \neq j_0,\hat{l}_0 \neq l_0,\hat{t}_0 \neq t_0}\!\!\!\!\!\!\!\!P\left(E_{\hat{j}_0\hat{l}_0\hat{t}_0t_3}\right)+\\ \nonumber
\sum_{\hat{j}_0 \neq j_0,\hat{l}_0 \neq l_0,\hat{t}_3 \neq t_3}\!\!\!\!\!\!\!\!P\left(E_{\hat{j}_0\hat{l}_0t_0\hat{t}_3}\right)+\sum_{\hat{j}_0 \neq j_0,\hat{t}_0 \neq t_0,\hat{t}_3 \neq t_3}\!\!\!\!\!\!\!\!P\left(E_{\hat{j}_0l_0\hat{t}_0\hat{t}_3}\right)+\sum_{\hat{l}_0 \neq l_0,\hat{t}_0 \neq t_0,\hat{t}_3 \neq t_3}\!\!\!\!\!\!\!\!P\left(E_{j_0\hat{l}_0\hat{t}_0\hat{t}_3}\right)
+\sum_{\hat{j}_0 \neq j_0,\hat{l}_0 \neq l_0,\hat{t}_0 \neq t_0,\hat{t}_3 \neq t_3}\!\!\!\!\!\!\!\!\!\!\!\!\!\!\!\!P\left(E_{\hat{j}_0\hat{l}_0\hat{t}_0\hat{t}_3}\right)\\\nonumber\\ 
\nonumber \leq P\left(E_{j_0l_0t_0t_3}^c\right)+ 2^{n(R_{30}+I(W_0,W_1,U_0,U_2;V_0|Q)+4\epsilon)}P(E_{j_0l_0\hat{t}_0t_3})\\ \nonumber +2^{n(R_{33}+I(W_0,W_1,U_0,U_2;V_3|Q)+4\epsilon)}P(E_{j_0l_0t_0\hat{t}_3}) \\ \nonumber + 2^{n(R_{30}+R_{33}+I(W_0,W_1;V_0|Q)+4\epsilon+I(W_0,W_1;V_3|Q)+4\epsilon)}P(E_{j_0l_0\hat{t}_0\hat{t}_3})+\\ \nonumber
2^{n(R_{10}+R_{30}+I(W_0,W_1;V_0|Q)+4\epsilon)}P(E_{\hat{j}_0l_0\hat{t}_0t_3})+\\ \nonumber
2^{n(R_{10}+R_{33}+I(W_0,W_1;V_3|Q)+4\epsilon)}P(E_{\hat{j}_0l_0t_0\hat{t}_3})+\\ \nonumber
2^{n(R_{20}+R_{30}+I(W_0,W_1;U_0|Q)+4\epsilon+I(W_0,W_1;V_0|Q)+4\epsilon)}P(E_{j_0\hat{l}_0\hat{t}_0t_3})+\\ \nonumber
2^{n(R_{20}+R_{33}+I(W_0,W_1;U_0|Q)+4\epsilon+I(W_0,W_1;V_3|Q)+4\epsilon)}P(E_{j_0\hat{l}_0t_0\hat{t}_3})+\\ \nonumber
2^{n(R_{10}+R_{20}+R_{30}+I(W_0,W_1;U_0|Q)+4\epsilon+I(W_0,W_1;V_0|Q)+4\epsilon)}P(E_{\hat{j}_0\hat{l}_0\hat{t}_0t_3})+\\ \nonumber
2^{n(R_{10}+R_{20}+R_{33}+I(W_0,W_1;U_0|Q)+4\epsilon+I(W_0,W_1;V_3|Q)+4\epsilon)}P(E_{\hat{j}_0\hat{l}_0t_0\hat{t}_3})+\\ \nonumber
2^{n(R_{10}+R_{30}+R_{33}+I(W_0,W_1;V_0|Q)+4\epsilon+I(W_0,W_1;V_3|Q)+4\epsilon)}P(E_{\hat{j}_0l_0\hat{t}_0\hat{t}_3})+\\ \nonumber
2^{n(R_{20}+R_{30}+R_{33}+I(W_0,W_1;U_0|Q)+4\epsilon+I(W_0,W_1;V_0|Q)+4\epsilon+I(W_0,W_1;V_3|Q)+4\epsilon)}P(E_{j_0\hat{l}_0\hat{t}_0\hat{t}_3})+\\ \nonumber 2^{n(R_{10}+R_{20}+R_{30}+R_{33}+I(W_0,W_1;U_0|Q)+4\epsilon+I(W_0,W_1;V_0|Q)+4\epsilon+I(W_0,W_1;V_3|Q)+4\epsilon)}P(E_{\hat{j}_0\hat{l}_0\hat{t}_0\hat{t}_3}) \end{eqnarray}
\end{center}

The probabilities of error events can be upper bounded as follows.
\begin{eqnarray}
\nonumber P(E_{j_0l_0\hat{t}_0t_3}) \leq 2^{-n(I(V_0;W_0,U_0,V_3,Y_3|Q)-3\epsilon)},\\
\nonumber P(E_{j_0l_0t_0\hat{t}_3})\leq 2^{-n(I(V_3;W_0,U_0,V_0,Y_3|Q)-3\epsilon)},\\
\nonumber P(E_{j_0l_0\hat{t}_0\hat{t}_3})\leq 2^{-n(I(V_0,V_3;W_0,U_0,Y_3|Q)+I(V_0;V_3|Q)-4\epsilon)},\\
\nonumber P(E_{\hat{j}_0l_0\hat{t}_0t_3}) \leq 2^{-n(I(W_0,V_0;U_0,V_3,Y_3|Q)+I(W_0;V_0|Q)-4\epsilon)},\\
\nonumber P(E_{\hat{j}_0l_0t_0\hat{t}_3})\leq 2^{-n(I(W_0,V_3;U_0,V_0,Y_3|Q)+I(W_0;V_3|Q)-4\epsilon)},\\
\nonumber P(E_{j_0\hat{l}_0\hat{t}_0t_3})\leq 2^{-n(I(U_0,V_0;W_0,V_3,Y_3|Q)+I(U_0;V_0|Q)-4\epsilon)},\\
\nonumber P(E_{j_0\hat{l}_0\hat{t}_0t_3})\leq 2^{-n(I(U_0,V_3;W_0,V_0,Y_3|Q)+I(U_0;V_3|Q)-4\epsilon)},\\
\nonumber P(E_{\hat{j}_0\hat{l}_0\hat{t}_0t_3})\leq 2^{-n(I(W_0,U_0,V_0;V_3,Y_3|Q)+I(W_0,U_0;V_0|Q)+I(W_0;U_0|Q)-5\epsilon)},\\
\nonumber P(E_{\hat{j}_0\hat{l}_0t_0\hat{t}_3})\leq 2^{-n(I(W_0,U_0,V_3;V_0,Y_3|Q)+I(W_0,U_0;V_3|Q)+I(W_0;U_0|Q)-5\epsilon)},\\
\nonumber P(E_{\hat{j}_0l_0\hat{t}_0\hat{t}_3})\leq 2^{-n(I(W_0,V_0,V_3;U_0,Y_3|Q)+I(W_0,V_0;V_3|Q)+I(W_0;V_0|Q)-5\epsilon)},\\
\nonumber P(E_{\hat{j}_0l_0\hat{t}_0\hat{t}_3})\leq 2^{-n(I(U_0,V_0,V_3;W_0,Y_3|Q)+I(U_0,V_0;V_3|Q)+I(U_0;V_0|Q)-5\epsilon)},\\
\nonumber P(E_{\hat{j}_0l_0\hat{t}_0\hat{t}_3})\leq 2^{-n(I(W_0,U_0,V_0,V_3;Y_3|Q)+I(W_0,U_0,V_0;V_3|Q)+I(W_0,U_0;V_0|Q)+I(W_0;U_0|Q)-6\epsilon)}.
\end{eqnarray}
Substituting these in the probability of decoding error at $\mathcal{R}_3$, we note that $P_{e,\mathcal{R}_3}^{(n)} \rightarrow 0$ as $n \rightarrow \infty$ if the following constraints are satisfied:
\begin{align*}
R_{30} \leq I(V_0;W_0,U_0,V_3,Y_3|Q)-I(W_0,W_1;V_0|Q),\\
R_{33} \leq I(V_3;W_0,U_0,V_0,Y_3|Q)-I(W_0,W_1;V_3|Q),\\
\nonumber R_{30}+R_{33} \leq I(V_0,V_3;W_0,U_0,Y_3|Q)+I(V_0;V_3|Q)\\ -I(W_0,W_1;V_0|Q)-I(W_0,W_1;V_3|Q),\\
R_{10}+R_{30} \leq I(W_0,V_0;U_0,V_3,Y_3|Q)+I(W_0;V_0|Q) - I(W_0,W_1;V_0|Q),\\
R_{10}+R_{33} \leq I(W_0,V_3;U_0,V_0,Y_3|Q)+I(W_0;V_3|Q) - I(W_0,W_1;V_3|Q),\\
R_{20}+R_{30} \leq I(U_0,V_0;W_0,V_3,Y_3|Q)+I(U_0;V_0|Q) - I(W_0,W_1;U_0|Q) - I(W_0,W_1;V_0|Q),\\
R_{20}+R_{33} \leq I(U_0,V_3;W_0,V_0,Y_3|Q)+I(U_0;V_3|Q) - I(W_0,W_1;U_0|Q) - I(W_0,W_1;V_3|Q),\\
\nonumber R_{10}+R_{20}+R_{30} \leq I(W_0,U_0,V_0;V_3,Y_3|Q)+I(W_0,U_0;V_0|Q)+I(W_0;U_0|Q) \\ - I(W_0,W_1;U_0|Q) - I(W_0,W_1;V_0|Q),\\
\nonumber R_{10}+R_{20}+R_{33} \leq I(W_0,U_0,V_3;V_0,Y_3|Q)+I(W_0,U_0;V_3|Q)+I(W_0;U_0|Q) \\ - I(W_0,W_1;U_0|Q) - I(W_0,W_1;V_3|Q),\\
\nonumber R_{10}+R_{30}+R_{33} \leq I(W_0,V_0,V_3;U_0,Y_3|Q)+I(W_0,V_0;V_3|Q)+I(W_0;V_0|Q) \\ -I(W_0,W_1;V_0|Q)-I(W_0,W_1;V_3|Q),\\
\nonumber R_{20}+R_{30}+R_{33} \leq I(U_0,V_0,V_3;W_0,Y_3|Q)+I(U_0,V_0;V_3|Q)+I(U_0;V_0|Q)\\ - I(W_0,W_1;U_0|Q)- I(W_0,W_1;V_0|Q)-I(W_0,W_1;V_3|Q),\\
\nonumber R_{10}+R_{20}+R_{30}+R_{33} \leq I(W_0,U_0,V_0,V_3;Y_3|Q)+I(W_0,U_0,V_0;V_3|Q)+I(W_0,U_0;V_0|Q)+I(W_0;U_0|Q)\\ - I(W_0,W_1;U_0|Q)- I(W_0,W_1;V_0|Q)-I(W_0,W_1;V_3|Q).
\end{align*}

\section{}
Here, we provide the proofs for some of the corollaries stated in Section IV. The proofs for the remaining corollaries are similar and therefore are omitted.

\subsection*{Proof of Corollary \ref{corol1}:}
In the case of $\mathcal{C}^t_{G,\text{CuMS}}$, when senders $\mathcal{S}_{2}$ and $\mathcal{S}_{3}$ do not have any message of their own to transmit, they can use their noncausal message knowledge to entirely help sender $\mathcal{S}_{1}$. The rate tuple $(R_{1}^{*},0,0)$ is therefore achievable, where $R_{1}^{*}$ is the capacity of the vector channel $(\mathcal{S}_{1}, \mathcal{S}_{2}, \mathcal{S}_{3}) \rightarrow \mathcal{R}_{1}$, given by
\begin{eqnarray}
R_{1}^{*} = \dfrac{1}{2}\log_2\lb1 + \dfrac{\lb\sqrt P_{1} + |a_{12}|\sqrt P_{2} + \abs{a_{13}}\sqrt P_{3}\rb^2}{Q_{1}}\rb. \label{eq:crhelp1}
\end{eqnarray}
Next, when the rate achieved by sender $\mathcal{S}_{1}$ is zero, $\mathcal{S}_{2}$ can cancel the interference from $\mathcal{S}_{1}$ completely by employing dirty paper coding. However, due to the the message splitting model assumed here, $\mathcal{R}_2$ sees interference from $\mathcal{S}_{3}$ regardless of the $R_3$ achieved\footnote{Except in the case where $\mathcal{S}_3$ helps $\mathcal{R}_2$ in receiving its message. This case is dealt with in  corollary \ref{corol2}.}. Hence, the rate achievable by $(\mathcal{S}_{2},\mathcal{R}_{2})$ is
\begin{eqnarray}
& & R_{2}^{*} = \dfrac{1}{2}\log_2\lb1 + \dfrac{P_{2}}{Q_{2} + \abs{a_{23}}^2P_{3}}\rb. \label{eq:crhelp2}
\end{eqnarray}
When $R_{1} = 0$ and $R_{2} = R_{2}^{*}$, due to the noncausal knowledge of $\mathcal{S}_{1}$ and $\mathcal{S}_{2}$'s messages, $\mathcal{S}_{3}$ can completely mitigate the effect of interference and achieve the interference free rate, $R_{3}^{*}$, given by
\begin{eqnarray}
R_{3}^{*} = \dfrac{1}{2}\log_2\lb1 + \dfrac{P_{3}}{Q_{3}}\rb. \label{eq:crhelp3}
\end{eqnarray}
Hence, the rate tuple $(0,R_{2}^{*},R_{3}^{*})$ is achievable. Finally, the convex hull of the rate region $\mathfrak{G}_{\text{CuMS}}^2$ with these points is achievable by standard time-sharing arguments.

\subsection*{Proof of Corollary \ref{corol2}:}
As $\mathcal{S}_{3}$ has noncausal knowledge of $\mathcal{S}_{1}$ and $\mathcal{S}_{2}$, it can completely mitigate the effect of interference and achieve a rate of
\begin{eqnarray}
& & R_{3}^{*} = \dfrac{1}{2}\log_2\left(1 + \dfrac{P_{3}}{Q_{3}}\right). \label{eq:crhelp4}
\end{eqnarray}
If $\mathcal{S}_{3}$ achieves a rate less than the interference free rate, then it can use its remaining power to help either $\mathcal{S}_{1}$ or $\mathcal{S}_{2}$. The power required for $\mathcal{S}_{3}$ to achieve a rate of $r$ $(r \leq R_{3}^{*})$ is $P_{3}^{\mathcal{S}_3} = (2^{2r}-1)Q_{3}$. The power that can be used to help $\mathcal{S}_{1}$ or $\mathcal{S}_{2}$ is
\begin{eqnarray}
& & P_{3}^{\mathcal{S}_1} = P_{3}^{\mathcal{S}_2} = P_{3} - P_{3}^{\mathcal{S}_3}. \nonumber
\end{eqnarray}
When $\mathcal{S}_{2}$ achieves a rate of zero, then $\mathcal{S}_{2}$ can completely help $\mathcal{S}_{1}$. Further, $\mathcal{S}_{3}$ can use the power of $P_{3}^{\mathcal{S}_1}$ to help $\mathcal{S}_{1}$. Therefore, $\mathcal{S}_{1}$ can achieve a rate
\begin{eqnarray}
& & R_{1}^{*} = \dfrac{1}{2}\log_2\left(1 + \dfrac{\left(\sqrt{P_1} + |a_{13}|\sqrt{P_{3}^{\mathcal{S}_1}} + |a_{12}|\sqrt{P_{2}}\right)^2}{Q_{1} + \abs{a_{13}}^{2}P_{3}^{\mathcal{S}_3}}\right). \label{eq:crhelp5}
\end{eqnarray}
When $R_{2}^{*} = 0$ and $R_{3} = r$, then $S_{3}$ can use the power of $P_{3}^{\mathcal{S}_2}$ to transmit the message of $\mathcal{S}_{2}$, and $\mathcal{S}_{2}$ can cancel the interference from $\mathcal{S}_{1}$ by employing dirty paper coding and achieve a rate
\begin{eqnarray}
&  & R_{2}^{*} = \dfrac{1}{2}\log_2\left(1 + \dfrac{\left(\sqrt{P_2} + |a_{23}|\sqrt{P_{3}^{\mathcal{S}_1}}\right)^2}{Q_{2} + \abs{a_{23}}^{2}P_{3}^{\mathcal{S}_3}}\right). \label{eq:crhelp6}
\end{eqnarray}
Hence the rate tuples $(R_{1}^{*}, 0, r)$ and $(0, R_{2}^{*}, r)$ are achievable. The convex hull of the region $\mathfrak{G}_{\text{CuMS}}^2$ with these rate tuples is also achievable by time-sharing arguments.

\subsection*{Proof of Corollary \ref{corol3}:}
Due to the knowledge of $\mathcal{S}_{1}$ message, $\mathcal{S}_{2}$ can completely cancel the effect of interference from primary but it will always see the interference from $\mathcal{S}_{3}$. Hence $\mathcal{S}_{2}$ can achieve a rate
\begin{eqnarray}
& & R_{2}^{*} = \dfrac{1}{2}\log\left(1 + \dfrac{P_{2}}{Q_{2} + |a_{23}|^{2}P_{3}}\right). \nonumber
\end{eqnarray}
When $\mathcal{S}_{2}$ achieves certain specific rate of $r$ $(r \leq R_{3}^{*})$, the power required is
\begin{eqnarray}
P_{2}^{\mathcal{S}_2} = (2^{2r}-1)(Q_{2} + |a_{23}|^{2}P_{3}). \label{eq:crhelp7}
\end{eqnarray}
The remaining power $P_{2}^{\mathcal{S}_1} = P_{2} - P_{2}^{\mathcal{S}_2} $ can be used to help $\mathcal{S}_{1}$'s transmission. When $R_{3} = 0$ by similar arguments as in the previous Corollary, $\mathcal{S}_{1}$ can achieve a rate
\begin{eqnarray}
& & R_{1}^{*} = \dfrac{1}{2}\log_2\left(1 + \dfrac{\left(\sqrt{P_1} + |a_{12}|\sqrt{P_{2}^{\mathcal{S}_1}} + |a_{13}|\sqrt{P_{3}}\right)^2}{Q_{1} + |a_{12}|^{2}P_{2}^{\mathcal{S}_2}}\right). \label{eq:crhelp8}
\end{eqnarray}
When $R_{1}^{*} = 0$, the sender $\mathcal{S}_{3}$ can mitigate the interference from $\mathcal{S}_{1}$ and achieves the interference free rate i.e.
\begin{eqnarray}
& & R_{3}^{*} = \dfrac{1}{2}\log\left(1 + \dfrac{P_{3}}{Q_{3}}\right). \label{eq:crhelp9}
\end{eqnarray}
From (\ref{eq:crhelp8}) and (\ref{eq:crhelp9}), the rate tuples $(R_{1}^{*},r,0)$ and $(0,r,R_{3}^{*})$ are achievable. The convex hull is achieved using time-sharing arguments.

\subsection*{Proof of Corollary \ref{corol5}:}
In case of $\mathcal{C}^t_{G,\text{PrMS}}$ channel model, the senders $\mathcal{S}_{2}$ and $\mathcal{S}_{3}$ can only help sender $\mathcal{S}_1$ in its transmission. When $R_{2}=0$ and $R_{3}$ = 0, $\mathcal{S}_{1}$ can achieve a rate
\begin{eqnarray}
R_{1}^{*} = \dfrac{1}{2}\log\lb1 + \dfrac{\lb\sqrt P_{1} + \abs{a_{12}}\sqrt P_{2} + \abs{a_{13}}\sqrt P_{3}\rb^2}{Q_{1}}\rb. \label{eq:crhelp10}
\end{eqnarray}
When $R_{1} =0$, both $\mathcal{S}_{2}$ and $\mathcal{S}_{3}$ can completely eliminate the effect of interference from $\mathcal{S}_{1}$. However, they will experience the interference from each other. Hence, $\mathcal{S}_{2}$ and $\mathcal{S}_{3}$ can achieve a rate of $R_{2}^{*}$ and $R_{3}^{*}$ given by
\begin{eqnarray}
R_{2}^{*} = \dfrac{1}{2}\log\lb1 + \dfrac{P_{2}}{Q_{2} + \abs{a_{23}}^{2}P_{3}}\rb, \label{eq:crhelp11}\\
R_{3}^{*} = \dfrac{1}{2}\log\lb1 + \dfrac{P_{3}}{Q_{3} + \abs{a_{32}}^{2}P_{2}}\rb. \label{eq:crhelp12}
\end{eqnarray}
From (\ref{eq:crhelp10}) - (\ref{eq:crhelp12}), it is clear that the rate tuples $(R_{1}^{*},0,0)$ and $(0,R_{2}^{*},R_{3}^{*})$ are achievable. By standard time-sharing arguments, the convex hull is achievable.

\subsection*{Proof of Corollary \ref{corol6}:}
As $\mathcal{S}_{3}$ has noncausal knowledge of primary message, it can employ dirty paper coding to completely mitigate the effect of interference from $\mathcal{S}_{1}$. However it sees interference from $\mathcal{S}_{2}$ due to the rate splitting. Hence, $\mathcal{S}_{3}$ can achieve a rate
\begin{eqnarray}
R_{3}^{*} = \dfrac{1}{2}\log_2\lb1 + \dfrac{P_{3}}{Q_{3} + \abs{a_{32}}^{2}P_{2}}\rb. \label{eq:crhelp13}
\end{eqnarray}
In order to achieve a rate $r$ $(r \leq R_{3}^{*})$, the power required by $\mathcal{S}_{3}$ is
\begin{eqnarray}
P_{3}^{\mathcal{S}_3} = (1 + (2^{2r}-1)\abs{a_{32}}^{2}P_{2}). \nonumber
\end{eqnarray}
The remaining power $P_{3}^{\mathcal{S}_1} = P_{3} - P_{3}^{\mathcal{S}_3}$ can be used to help $\mathcal{S}_{1}$. When $R_{2} = 0$ and $R_{3} = r$, $\mathcal{S}_{2}$ and $\mathcal{S}_{3}$ can use the power of $P_{2}$ and $P_{3}^{\mathcal{S}_1}$, respectively, to help the primary. The rate achieved by $\mathcal{S}_{1}$ is
\begin{equation}
R_{1}^{*} = \dfrac{1}{2}\log_2\lb1 + \dfrac{\lb\sqrt{P_{1}} + \abs{a_{12}}\sqrt{P_{2}} + \abs{a_{13}}\sqrt{P_{3}^{\mathcal{S}_1}}\rb^2}{Q_{1} + \abs{a_{13}}^{2}P_{3}^{\mathcal{S}_3}}\rb, \label{eq:crhelp14}
\end{equation}
where the $\abs{a_{13}}^{2}P_{3}^{\mathcal{S}_3}$ in the denominator arises because of sender $\mathcal{S}_3$ transmitting the message to its pairing receiver. When $R_{1} = 0$ and $R_{3} = r$, $\mathcal{R}_{2}$ can achieve a rate of
\begin{eqnarray}
R_{2}^{*} = \dfrac{1}{2}\log_2\lb 1 + \dfrac{P_{2}}{Q_{2} + \abs{a_{23}}^{2}P_{3}}\rb. \label{eq:crhelp15}
\end{eqnarray}
From (\ref{eq:crhelp14}) and (\ref{eq:crhelp15}), the rate tuples $(R_{1}^{*},0,r)$ and $(0,R_{2}^{*},r)$ are achievable. The convex hull can be achieved by time-sharing.

\bibliographystyle{IEEEtran}
\bibliography{IEEEabrv,cr_journal2}
\begin{table}[h]
  \centering
  \begin{tabular}{|c|c|c|}
  \hline
  Sub-message & Rate & Description \\ \hline
  $m_{10} \in \{1,...,2^{nR_{10}}\}$ & $R_{10}$ & Rate achieved: $\mathcal{S}_1 \rightarrow (\mathcal{R}_1, \mathcal{R}_2, \mathcal{R}_3)$ \\ \hline
  $m_{11}\in \{1,...,2^{nR_{11}}\}$ & $R_{11}$ & Rate achieved: $\mathcal{S}_1 \rightarrow \mathcal{R}_1$ \\ \hline
  $m_{20}\in \{1,...,2^{nR_{20}}\}$ & $R_{20}$ & Rate achieved: $\mathcal{S}_2 \rightarrow (\mathcal{R}_1, \mathcal{R}_2, \mathcal{R}_3)$ \\ \hline
  $m_{21}\in \{1,...,2^{nR_{21}}\}$ & $R_{21}$ & Rate achieved: $\mathcal{S}_2 \rightarrow (\mathcal{R}_1, \mathcal{R}_2)$ \\ \hline
  $m_{22}\in \{1,...,2^{nR_{22}}\}$ & $R_{22}$ & Rate achieved: $\mathcal{S}_2 \rightarrow \mathcal{R}_2$ \\ \hline
  $m_{30}\in \{1,...,2^{nR_{30}}\}$ & $R_{30}$ & Rate achieved: $\mathcal{S}_3 \rightarrow (\mathcal{R}_1, \mathcal{R}_2, \mathcal{R}_3)$ \\ \hline
  $m_{31}\in \{1,...,2^{nR_{31}}\}$ & $R_{31}$ & Rate achieved: $\mathcal{S}_3 \rightarrow (\mathcal{R}_1, \mathcal{R}_3)$  \\ \hline
  $m_{33}\in \{1,...,2^{nR_{33}}\}$ & $R_{33}$ & Rate achieved: $\mathcal{S}_3 \rightarrow \mathcal{R}_3$ \\ \hline
  $m_{1} \in \{1,...,2^{nR_{1}}\}$ & $R_{1}$ & Rate achieved: $\mathcal{S}_1 \rightarrow \mathcal{R}_1$ \\ \hline
  $m_{2} \in \{1,...,2^{nR_{2}}\}$ & $R_{2}$ & Rate achieved: $\mathcal{S}_2 \rightarrow \mathcal{R}_2$ \\ \hline
\end{tabular}
  \caption{Achievable rates and their description. For ex., $R_{11}$ is the rate achieved between $\mathcal{S}_1$ and $\mathcal{R}_1$, while $R_{21}$ is the rate achieved between $\mathcal{S}_2$, and $\mathcal{R}_2$, $\mathcal{R}_1$, etc. The last two rows correspond to the channel $\mathcal{C}_{spc}$, wherein the senders $\mathcal{S}_1$ and $\mathcal{S}_2$ do not perform rate-splitting.}\label{table1}
\end{table}
\begin{table}
  \centering
  \begin{tabular}{|c|c|}
  \hline
  Receiver & Decoding capability \\ \hline
  $\mathcal{R}_1$ & $m_{10}$, $m_{11}$, $m_{20}$, $m_{30}$ \\ \hline
  $\mathcal{R}_2$ & $m_{10}$, $m_{20}$, $m_{22}$, $m_{30}$ \\ \hline
  $\mathcal{R}_3$ & $m_{10}$, $m_{20}$, $m_{30}$, $m_{33}$  \\ \hline
\end{tabular}
  \caption{Effect of rate-splitting on the decoding capability of receivers for the channels $\mathcal{C}^1_{cms}$, $\mathcal{C}^1_{pms}$. For ex., receiver $\mathcal{R}_2$ can decode messages $m_{10}$, $m_{20}$, $m_{22}$, $m_{30}$}\label{table2}\vspace{-0.5cm}
\end{table}
\begin{table}
  \centering
  \begin{tabular}{|c|c|}
  \hline
  Receiver & Decoding capability \\ \hline
  $\mathcal{R}_1$ & $m_{11}$, $m_{21}$, $m_{31}$ \\ \hline
  $\mathcal{R}_2$ & $m_{21}$, $m_{22}$ \\ \hline
  $\mathcal{R}_3$ & $m_{31}$, $m_{33}$ \\ \hline
\end{tabular}
  \caption{Effect of rate-splitting on the decoding capability of receivers for the channels $\mathcal{C}^2_{cms}$, $\mathcal{C}^2_{pms}$. For ex., receiver $\mathcal{R}_3$ can decode messages $m_{31}$, $m_{33}$}\label{table3}
\end{table}
\begin{table}[h]
  \centering
\begin{tabular}{|c|c|}
  \hline
  Receiver & Can decode \\ \hline
  $\mathcal{R}_1$ & $m_1$, $m_{31}$ \\ \hline
  $\mathcal{R}_2$ & $m_2$, $m_{31}$ \\ \hline
  $\mathcal{R}_3$ & $m_{31}$, $m_{33}$ \\ \hline
\end{tabular}
  \caption{Effect of rate-splitting on the decoding capability of receivers for the channel $\mathcal{C}_{spc}$ . For ex. the receiver denoted $\mathcal{R}_2$ can decode messages $m_2$ and $m_{31}$. Note that, there is no rate-splitting at the senders $\mathcal{S}_1$ and $\mathcal{S}_2$.}\label{table4}
\end{table}

\begin{table}
  \centering
  \begin{tabular}{|c|c|}
  \hline
  Variable & Description \\ \hline
  $W_0 \in \mathcal{W}_0$ & Public Information: $\mathcal{S}_1 \rightarrow (\mathcal{R}_1, \mathcal{R}_2, \mathcal{R}_3)$ \\ \hline
  $W_1 \in \mathcal{W}_1$ & Private Information: $\mathcal{S}_1 \rightarrow \mathcal{R}_1$ \\ \hline
  $U_0 \in \mathcal{U}_0$ & Public Information: $\mathcal{S}_2 \rightarrow (\mathcal{R}_1, \mathcal{R}_2, \mathcal{R}_3)$ \\ \hline
  $U_1 \in \mathcal{U}_1$ & Public information: $\mathcal{S}_2 \rightarrow (\mathcal{R}_1, \mathcal{R}_2)$ \\ \hline
  $U_2 \in \mathcal{U}_2$ & Private information: $\mathcal{S}_2 \rightarrow \mathcal{R}_2$  \\ \hline
  $V_0 \in \mathcal{V}_0$ & Public information: $\mathcal{S}_3 \rightarrow (\mathcal{R}_1, \mathcal{R}_2, \mathcal{R}_3)$ \\ \hline
  $V_1 \in \mathcal{V}_1$ & Public information: $\mathcal{S}_3 \rightarrow (\mathcal{R}_1, \mathcal{R}_3)$ \\ \hline
  $V_3 \in \mathcal{V}_3$ & Private information: $\mathcal{S}_3 \rightarrow \mathcal{R}_3$ \\ \hline
\end{tabular}
  \caption{Auxiliary Random variables and their description. For ex., $U_1$ denotes public information from $\mathcal{S}_2$ decodable at $\mathcal{R}_1$ and $\mathcal{R}_2$ }\label{table5}\vspace{-0.5cm}
\end{table}
\begin{figure}[h]
\includegraphics[height=3.4in,width=5.9in]{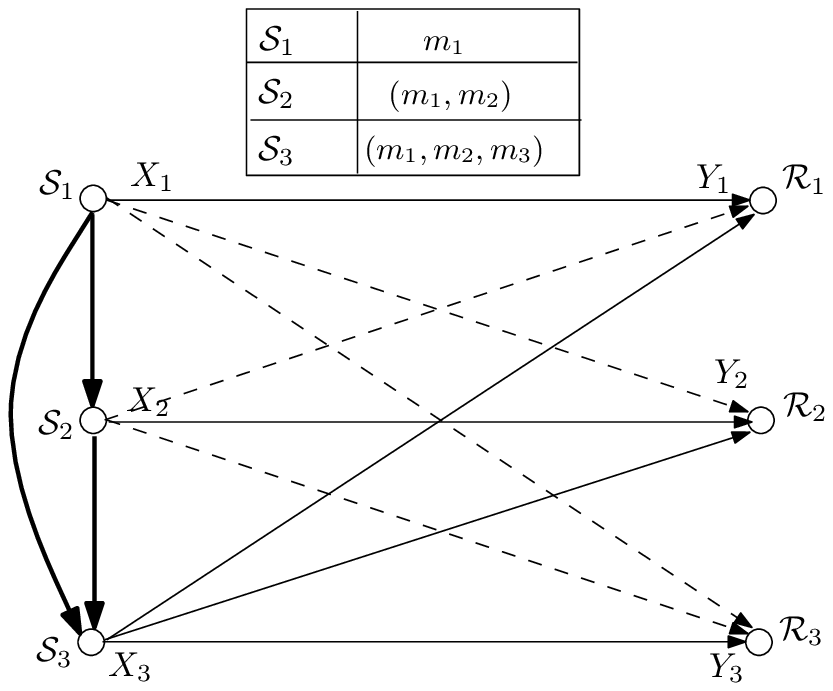}
\caption{Three-user cognitive channel with CuMS}\label{fig:fig1}
\end{figure}
\begin{figure}
\includegraphics[height=3.4in,width=5.9in]{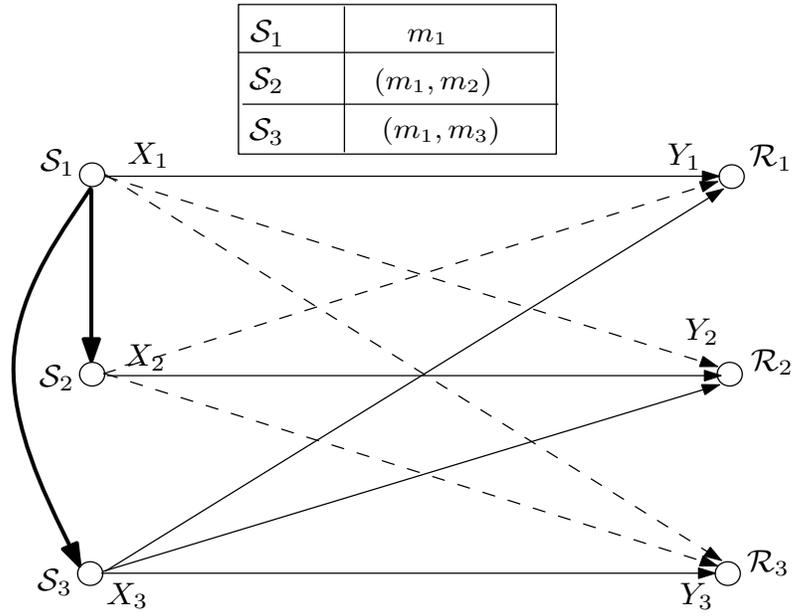}
\caption{Three-user cognitive channel with PrMS}\label{fig:fig2}
\end{figure}
\begin{figure}
\includegraphics[height=3.4in,width=5.9in]{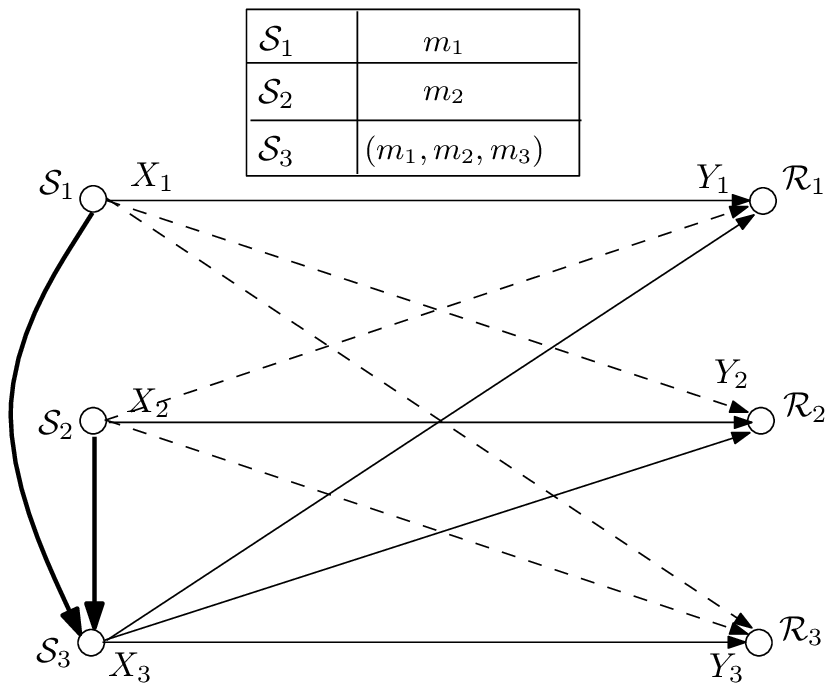}
\caption{Three-user cognitive channel with CoMS}\label{fig:fig3}
\end{figure}
\begin{figure}
\centering
\includegraphics[height=3.4in,width=5.9in]{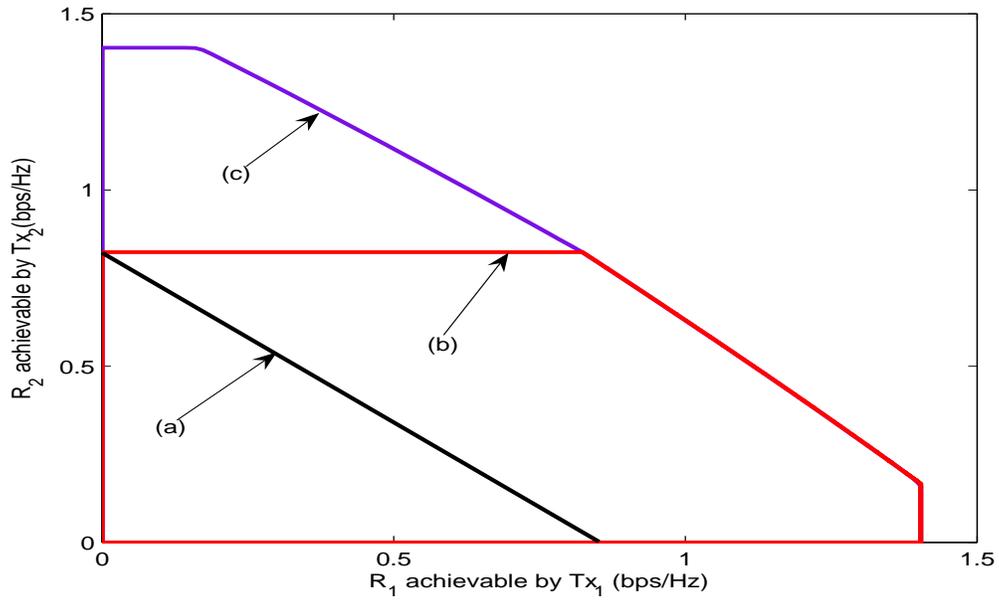}
\caption{Two-user interference channels with different rate-splitting strategies. In (a), neither transmitter performs rate-splitting. In (b), one of the transmitters performs rate-splitting. In (c), both the transmitters perform rate-splitting. The power at the transmitters are 7.8dB.}\label{fig:fig4}
\end{figure}
\begin{figure}
\centering
\includegraphics[height=3.4in,width=5.9in]{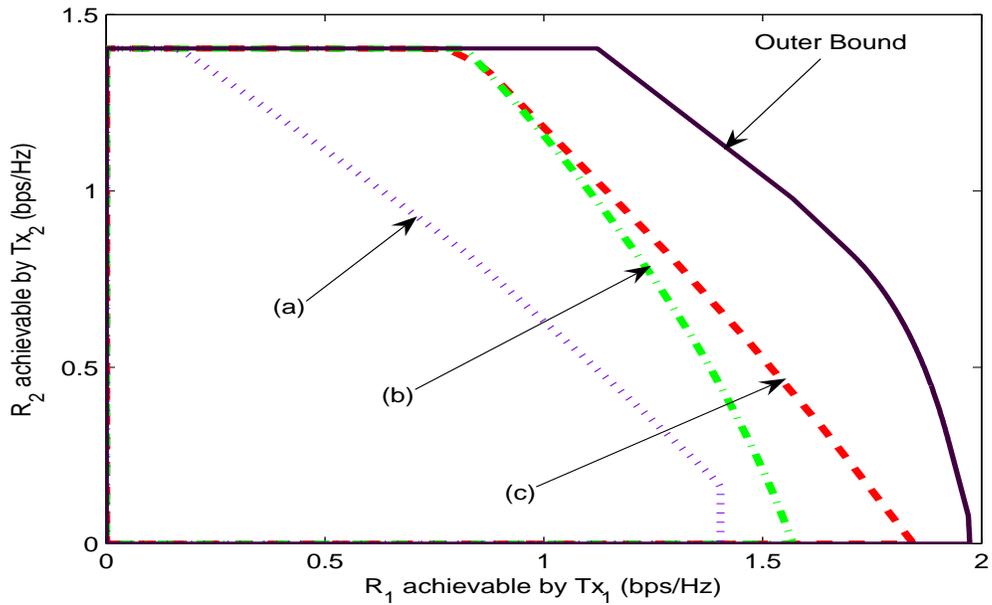}
\caption{Two-user CR and interference channels. (a) is the Han-Kobayashi rate region, (b) is the rate region of a CR channel where only the cognitive transmitter performs rate-splitting \cite{refjiang} and (c) is the rate region of a CR channel where both transmitters perform rate-splitting \cite{refnatasha1}. The power at the transmitters are 7.8dB.}\label{fig:fig5}
\end{figure}
\clearpage
\begin{figure}
\includegraphics[height=3.4in,width=5.9in]{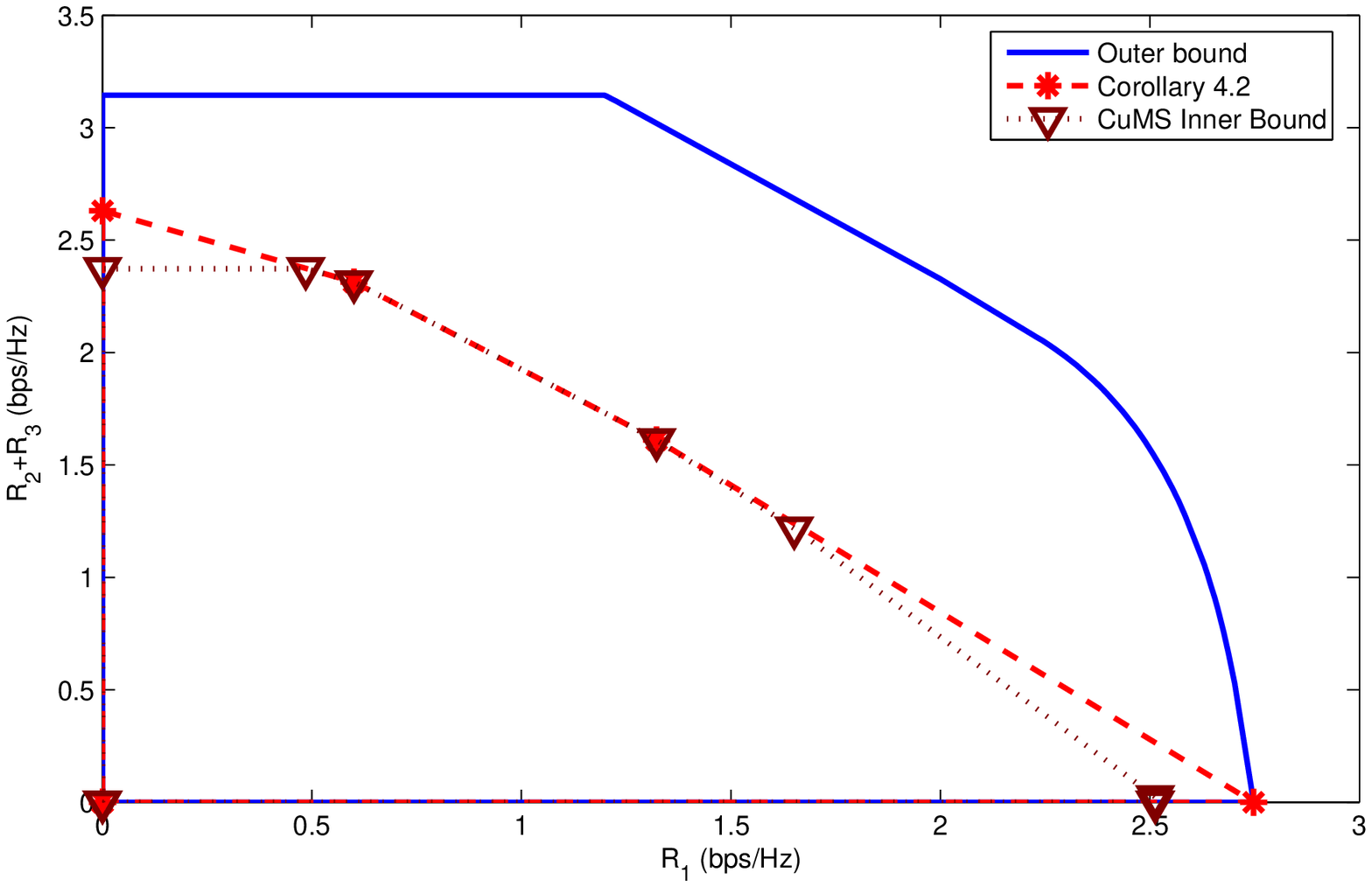}
\caption{Rate of $\mathcal{S}_1$ ($R_{1}$) versus the sum rate of $\mathcal{S}_2$ and $\mathcal{S}_3$ ($R_{2} + R_{3}$) for the channel $\mathcal{C}_\text{CuMS}^2$. The power at the transmitters is 10dB.}\label{fig:fig6}
\end{figure}
\begin{figure}
\includegraphics[height=3.4in,width=5.9in]{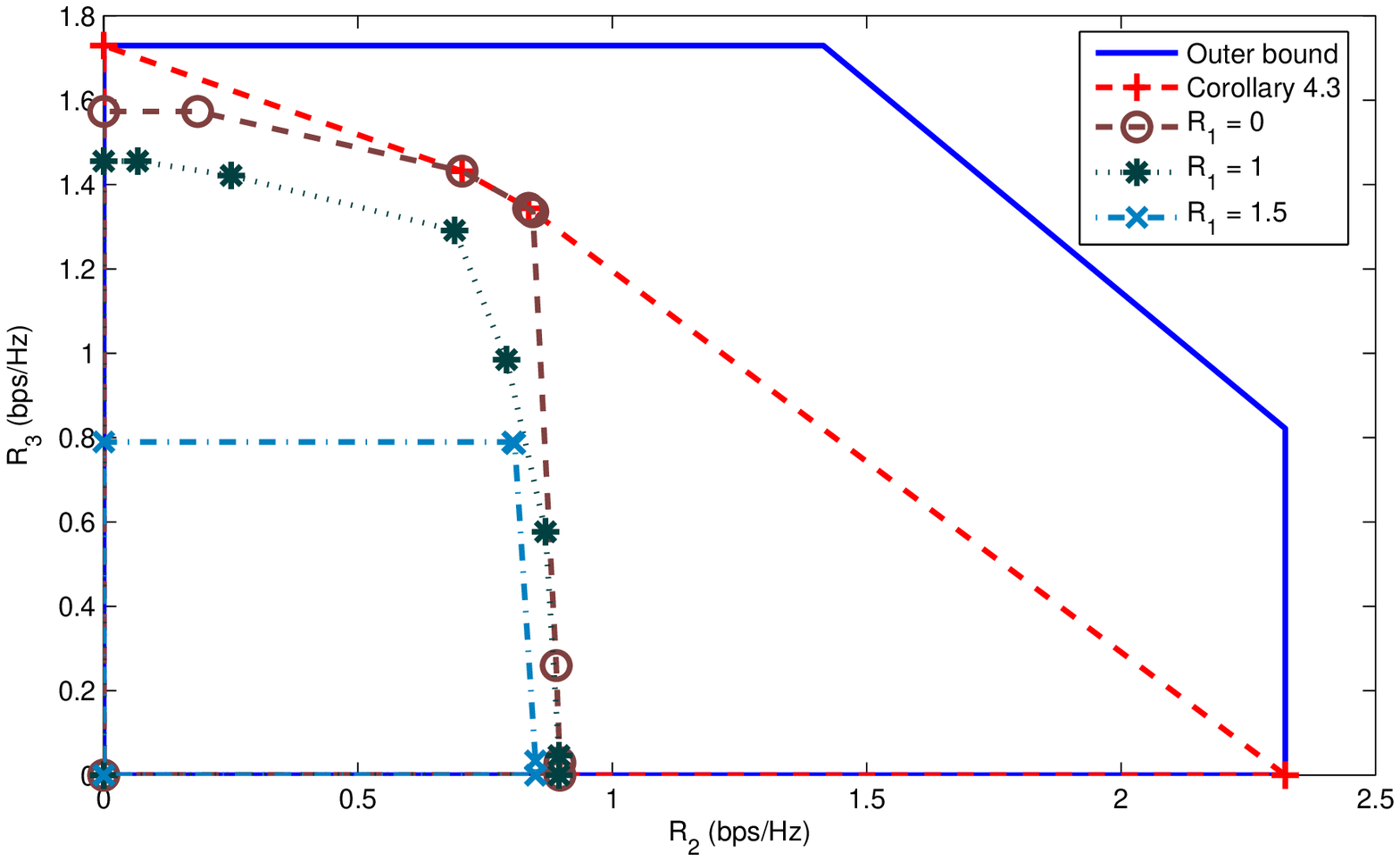}
\caption{Rate of $\mathcal{S}_2$ ($R_{2}$) versus the rate of $\mathcal{S}_3$ ($R_{3}$) when $\mathcal{S}_1$ is guaranteed to achieve a minimum rate $R_{1} = 0, 1 \text{ and } 1.5$ bps/Hz, for the channel $\mathcal{C}_\text{CuMS}^2$. The power at the transmitters is 10dB. }\label{fig:fig7}
\end{figure}\clearpage
\begin{figure}
\includegraphics[height=3.4in,width=5.9in]{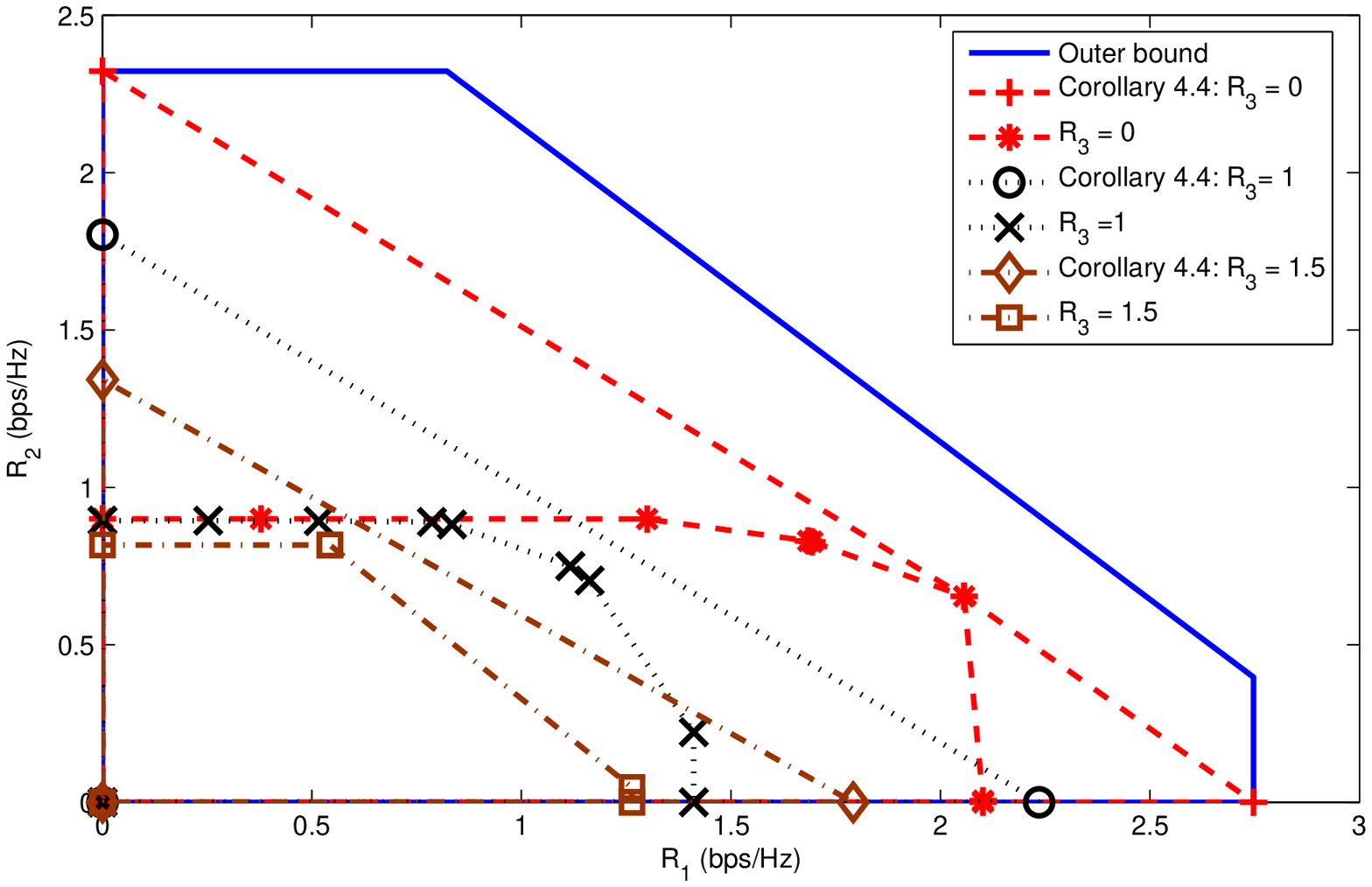}
\caption{Rate of $\mathcal{S}_1$ ($R_{1}$) versus the rate of $\mathcal{S}_2$ ($R_{2}$) when $\mathcal{S}_3$ is guaranteed to achieve a minimum rate $R_{3} = 0, 1 \text{ and } 1.5$ bps/Hz, for the channel $\mathcal{C}_\text{CuMS}^2$. The power at the transmitters is 10dB. }\label{fig:fig8}
\end{figure}
\begin{figure}
\includegraphics[height=3.4in,width=5.9in]{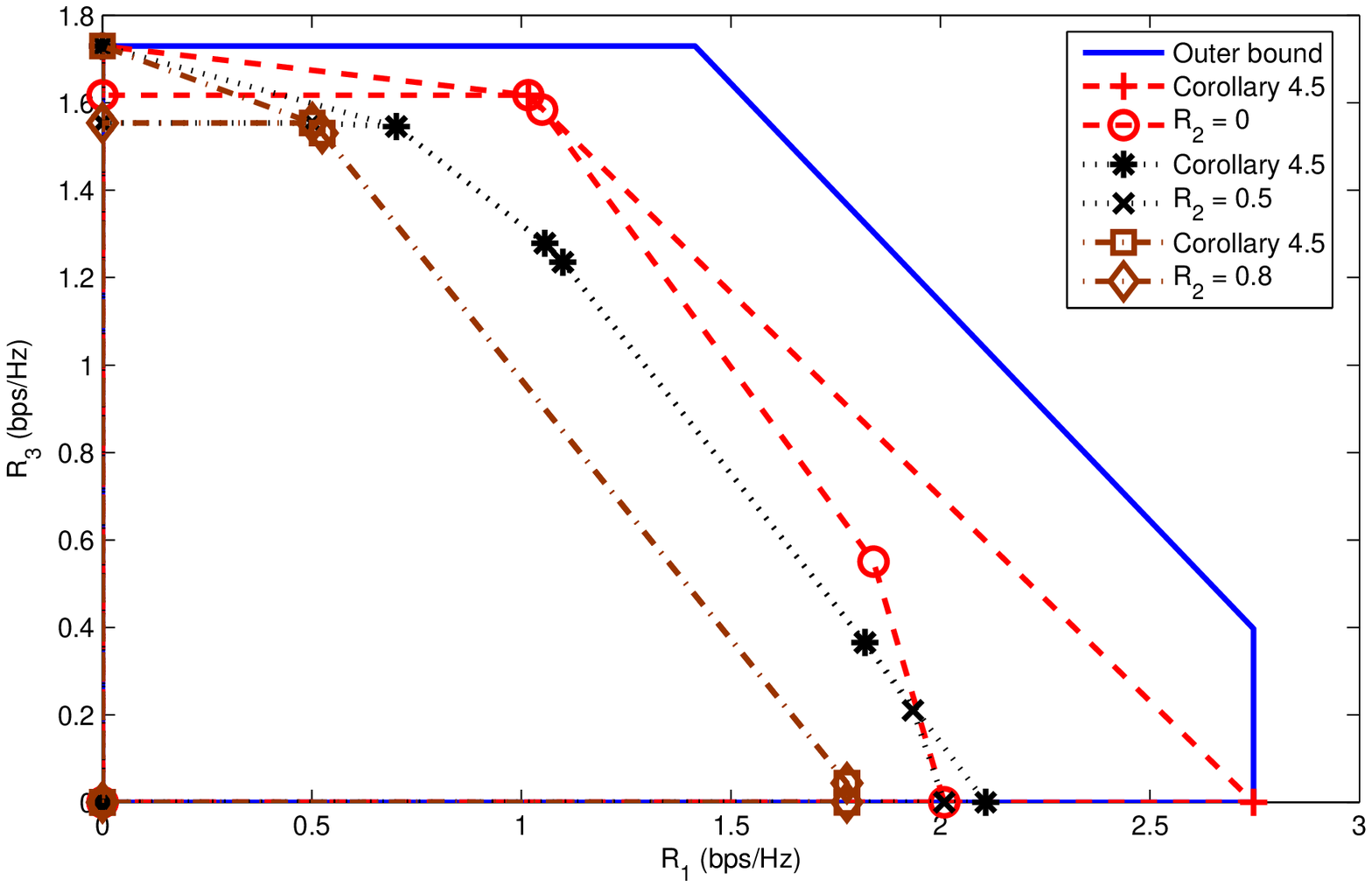}
\caption{Rate of $\mathcal{S}_1$ ($R_{1}$) versus the rate of $\mathcal{S}_3$ ($R_{3}$) when $\mathcal{S}_2$ is guaranteed to achieve a minimum rate $R_{2} = 0, 0.5 \text{ and } 0.8$ bps/Hz, for the channel $\mathcal{C}_\text{CuMS}^2$. The power at the transmitters is 10dB.}\label{fig:fig9}
\end{figure}
\begin{figure}
\includegraphics[height=3.4in,width=5.9in]{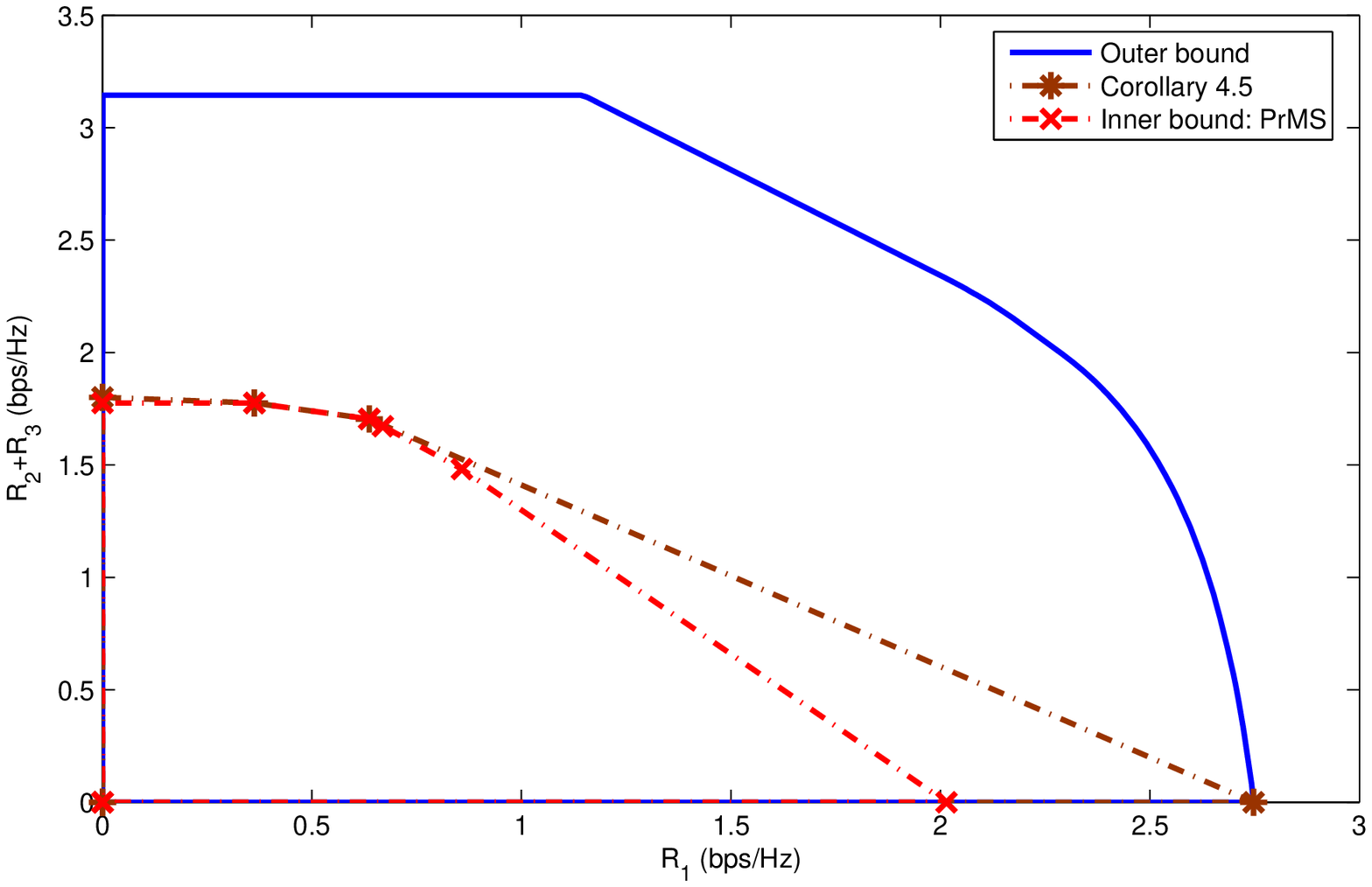}
\caption{Rate of $\mathcal{S}_1$ ($R_{1}$) versus the sum rate of $\mathcal{S}_2$ and $\mathcal{S}_3$ ($R_{2} + R_{3}$) for the channel $\mathcal{C}_\text{PrMS}^2$. The power at the transmitters is 10dB.}\label{fig:fig10}
\end{figure}
\begin{figure}
\includegraphics[height=3.4in,width=5.9in]{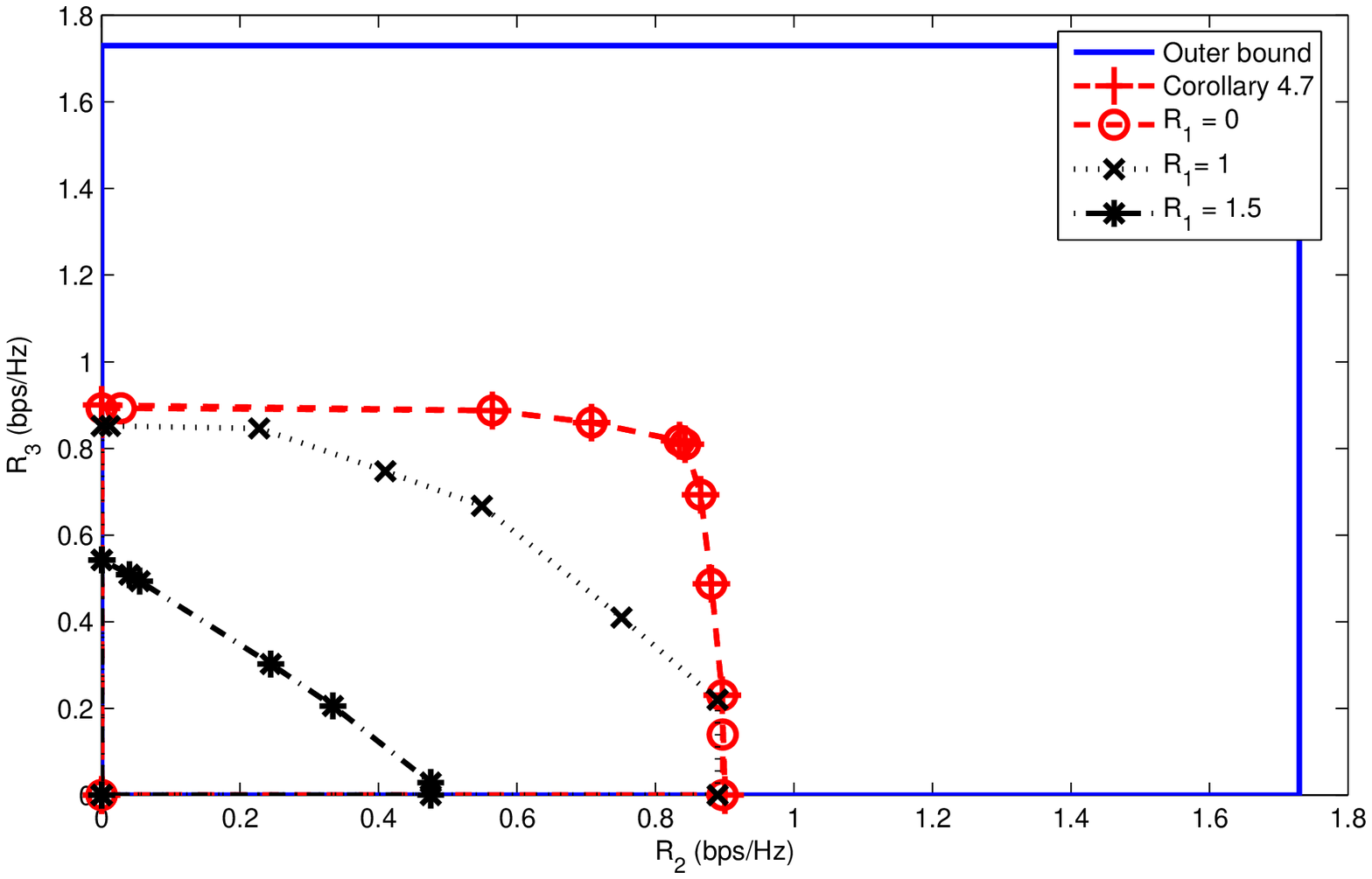}
\caption{Rate of $\mathcal{S}_2$ ($R_{2}$) versus the rate of $\mathcal{S}_3$ ($R_{3}$) when $\mathcal{S}_1$ is guaranteed to achieve a minimum rate $R_{1}=0,  1 , \text{ and } 1.5$ bps/Hz, for the channel $\mathcal{C}_\text{PrMS}^2$. The power at the transmitters is 10dB. }\label{fig:fig11}
\end{figure}
\begin{figure}
\includegraphics[height=3.4in,width=5.9in]{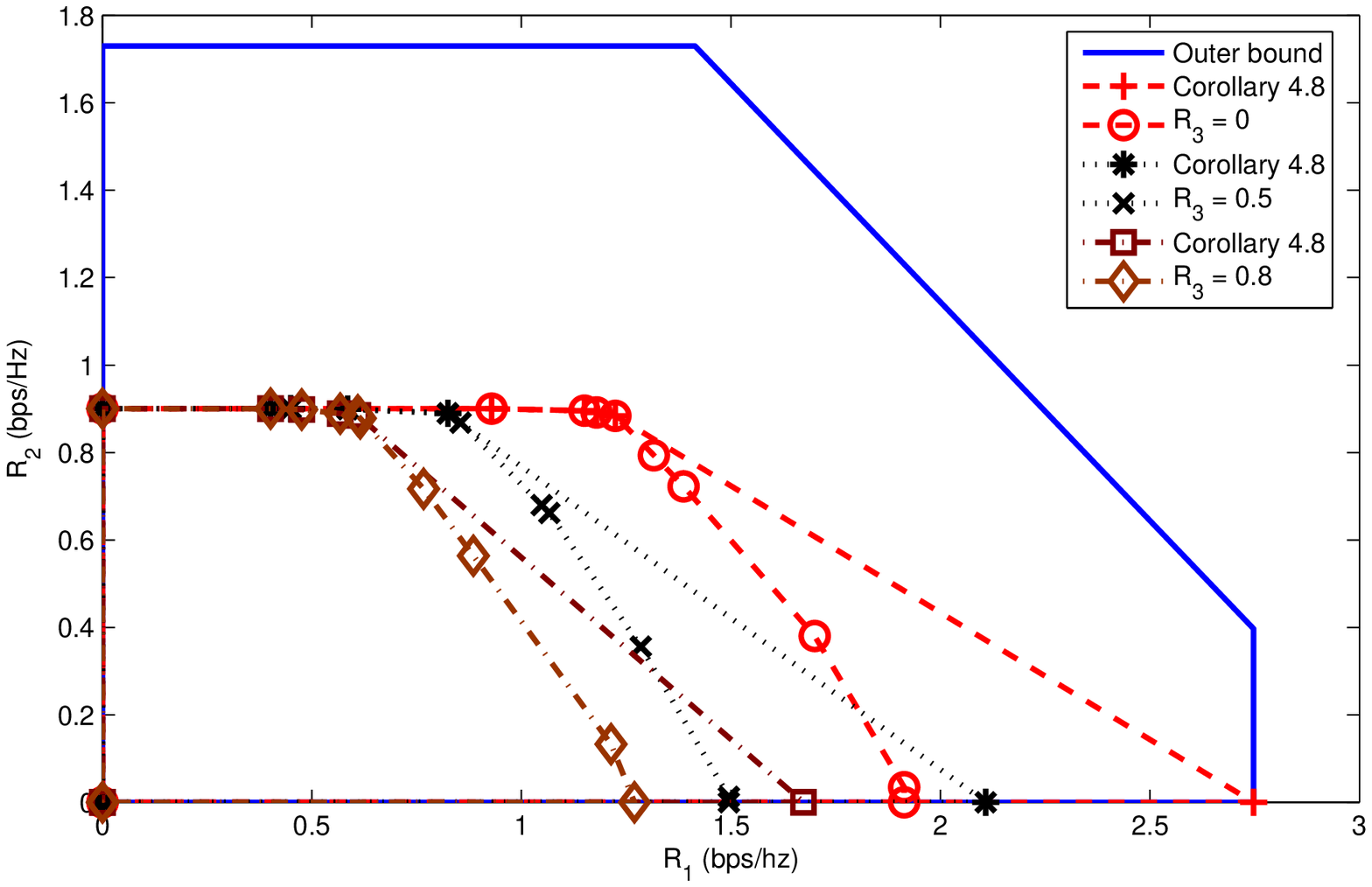}
\caption{Rate of $\mathcal{S}_1$ ($R_{1}$) versus the rate of $\mathcal{S}_2$ ($R_{2}$) when $\mathcal{S}_3$ is guaranteed to achieve a minimum rate $R_{3} = 0, 0.5 \text{ and } 0.8$ bps/Hz, for the channel $\mathcal{C}_\text{PrMS}^2$. The power at the transmitters is 10dB. }\label{fig:fig12}
\end{figure}
\begin{figure}
\includegraphics[height=3.4in,width=5.9in]{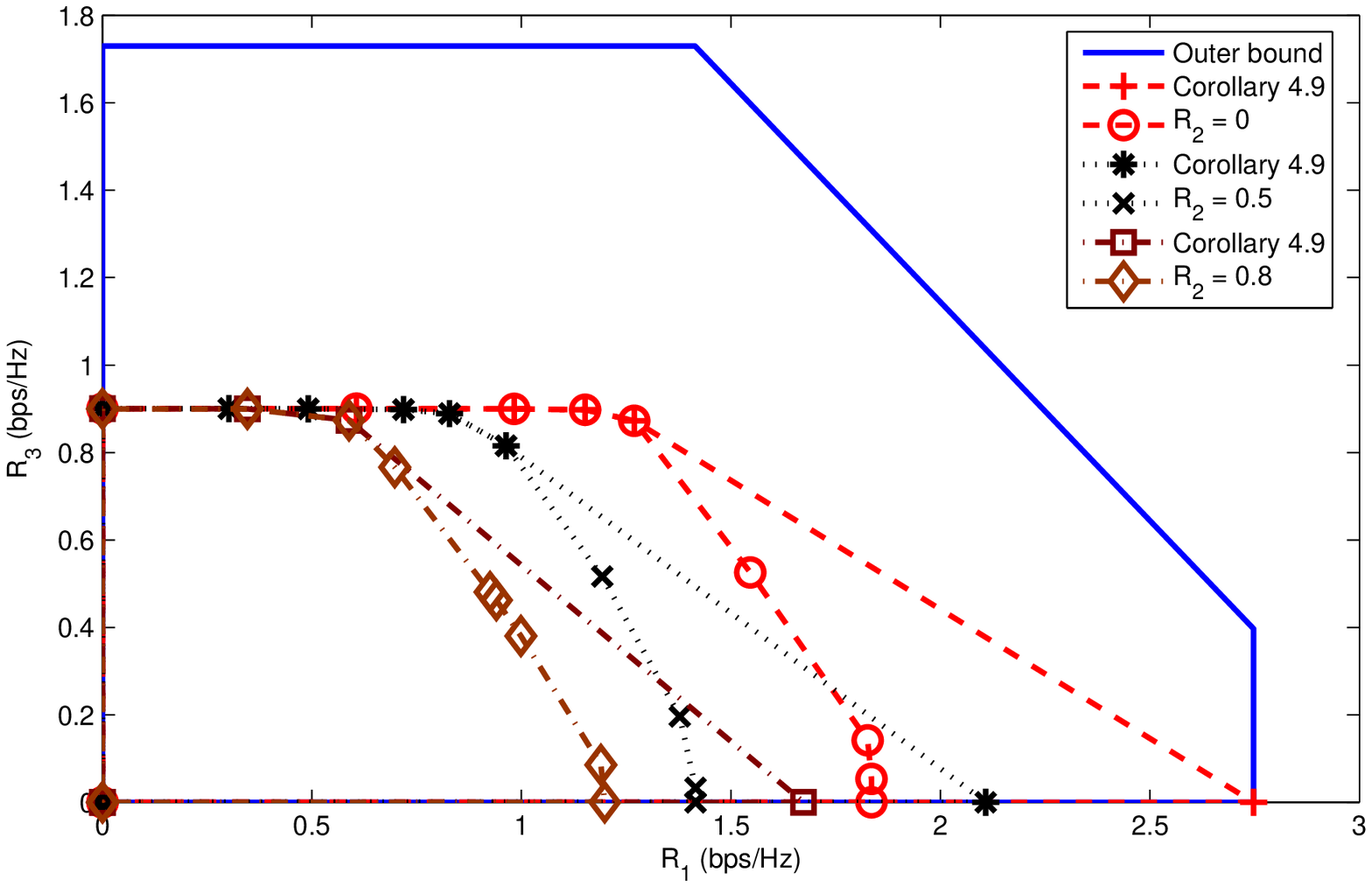}
\caption{Rate of $\mathcal{S}_1$ ($R_{1}$) versus the rate of $\mathcal{S}_3$ ($R_{3}$) when $\mathcal{S}_2$ is guaranteed to achieve a minimum rate $R_{2} = 0, 0.5 \text{ and } 0.8$ bps/Hz, for the channel $\mathcal{C}_\text{PrMS}^2$. The power at the transmitters is 10dB. }\label{fig:fig13}
\end{figure}
\begin{figure}
\includegraphics[height=3.4in,width=5.9in]{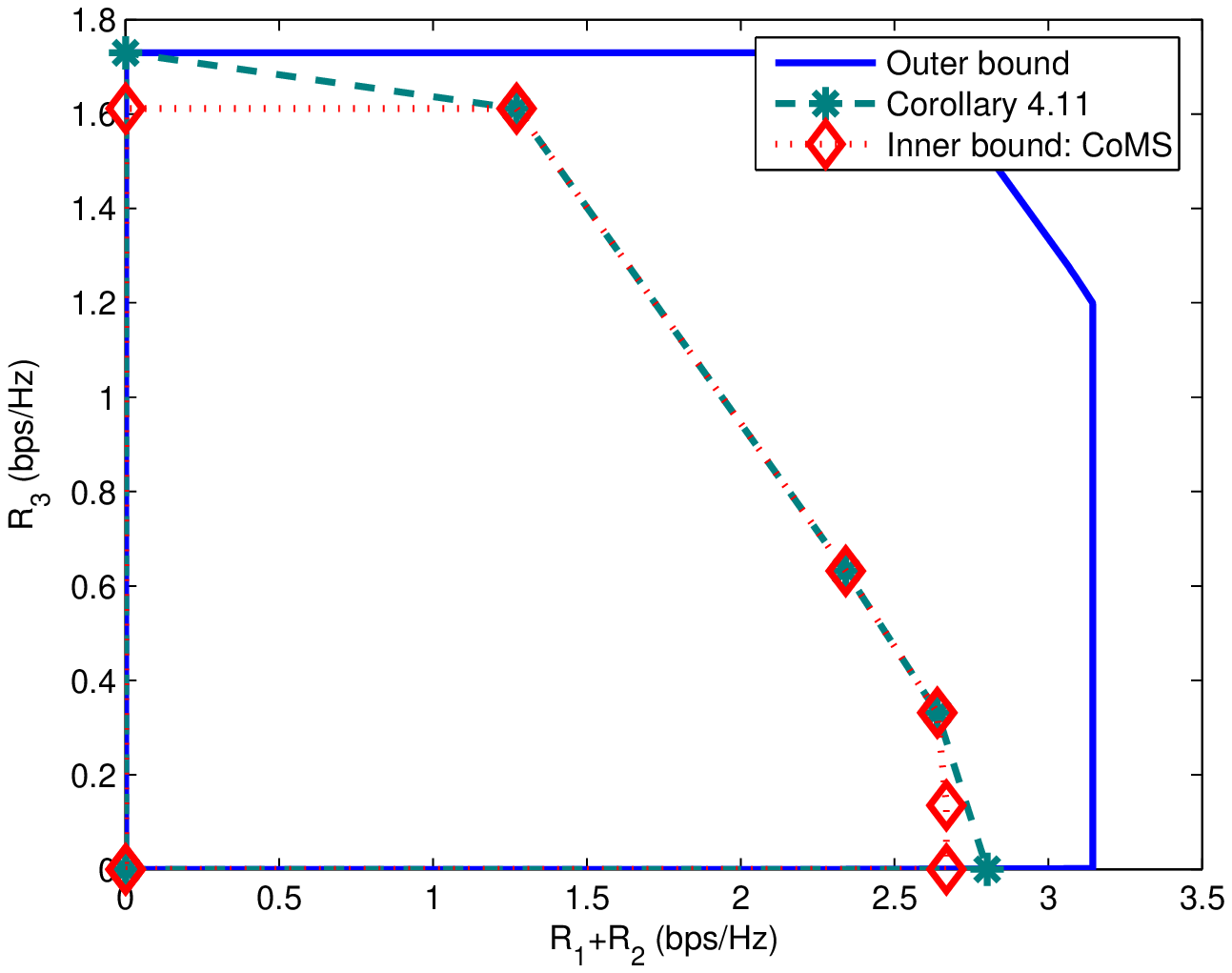}
\caption{Rate of $\mathcal{S}_3$ ($R_{3}$) versus the sum rate of $\mathcal{S}_1$ and $\mathcal{S}_2$ ($R_1+R_2$) for the channel $\mathcal{C}_\text{CoMS}^2$. The power at the transmitters is 10dB. }\label{fig:fig14}
\end{figure}
\begin{figure}
\includegraphics[height=3.4in,width=5.9in]{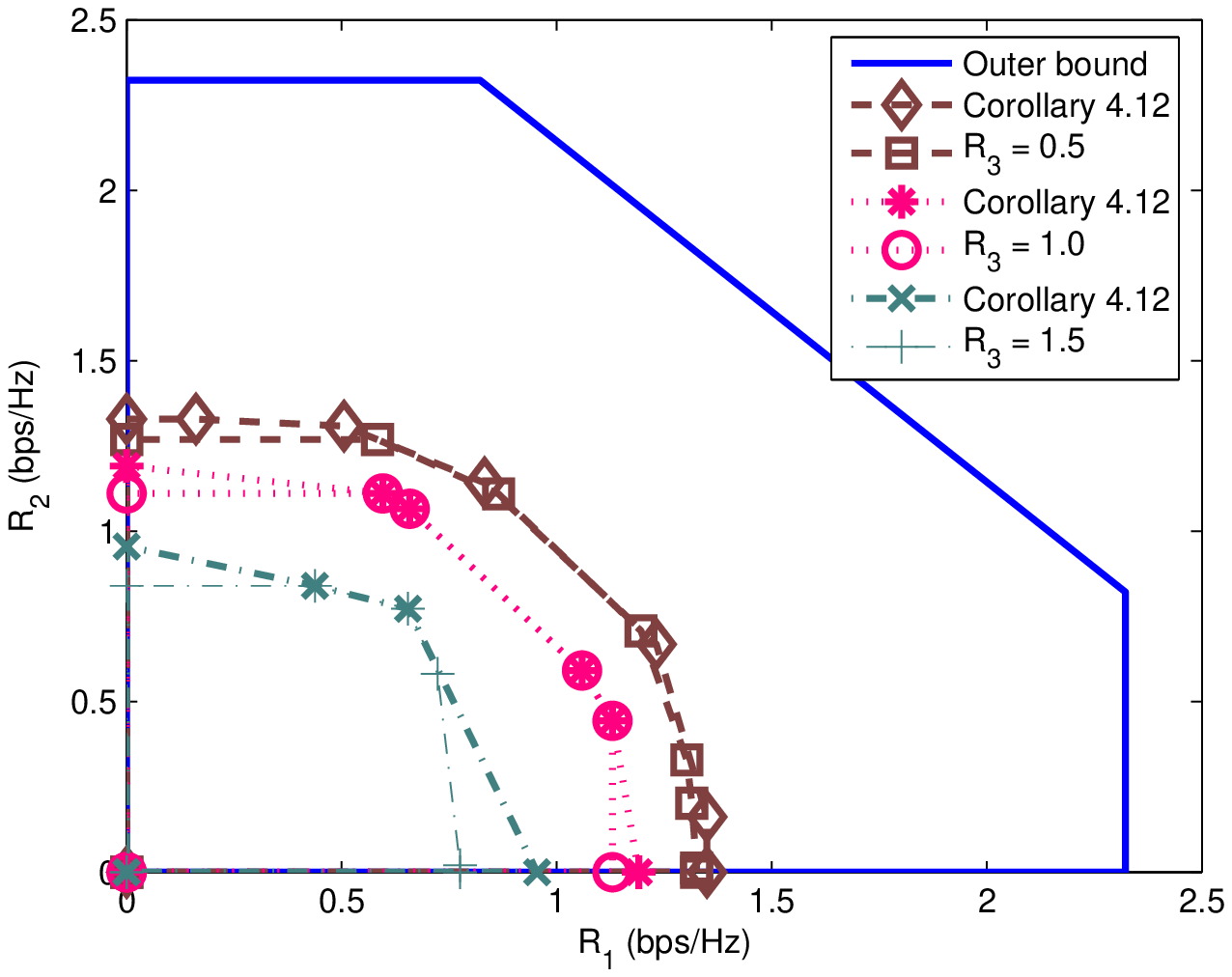}
\caption{Rate of $\mathcal{S}_1$ ($R_{1}$) versus the rate of $\mathcal{S}_2$ ($R_{2}$) when $\mathcal{S}_3$ is guaranteed to achieve a minimum rate $R_{3} = 0.5, 1 \text{ and } 1.5$ bps/Hz for the channel $\mathcal{C}_\text{CoMS}^2$. The power at the transmitters is 10dB. }\label{fig:fig15}
\end{figure}

\end{document}